\documentclass[12pt]{article}
\usepackage{amsfonts,amsthm,amsmath,amssymb,upgreek,bm}
\usepackage{dsfont}
\usepackage[paper=letterpaper,margin=.85in]{geometry}
\usepackage{graphicx}
\usepackage{units}
\usepackage{float}
\usepackage{wrapfig}
\usepackage[export]{adjustbox}
\usepackage{bbold}
\usepackage[T1]{fontenc}
\usepackage[shortlabels]{enumitem}

\usepackage{color}
\usepackage[dvipsnames]{xcolor}
\definecolor{darkblue}{rgb}{0.1,0.1,.7}
\definecolor{purple}{rgb}{0.6,0,0.6}
\definecolor{orange}{rgb}{0.9,0.6,0}
\usepackage[colorlinks, linkcolor=darkblue, citecolor=darkblue, urlcolor=darkblue, linktocpage]{hyperref} 
\usepackage[square, comma, sort&compress, numbers]{natbib}
\usepackage{physics}
\usepackage{tensor}
\usepackage{psfrag}
\usepackage{footmisc}
\usepackage{url}
\usepackage{mathtools}
\usepackage{tikz}
    \usepackage{amssymb,amsfonts,amsmath}
    \usepackage{tkz-euclide}
        \usetikzlibrary{arrows,calc,patterns}
\usepackage{pgfplots}
\pgfplotsset{compat=1.18}
\usepackage[]{amsmath}
\usepackage[]{graphicx}
\usepackage[]{latexsym}
\usepackage[utf8]{inputenc}
\usepackage{geometry}
\usepackage{amscd}
\usepackage[all,cmtip]{xy}
\usepackage{mathrsfs}
\usepackage{hyperref}
\usepackage{cleveref}

\usepackage[most]{tcolorbox}
\usepackage{tikz}
\usepackage{subcaption}

\numberwithin{equation}{section}
\newcommand{\be}{\begin{equation}}
\newcommand{\ee}{\end{equation}}

\usetikzlibrary{arrows,positioning}

\def\O{{\cal O}}

\def\SL2{\widetilde{SL}(2,\mathbb R)}

\newcommand\lpl{\ell_\text{Pl}}

\numberwithin{equation}{section}

\newcommand {\bes} {\begin {equation*}}
\newcommand {\ees} {\end {equation*}}
\newcommand {\beq} {\begin {equation}}
\newcommand {\eeq} {\end {equation}}
\newcommand {\bea} {\begin {eqnarray}}
\newcommand {\ea} {\end {eqnarray}}
\newcommand {\eea} {\end {eqnarray}}

\newcommand{\cA}{\mathcal{A}}
\newcommand{\cC}{\mathcal{C}}
\newcommand{\cD}{\mathcal{D}}

\newcommand{\cH}{\mathcal{H}}

\newcommand{\cQ}{\mathcal{Q}}
\newcommand{\cS}{\mathcal{S}}

\renewcommand{\dd}{\mathrm{d}}

\numberwithin{equation}{section}

\renewcommand{\i}{{\rm i}}

\newcommand{\CN}{\mathcal{N}}

\def\<{\langle}
\def\>{\rangle}

\tikzset{
    >=stealth',
    punkt/.style={
           rectangle,
           rounded corners,
           draw=black, very thick,
           text width=15em,
           minimum height=2em,
           text centered},
    pil/.style={
           ->,
           thick,
           shorten <=2pt,
           shorten >=2pt,}
}

\def\ie{\begin{equation}\begin{aligned}}
\def\fe{\end{aligned}\end{equation}}

\begin{document}
\thispagestyle{empty}
\vspace*{2.5cm}
\begin{center}

{\LARGE\bfseries
   An Argument for the Weak Gravity Conjecture from Constrained Wormholes
}
\begin{center} 
\vspace{1cm}
 {Gabriele Di Ubaldo$^{1,2}$, Luca V. Iliesiu$^{1}$, Dawid Maskalaniec$^{1}$}\\
\end{center}
\bigskip 
${}^1$ Leinweber Institute for Theoretical Physics and Department of Physics, \\
University of California, Berkeley, CA 94720, USA\\[0.15cm]
${}^2$ RIKEN iTHEMS Center for Interdisciplinary Theoretical and Mathematical Sciences,\\ 2-1 Hirosawa, Wako, Saitama 351-0198 Japan

\vspace{1.5cm}
{\bf Abstract}

\end{center}

The gravitational path integral (GPI) computes inner products between states of open and closed universes in any low-energy effective theory. A well-defined Hilbert space requires these inner products to be positive semidefinite, but this positivity is not manifest in the effective theory. In this paper, we study the constraints that positivity imposes on Einstein--Maxwell theory coupled to charged matter in asymptotically flat space in $D\geq 4$ dimensions. We start by constructing a new class of constrained instantons that compute multi-boundary wormhole amplitudes. We show that, under mild assumptions, if no charged particle satisfies the weak gravity conjecture (WGC), these amplitudes violate positivity of the GPI. If instead some particle satisfies the WGC, the wormholes develop an instability that allows positivity to be restored. The same argument shows that positivity can only be restored if the magnetic WGC holds and the convex hull condition applies to theories with several $\mathrm{U}(1)$ gauge fields.

\setcounter{page}{0}
\setcounter{tocdepth}{2}
\setcounter{footnote}{0}
\newpage

\newpage
\tableofcontents

\section{Introduction}\label{sec:intro}

The gravitational path integral (GPI) can compute inner products between different states of open and closed universes. To have a well-defined Hilbert space, every state must have a non-negative norm, which requires that inner products be positive semidefinite.  This requirement is not manifest when the path integral is evaluated in a low-energy gravitational effective field theory (EFT).  Positivity of the GPI can therefore constrain which gravitational EFTs admit a consistent quantum completion. 

As a first application of these constraints, positivity of the GPI \cite{DiUbaldo:2026rly,DiUbaldo:2026tkx} was shown to imply the axionic weak gravity conjecture (WGC) \cite{Arkani-Hamed:2006emk,Rudelius:2015xta,Brown:2015iha,Heidenreich:2015wga,Reece:2023NoGlobalSymmetries,Arvanitaki:2010StringAxiverse,Svrcek:2006AxionsStringTheory,Cicoli:2012IIBAxiverse,Montero:2015ofa}  in theories of axions coupled to gravity. The inner product between closed universes with different values for the axion was computed via semiclassical wormhole saddles.  The axion wormhole amplitudes, however, were shown \cite{DiUbaldo:2026rly,Maldacena:2026jqd} to diverge at a critical imaginary value of the axion difference, leading to the Imaginary Distance Bound.\footnote{Further relevant work on the Imaginary Distance Bound includes \cite{Licciardello:2026roi,Mallik:2026man,Kehagias:2026hca,Lin:2026ztf,Hallsten:2026ptj,Chandra:2026ynq}.} This divergence was shown to violate positivity of the GPI, and requiring that instanton effects restore positivity yielded the axionic weak gravity bound with precise coefficients. It is then reasonable to expect that, when other matter fields are coupled to gravity, positivity of the GPI implies additional bounds on EFTs, both new and previously conjectured. 

In this paper, we consider coupling an electromagnetic field to gravity and show, under mild assumptions, that positivity of the GPI in asymptotically flat space implies the existence of charged particles that satisfy the WGC \cite{Arkani-Hamed:2006emk}\footnote{See \cite{Harlow:2022ich} for a review and \cite{Kats:2006xp,Cheung:2014vva,Heidenreich:2015nta,Harlow:2015lma,Heidenreich:2016aqi,Montero:2016tif,Cheung:2018cwt,Hamada:2018dde,Heidenreich:2019zkl,Palti:2019pca,Arkani-Hamed:2021ajd} for a partial sample of relevant previous work on the WGC.}:
\begin{equation}
 \frac{|q|}{m}\geq \sqrt{\frac{8\pi(D-3)G_D}{D-2}}.
 \label{eq:intro-wgc}
\end{equation}
Here $m$ and $q$ are the particle mass and charge, and $G_D$ is Newton's constant in $D\geq4$ spacetime dimensions with zero cosmological constant. The bound requires at least one particle to have a charge-to-mass ratio no smaller than that of an extremal Reissner--Nordstr\"om black hole. Using the same argument in $D=4$, we also show that positivity implies  the magnetic and dyonic WGC: 
\begin{align}
 \frac{|q_m|}{m}&\geq\sqrt{4\pi G_4},
 \label{eq:intro-magnetic-wgc}\\
 \frac{\sqrt{q_e^2+q_m^2}}{m}&\geq\sqrt{4\pi G_4},
 \label{eq:intro-dyonic-wgc}
\end{align}
where $q_e$ and $q_m$ are the electric and magnetic charges.

Considering several $\mathrm{U}(1)$ gauge fields yields the convex hull condition \cite{Cheung:2014vva}, which states that  the convex hull of particle and antiparticle charge-to-mass vectors, normalized so that extremal black holes have unit length, must contain the unit ball:
\begin{equation}
 B_N\subseteq\operatorname{Conv}\{\pm\vec z_i\},
 \qquad
 \vec z_i=\sqrt{\frac{D-2}{8\pi(D-3)G_D}}\,
 \frac{\vec q_i}{m_i}.
 \label{eq:intro-convex-hull}
\end{equation}
Here $i$ labels particle species, $B_N$ is the unit ball in charge space, and antiparticles supply the vectors $-\vec z_i$. Equivalently, for every unit vector $\hat n$, some species satisfies $|\hat n\cdot\vec z_i|\geq1$. In four dimensions, including both electric and magnetic charges gives the analogous dyonic condition in the combined $2N$-dimensional charge space.

Let us give an outline of the derivation.  Consider the GPI with $n$ asymptotic boundaries of topology $S^1\times S^{D-2}$, with inverse temperature $\beta$ and electric potential $\Phi$ fixed at each boundary. The connected $n$-boundary amplitudes with identical boundary data are denoted by $\kappa_n$. Positivity of the open- and closed-universe inner products gives an infinite set of inequalities \cite{DiUbaldo:2026rly,DiUbaldo:2026tkx} on $\kappa_n$, the first few being:
\begin{equation}
 \kappa_1\geq0,\qquad \kappa_2\geq0,\qquad
 \kappa_1\kappa_3\geq\kappa_2\bigl(\kappa_2-\kappa_1^2\bigr).
 \label{eq:intro-positivity}
\end{equation}
These are the inequalities satisfied by the cumulants of a non-negative random variable in the (Stieltjes) moment problem: positivity implies that the $n$-boundary amplitudes $\overline{Z^n}$ are the moments of a random variable $Z\geq0$, and the $\kappa_n$ are its cumulants. 
The third inequality above states that the two-boundary wormhole amplitude cannot grow independently of the one- and three-boundary amplitudes. In particular, if $\kappa_2$ grows much faster than $\kappa_1^2$, the inequality requires $\kappa_3\gtrsim\kappa_2^2/\kappa_1$. Conversely, if $\kappa_1\kappa_3\ll\kappa_2^2$, the growth of the two-boundary amplitude $\kappa_2$ violates positivity. This is the mechanism used in \cite{DiUbaldo:2026rly}, where the axion wormhole amplitudes were shown to diverge, violating this inequality. We will show that the same inequality is violated by wormholes in Einstein--Maxwell theory. 

We compute two-boundary wormhole amplitudes $\kappa_2$ in asymptotically flat and AdS spacetimes in $D\geq 4$ dimensions. Since there are no saddles with the required boundary conditions, we compute $\kappa_2$ by using constrained instantons. These are saddles of the path integral in which one mode is held fixed by a constraint and is integrated over only at the end.
We construct a new family of fixed-charge constrained wormholes that satisfy the required boundary conditions. They are obtained by joining two Reissner--Nordstr\"om black hole exteriors across a throat of fixed area.
We clarify various general aspects of gravitational constrained instantons, such as how to correctly analyze their stability. We then establish the stability of our wormholes against spherically symmetric perturbations. We also perform an independent stability analysis in the minisuperspace approximation for the path integral over static, spherically symmetric geometries, confirming the stability of the asymptotically flat wormholes. We similarly construct families of constrained multi-boundary wormholes with more than two boundaries, obtained by gluing Reissner--Nordstr\"om exteriors to a core AdS$_2\times S^{D-2}$ geometry with multiple fixed-area boundaries.\footnote{As far as we know, these are the first constructions of Euclidean wormholes with more than two boundaries in $D\geq 4$, and they allow us to compute the amplitudes $\kappa_n$.}

To compute the amplitude at fixed electric potential $\Phi$, we sum over fixed charge wormhole amplitudes $Z_{\text{2-bdry}}(\beta,Q)$:
\begin{equation}
 \kappa_2(\beta,\Phi)=
 \sum_{Q\in e\mathbb Z}
 e^{2\beta\Phi Q} Z_{\text{2-bdry}}(\beta, Q).
 \label{eq:intro-charge-sum}
\end{equation}
In asymptotically flat space we find, after a careful analysis, that as $|\Phi|$ exceeds the critical value 
\begin{equation}
     \Phi_{\rm ext}=\sqrt{\frac{D-2}{8\pi(D-3)G_D}},
\end{equation}
the second cumulant $\kappa_2$ is enhanced as $\beta/\lpl\to \infty$ as
\begin{equation}
    \log \kappa_2 \sim \left(\frac{\beta}{\lpl}\right)^{\frac{(D-1)(D-2)}{2}},
\end{equation}
where $\lpl = (G_D)^{1/(D-2)}$ is the Planck length. By contrast, the single-boundary contribution $\kappa_1$ grows as $\log\kappa_1 \sim \left(\beta/\lpl\right)^{D-2} \ll \log\kappa_2$. We compute the replica-symmetric contribution to $\kappa_3$ and argue that the full three-boundary wormhole $\kappa_3$ grows only as fast as $\kappa_2$. Consequently, these amplitudes violate GPI positivity in flat-space Einstein--Maxwell theory.\footnote{In contrast to flat space, in AdS $\kappa_2$ scales in the same way as $\kappa_1$ in the $\beta/\lpl \to \infty$ limit, and the positivity constraint is not violated.}

Restoring positivity requires the two-boundary wormhole to develop a non-perturbative instability before $|\Phi|$ exceeds the critical value $\Phi_{\rm ext}$. Charged particles can trigger such an instability. A particle whose worldline wraps the Euclidean time circle at asymptotic infinity has an action
\begin{equation}
 I_p(r\to\infty)=\beta(m-q\Phi).
 \label{eq:intro-particle-action}
\end{equation}
For $q\Phi>m$, adding such particles lowers the action. The same effect occurs on the one-boundary black hole and three-boundary constrained wormhole backgrounds. A large number of particles can then backreact on these backgrounds and modify all three cumulants, changing the large-charge behavior responsible for the violation of the inner product positivity.

If no particle in the theory satisfies the charge-to-mass ratio  bound \eqref{eq:intro-wgc}, this instability is absent  and cannot resolve the problem. Conservatively requiring the instability to set in no later than the critical potential implies the WGC bound. Analogously, in the magnetically charged case the same argument implies the magnetic WGC (\ref{eq:intro-magnetic-wgc}) and, for multiple electric and magnetic charges, the convex hull condition (\ref{eq:intro-convex-hull}).

Along the way, we develop the general theory of constrained instantons in path integrals, clarifying how to analyze their stability. We show that stability must be tested only on fluctuations that preserve the constraint, and that the Lagrange-multiplier contribution must be included. The constraint values should then be integrated over with a measure that includes the Jacobians and fluctuation determinants. We test and exemplify the theory by using constrained instantons in three quantum systems: a free non-relativistic particle, the double well and massive $\phi^4$ theory in two dimensions. We recover the exact free-particle propagator using constrained instantons. In the double well, it is well known that the instanton--anti-instanton contribution cannot be obtained from a saddle-point approximation. We construct a constrained instanton that precisely reproduces this contribution. We also construct stable constrained saddles in massive two-dimensional $\phi^4$ theory at positive coupling, where no nontrivial real  saddle exists, and use them to compute the leading large-order behavior of the vacuum-energy perturbation series, which can be checked against known results. These examples test the constrained-instanton prescription in controlled settings before we apply it to gravitational wormholes.

Additionally, as a separate result, when calculating $\kappa_1$, we perform a Picard--Lefschetz analysis to determine the saddles that contribute to the path integral, possibly allowing for a conical singularity at the horizon.\footnote{We did such an analysis in order to ensure that the calculation of $\kappa_1$ is reliable.} Then, the single-boundary GPI becomes an integral over the horizon area~\cite{Carlip:1993sa,Marolf2022thermal}. In AdS, this integral converges, and its saddles were analyzed in~\cite{AMN2026,KM2026}. In flat space, however, the entropy grows faster than the energy at large horizon areas, and the integral naively diverges. Taking the flat space limit of the AdS integral at complex inverse temperature nevertheless defines a convergent contour. We determine its saddles using a Picard--Lefschetz analysis. At fixed charge, we find that the thermodynamically unstable large black holes do not contribute, leaving only the contribution of stable small black holes.

The remainder of this paper is organized as follows. Section~\ref{sec:constr-instantons} develops the constrained-instanton prescription, explains the appropriate stability test and integration measure, and illustrates them with a free particle, a double well and massive two-dimensional $\phi^4$ theory. Section~\ref{sec:grav-constr-saddles} constructs single-boundary saddles and two- and multi-boundary wormholes by fixing internal surface areas. Section~\ref{sec:stability} uses dimensional reduction on $S^{D-2}$ to analyze constraint-preserving fluctuations in the spherically symmetric sector, treating the two-boundary neck and multi-boundary cores separately. Section~\ref{sec:wgc-proof} evaluates the actions, the constrained area integrals and charge sums. It proceeds to test positivity inequalities to derive the electric WGC and subsequently extends the argument to magnetic and dyonic charges and to the convex hull condition for several gauge fields. Section~\ref{sec:discussion} provides a discussion of the results and directions for further work. Appendix~\ref{app:scalar-ci} provides a more detailed exposition of constrained instanton calculations in $\phi^4$ theory. Appendix \ref{app:cotler-jensen} provides a comparison with the constrained instanton wormholes of Cotler and Jensen. Appendix \ref{app:three-boundary-core} provides details on the  three-boundary wormhole construction. Appendix~\ref{app:timelike-reduction} describes the dimensional reduction along the time circle of the two-boundary geometries to constrained axio-dilaton wormholes.  Appendix~\ref{app:minisuperspace} develops the minisuperspace path integral for the two-boundary geometries and verifies the stability of the asymptotically flat saddles. Finally, in Appendix~\ref{app:picard-lefschetz}, we perform the Picard--Lefschetz analysis for asymptotically flat single-boundary saddles in the canonical and the grand canonical ensemble.

\textbf{Related work:} During the final stages of the preparation of this manuscript we became aware of related work \cite{Herderschee:2026mwgc} which uses the same positivity constraints on the gravitational path integral with finite IR cutoff to argue for the magnetic WGC.

\section{Constrained instantons}\label{sec:constr-instantons}
In this section, we review the formalism of constrained instantons, originally formulated in \cite{Affleck:1980mp} (see also reviews \cite{Dorey:2002ik,Bianchi:2007ft}). The method has been particularly useful for studying the Coulomb branch of $\mathcal N=2$ gauge theories, where constrained-instanton calculations reproduce the one- and two-instanton contributions to the Seiberg--Witten effective action and prepotential \cite{Dorey:2002ik,Bellisai:2000bc,Hollowood:2002ds}. More recently, Cotler and Jensen used this approach to construct constrained saddles of a gravitational path integral that describe two-boundary AdS wormholes and contribute to black-hole spectral statistics \cite{Cotler:2020lxj,Cotler:2021cqa,Cotler:2022rud}.

In all of these applications, the relevant configurations are not saddles of the full path integral, so the usual semiclassical expansion around a saddle does not apply directly. The constrained-instanton method handles such cases systematically: one fixes path integration variables $\zeta$ for which the saddle-point approximation does not apply, finds saddle-point values $\varphi_\star (\zeta)$ for the remaining modes $\varphi$ at each of its values, and performs the exact integral over the problematic modes $\zeta$ at the end. Denoting all the field modes in the theory as $\phi = (\zeta,\varphi)$, we can formally write
\begin{equation}
    Z
    =
    \int D\phi\,e^{-S[\phi]}
    \simeq
    \int \dd^n\zeta\,
    \exp\left[-S\bigl(\zeta,\varphi_\star(\zeta)\bigr)\right].
\end{equation}
If the number $n$ of problematic modes is reasonably small, it is relatively easy to calculate the integral over $\zeta$ without the saddle-point approximation. In the above equation we suppressed the loop corrections to the saddle-point integration over $\varphi$.

It is convenient to regard this choice of the order of integration as the insertion of a resolution of the identity into the path integral. We identify the problematic modes with real functionals $\mathcal{C}_i[\phi]$ of the fields and insert
\begin{equation}
    1
    =
    \int \dd^n\zeta
    \int_{c+\i\mathbb{R}}
    \frac{\dd^n\lambda}{(2\pi \i)^n}\,
    \exp\left[
        -\lambda^i\bigl(\mathcal{C}_i[\phi]-\zeta_i\bigr)
    \right]
\end{equation}
in the partition function. Formally, this gives us
\begin{equation}
    Z
    =
    \int \dd^n\zeta
    \int_{c+\i\mathbb{R}}
    \frac{\dd^n\lambda}{(2\pi \i)^n}
    \int D\phi\,
    \exp\left[
        -S[\phi]
        -\lambda^i\bigl(\mathcal{C}_i[\phi]-\zeta_i\bigr)
    \right].
\end{equation}
With a suitable choice of the constraints $\mathcal{C}_i[\phi]$, the integrals over $\phi$ and $\lambda$ can be evaluated by the saddle-point approximation. The saddle points are determined by equations of motion modified by the constraints:
\begin{equation}
    \frac{\delta S}{\delta\phi}
    +
    \lambda^i\frac{\delta\mathcal{C}_i}{\delta\phi}
    =
    0,
    \qquad
    \mathcal{C}_i[\phi]
    =
    \zeta_i.
\end{equation}
The saddle-point approximation is applicable because the problematic modes are held fixed at the prescribed values $\zeta_i$. This is an important point to keep in mind when analyzing stability of constrained saddles: one should investigate only fluctuations of the modes $\varphi$, rather than variations of the parameters $\zeta_i$. In terms of the constraint functionals $\mathcal{C}_i[\phi]$, this means that only fluctuations that preserve the constraints need to be included in the stability analysis.

This conclusion can be seen directly by expanding around saddle-point values $\phi = \phi_\star(\zeta)$, $\lambda = \lambda_\star(\zeta)$,
\begin{equation}
    \lambda^i=\lambda_\star^i+\xi^i,
    \qquad
    \phi=\phi_\star+\chi.
\end{equation}
The quadratic terms in $S[\phi]+\lambda^i\bigl(\mathcal{C}_i[\phi]-\zeta_i\bigr)$ are
\begin{align}
    I^{(2)}
    ={}&
    \frac{1}{2}
    \int \dd^d x\,\dd^d y\,
    \chi(x)
    \frac{
        \delta^2(S[\phi]+\lambda_\star^i\mathcal{C}_i[\phi])
    }{
        \delta\phi(x)\,\delta\phi(y)
    }
    \chi(y) +
    \xi^i
    \int \dd^d x\,
    \frac{\delta\mathcal{C}_i}{\delta\phi(x)}
    \chi(x).
\end{align}
Integrating over the fluctuations $\xi^i$ along the imaginary axes imposes
\begin{equation}
    \int \dd^d x\,
    \frac{\delta\mathcal{C}_i}{\delta\phi(x)}
    \chi(x)
    =
    0.
\end{equation}
The integration over $\xi^i$ therefore removes precisely those fluctuations that change the values of the constraints. Consequently, it is sufficient to verify that the quadratic form
\begin{equation}
    \frac{1}{2}
    \int \dd^d x\,\dd^d y\,
    \chi(x)
    \frac{
        \delta^2 (S[\phi]+\lambda_\star^i\mathcal{C}_i[\phi] )
    }{
        \delta\phi(x)\,\delta\phi(y)
    }
    \chi(y)
\end{equation}
is positive definite on the subspace of constraint-preserving fluctuations.

Clearly, the constrained-instanton method can also be applied when the path integral does have an ordinary saddle point. In that case, there exists a saddle-point value $\zeta_\star$ such that $\frac{\partial}{\partial \zeta} S[\phi_\star(\zeta)]|_{\zeta = \zeta_\star} = 0.$ Since $\mathcal{C}[\phi_\star(\zeta)]=\zeta$, the modified equations of motion set the Lagrange multiplier saddle-point value to zero, $\lambda_\star = -\frac{\partial}{\partial \zeta}S[\phi_\star(\zeta)]|_{\zeta_\star} = 0$. Consequently, $\phi_\star(\zeta_\star)$ satisfies the classical equations of motion, $\frac{\delta S}{\delta\phi}|_{\phi = \phi_\star (\zeta_\star)} = 0$. Performing the exact integral over $\zeta$ takes into account quantum corrections coming from the modes $\zeta$ running in loops.


In the remainder of this section, we illustrate the mechanism with three examples. First, to test the formalism on an analytically solvable system, we calculate the free non-relativistic particle propagator using constrained instantons (Section~\ref{sec:free_nonrel_particle}).  Next, we consider a quantum particle in a double well potential (Section~\ref{sec:double-well}), and verify that the constrained instantons reproduce the known instanton-anti-instanton contributions to the partition function. Finally, we discuss massive $\phi^4$ theory in two dimensions (Section~\ref{sec:scalar-ci}) in its symmetric phase. There, we use constrained instantons to compute the leading large-order behavior of the perturbative expansion of the vacuum energy, reproducing the results of~\cite{CIMPR}.

\subsection{Free non-relativistic particle}\label{sec:free_nonrel_particle}
In this subsection, we analyze constrained saddles in the Euclidean path integral of a free non-relativistic particle. We show explicitly that their stability should be tested only against constraint-preserving fluctuations. We then demonstrate that the one-loop integral over these fluctuations reproduces the known exact expression for the free-particle propagator.

Consider the Euclidean path integral of a free non-relativistic particle propagating from point $x_i$ at time $t=0$ to $x_f$ at $t= T$:
\begin{equation}
    K(x_f,T;x_i,0) = \int_{x(0)=x_i}^{x(T)=x_f}\mathcal{D}x(t)\, e^{-S[x]/\hbar}, \qquad S[x] = \frac{m}{2}\int_0^T\dd t\,\dot{x}(t)^2. \label{eq:2nd_toy_model}
\end{equation}
It is well-known that
\begin{equation}
    K(x',t';x,t) = \sqrt{\frac{m}{2\pi\hbar(t'-t)}}\, \exp\left[ -\frac{m(x'-x)^2}{2\hbar(t'-t)} \right],\label{eq:non-rel-propagator}
\end{equation}
and our goal is to reproduce this result using the constrained instanton formalism.

We introduce the constraint
\begin{equation}
    \mathcal{C}[x]=x(t_0)^2,
\end{equation}
with fixed $0<t_0<T$, and decompose the path integral as
\begin{equation}
    K(x_f,T;x_i,0) = \int_{\mathbb{R}}\dd\zeta\, K_\zeta(x_f,T;x_i,0),
\end{equation}
where
\begin{equation}
    K_\zeta(x_f,T;x_i,0) = \int_{x(0)=x_i}^{x(T)=x_f}\mathcal{D}x(t)\, e^{-S[x]/\hbar} \delta(x(t_0)^2-\zeta). \label{eq:2nd_toy_model_kernel}
\end{equation}
The constrained kernel has support only for $\zeta\geq0$. For $\zeta>0$, we use the integral representation of the delta function to write
\begin{equation}
    K_\zeta = \int_{c+\i\mathbb{R}} \frac{\dd\lambda}{2\pi \i\hbar} \int\mathcal{D}x(t)\, \exp\left[ -\frac{1}{\hbar} \left( S[x]+\lambda\bigl(x(t_0)^2-\zeta\bigr) \right) \right].\label{eq:non_rel_particle_kernel}
\end{equation}

\begin{figure}[t!]
\centering
\includegraphics[width=0.45\linewidth]{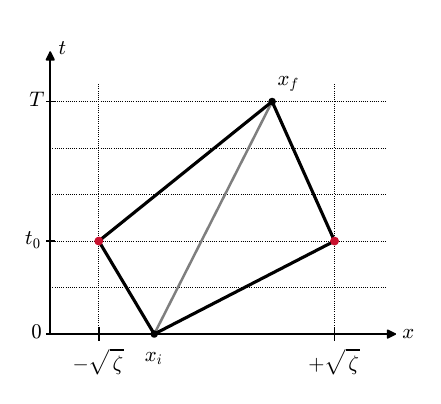}
\caption{Constrained saddle-point trajectories contributing to the kernel $K_{\zeta}$ in \eqref{eq:non_rel_particle_kernel}. For $\zeta > 0$, the constraint $x(t_0)^2 = \zeta$ gives rise to two saddles \eqref{eq:constr_saddle_trajectory} (black, solid lines). The particle moves with constant velocity from $x_i$ to $x = \pm \sqrt{\zeta}$, receives an instantaneous kick at $t=t_0$ (red dots) and then moves with constant velocity to $x_f$. The gray line represents the unconstrained classical path.}
\label{fig:free_particle}
\end{figure}

Equations of motion for the Lagrange multiplier $\lambda$ give $x(t_0)=\pm \sqrt{\zeta}$. We will denote $a = \pm \sqrt{\zeta}$. There are therefore two constrained saddles, corresponding to a particle moving with constant velocity $v_i = \frac{a - x_i}{t_0}$ from $x_i$ to $a$, experiencing an instantaneous force $F = 2\lambda a\,\delta(t-t_0)$ at the point $a$, and then moving with velocity $v_f = \frac{x_f - a}{T-t_0}$ to $x_f$ (see Fig.~\ref{fig:free_particle}),
\begin{equation}
    x_\star(t) = \begin{cases}x_i + \frac{a - x_i}{t_0}\, t\qquad \quad \;\,\text{for }0\leq t \leq t_0 \\
    a + \frac{x_f - a}{T-t_0}(t-t_0)\quad \text{for }t_0\leq t\leq T\end{cases}\label{eq:constr_saddle_trajectory}
\end{equation}
Change of the momentum of the particle at $t=t_0$ is $m(v_f - v_i) = 2a\lambda$, and thus, the saddle-point value of the Lagrange multiplier is
\begin{equation}
    \lambda_\star = \frac{m}{2a} \left(\frac{x_f-a}{T-t_0} - \frac{a-x_i}{t_0} \right).\label{eq:force_constr}
\end{equation}
The action evaluated on the constrained trajectory \eqref{eq:constr_saddle_trajectory} is:
\begin{equation}
    S_\star(a) = \frac{m}{2} \left[ \frac{(a-x_i)^2}{t_0} + \frac{(x_f-a)^2}{T-t_0} \right].\label{eq:pt_particle_os_action}
\end{equation}

Let us now analyze fluctuations around the constrained saddle,
\begin{equation}
    x(t)=x_\star(t)+\eta(t), \qquad \lambda=\lambda_\star +\xi,
\end{equation}
obeying Dirichlet boundary conditions
\begin{equation}
    \eta(t=0) = 0 = \eta(t=T).
\end{equation}
Expanding the action (together with the constraint) up to second order in fluctuations $\eta,\xi$, we obtain:
\begin{equation}
    S + \lambda(\mathcal C - a^2)= S_\star +\frac{m}{2}\int_0^T\dd t\,\dot{\eta}(t)^2 + \lambda_\star \eta(t_0)^2 + 2a\,\eta(t_0)\,\xi.
\end{equation}
To distinguish the mode which changes the constraint value, define the following function:
\begin{equation}
    h(t) = \begin{cases}
        \frac{t}{t_0}\qquad\,\text{for }t \leq t_0
        \\
        \frac{T-t}{T-t_0}\quad \text{for }t\geq t_0.
    \end{cases}
\end{equation}
Since $h(t)$ satisfies the boundary conditions $h(0) = h(T) = 0$, and it is nonzero at $t=t_0$, $h(t_0) = 1$, every fluctuation $\eta(t)$ can be decomposed as:
\begin{equation}
    \eta(t) = u\,h(t) + y(t),
\end{equation}
with $u = \eta(t_0)$ and $y(t_0) = y(0) = y(T) = 0$. Conveniently, modes $u$ and $y(t)$ are orthogonal with respect to the kinetic quadratic form, that is,
\begin{equation}
    \int_0^T\dd t\,\dot{h}(t)\dot{y}(t) = -\int_0^T\dd t\,\ddot h(t)\,y(t) \propto y(t_0) = 0,
\end{equation}
where in the second step we used the fact that the second derivative of $h(t)$ is proportional to Dirac delta function localized at $t=t_0$. Using
\begin{equation}
    \int_0^T \dd t\,\dot{h}(t)^2 = \frac{1}{t_0} + \frac{1}{T-t_0} = \frac{T}{t_0 (T-t_0)},
\end{equation}
we calculate:
\begin{equation}
    S + \lambda(\mathcal C- a^2) = S_\star + \frac{m}{2}\int_0^T\dd t\,\dot{y}(t)^2 + \left(\frac{mT}{2t_0(T-t_0)} + \lambda_\star\right) u^2 + 2a\,u\,\xi.
\end{equation}
All the infinitely many $y(t)$-modes are manifestly positive. 

Eventual instability could come from the $u$-mode. More explicitly, at fixed $\lambda = \lambda_\star$ given by \eqref{eq:force_constr}, the $u$-mode is unstable provided that:
\begin{equation}
    \frac{t_0 x_f + (T-t_0)x_i}{a} < 0.
\end{equation}
Since the constrained saddles have $a=\pm\sqrt{\zeta}$, there are two saddles with opposite signs of $a$, and for generic $t_0$, $T$, $x_f$ and $x_i$ one of them is unstable. However, this is not a true instability, as integration over $\xi$ sets the $u$-mode to zero, contributing
\begin{equation}
    \int_{\i\mathbb{R}}\frac{\dd \xi}{2\pi\i\hbar}\,e^{-2a\xi u/\hbar} = \frac{\delta(u)}{2|a|}.\label{eq:xi-integral}
\end{equation}
Indeed, we can explicitly check that to obtain the correct formula for \eqref{eq:non_rel_particle_kernel}, we have to include these saddles. To this end, note that because $y(t_0)=0$, the integral over the modes $y(t)$ splits into the integral over early-time fluctuations, $y(t< t_0)$ and late-time fluctuations $y(t > t_0)$. Upon the standard zeta-function regularization, they give us two determinants,
\begin{equation}
    D_{\rm early} = \sqrt{\frac{m}{2\pi \hbar\,t_0}},\qquad D_{\rm late} = \sqrt{\frac{m}{2\pi \hbar (T-t_0)}}.
\end{equation}
Combining these determinants with the classical action
\eqref{eq:pt_particle_os_action} and \eqref{eq:xi-integral}, the contribution from the saddle passing through $a$ is
\begin{equation}
    K_\zeta^{(a)} = \frac{1}{2|a|}\sqrt{\frac{m}{2\pi\hbar(T-t_0)}} \sqrt{\frac{m}{2\pi\hbar\,t_0}}\, e^{-S_\star(a)/\hbar} = \frac{1}{2|a|}K(x_f,T;a,t_0) K(a,t_0;x_i,0),
\end{equation}
with $K(x',t';x,t)$ given by \eqref{eq:non-rel-propagator}. Summing over the two constrained saddles gives
\begin{align}
    K_\zeta = \frac{1}{2\sqrt{\zeta}} \Big[K(x_f,T;\sqrt{\zeta},t_0) K(\sqrt{\zeta},t_0;x_i,0) + K(x_f,T;-\sqrt{\zeta},t_0) K(-\sqrt{\zeta},t_0;x_i,0) \Big].
\end{align}
Finally, we obtain the standard result by integrating over $\zeta$ and changing variables from $\zeta$ to $a=\pm\sqrt{\zeta}$,
\begin{align}
    \int_0^\infty\dd\zeta\,K_\zeta = \int_{-\infty}^{\infty}\dd a\, K(x_f,T;a,t_0) K(a,t_0;x_i,0) = K(x_f,T;x_i,0),
\end{align}
where the last equality follows from the composition law for the free-particle kernels.

\subsection{Particle in a double-well potential}\label{sec:double-well}
A particle in a double-well potential provides a less trivial application of the constrained-instanton formalism. It is well known that its partition function receives nonperturbative contributions from instantons, which interpolate between the two minima of the potential. Strictly speaking, however, instanton--anti-instanton pairs and higher multi-instanton configurations are not exact saddles of the path integral. A classical zero-energy trajectory takes an infinite time to travel from one minimum to the other, so it can never return to the first one. In this subsection, we show that such configurations are naturally described as constrained saddles. We construct an exact constrained instanton--anti-instanton pair, verify its stability against constraint-preserving fluctuations, and recover the leading two-instanton energy shift. This method was originally used in Appendix~C.3 of \cite{DWZinnJustinJentschuraI}.

Setting the particle mass and the frequency at either minimum to one, we consider
\begin{equation}
    S[x]=\int\dd t\left[\frac12\dot x^2+V(x)\right],\qquad V(x)=\frac18(x^2-1)^2.
    \label{eq:dw-model}
\end{equation}
The ordinary instanton $x_I(t)=\tanh[(t-t_0)/2]$ connects the two minima and has action $S_I=2/3$. It produces the leading splitting of the lowest even and odd levels, denoted by $E_{0,+}$ and $E_{0,-}$. Their center, $\bar E_0=(E_{0,+}+E_{0,-})/2$, is a particularly useful observable: odd-instanton contributions cancel, leaving a leading nonperturbative correction of order $e^{-2S_I/\hbar}$ \cite{DWZinnJustinJentschuraI,DWZinnJustinJentschuraII}.

To see explicitly that no exact instanton--anti-instanton saddle exists, note that a finite-action real solution with $x(\pm\infty)=1$ has conserved Euclidean energy
\begin{equation}
    \mathcal E=\frac12\dot x^2-V(x)=0.
    \label{eq:dw-zero-energy}
\end{equation}
A turning point requires $V(x)=0$, so the trajectory can reverse only at a minimum. Reaching $x=-1$ takes infinite time, since $\dot x$ vanishes linearly there. Consequently, there is no nonconstant real saddle at finite pair separation. The usual semiclassical treatment uses approximate superpositions of widely separated instantons and an integral over their separation. Constraints offer a way to construct an exact saddle before performing that remaining integral.

We fix a smooth measure of the time spent in the left well,
\begin{equation}
    \mathcal C[x]=\int_{\mathbb R}\dd t\,F(x(t))=R,
    \qquad F(x)=\frac14(x-1)^2(x+2).
    \label{eq:dw-constraint}
\end{equation}
On the interval between the wells, $F$ decreases from $F(-1)=1$ to $F(1)=0$, with $F'(\pm1)=0$. Thus, for a widely separated pair, $R$ approaches the time between its two transitions. The vanishing derivatives of $F$ at $x = \pm 1$ ensure that the constraint does not exert a force at either minimum. Inserting $1=\int_{\mathbb R}\dd R\,\delta(\mathcal C[x]-R)$, with $\delta(\mathcal C[x]-R)=\int_{c+\i\mathbb R}\frac{\dd\lambda}{2\pi\i\hbar}\,e^{-\lambda(\mathcal C[x]-R)/\hbar}$, into the path integral gives
\begin{equation}
\int\mathcal Dx\,e^{-S[x]/\hbar}
=\int_{\mathbb R}\dd R\int_{c+\i\mathbb R}\frac{\dd\lambda}{2\pi\i\hbar}\int\mathcal Dx\,
\exp\Bigl\{-\frac{1}{\hbar}\Bigl[S[x]+\lambda\bigl(\mathcal C[x]-R\bigr)\Bigr]\Bigr\}.
\label{eq:dw-constrained-path-integral}
\end{equation}
Its saddle points, with respect to both $x$ and $\lambda$, obey the modified equations of motion
\begin{equation}
    \ddot x=V'(x)+\lambda_\star F'(x),
    \qquad \mathcal C[x]=R.
    \label{eq:dw-constrained-eom}
\end{equation}
Writing $\lambda_\star=-\epsilon$, the trajectory is a zero-energy solution in the auxiliary potential $W_\epsilon=V-\epsilon F$. The left minimum is lowered, allowing a turning point before it is reached. For $0<\epsilon<2/3$, the resulting trajectory is
\begin{align}
    x_\epsilon(t)&=1-\frac{4w^2}{a+b\cosh[w(t-t_0)]},\nonumber\\
    a&=2-\epsilon,\qquad b=\sqrt{\epsilon(\epsilon+2)},
    \qquad w=\sqrt{1-\frac32\epsilon}.
    \label{eq:dw-constrained-pair}
\end{align}
It approaches $x=1$ at both ends and turns at $x=\epsilon-1+b$. The occupation time $R(\epsilon)$ decreases monotonically from infinity to zero, so this family supplies a constrained saddle for every $R>0$. At small $\epsilon$, the turning point approaches $x=-1$ and the two transitions become widely separated.

On the constrained saddle, the constraint is satisfied, $\mathcal C[x_\epsilon]=R$, so the term $\lambda(\mathcal C[x]-R)$ in the exponent of \eqref{eq:dw-constrained-path-integral} vanishes. The saddle therefore contributes with the weight $e^{-S_c(R)/\hbar}$, where $S_c(R)=S[x_\epsilon]$ is the original action evaluated on the constrained trajectory. Direct calculation gives
\begin{equation}
    R=\log\frac8\epsilon+O(\epsilon),\qquad
    \frac{\dd S_c}{\dd R}=\epsilon,\qquad
    S_c(R)=2S_I-8e^{-R}+O(e^{-2R}).
    \label{eq:dw-pair-action}
\end{equation}
The exact expressions are elementary functions of $\epsilon$.\footnote{With $\ell=\operatorname{arcosh}(a/b)$, one finds $R(\epsilon)=\bigl(2-\frac32\epsilon^2\bigr)\ell-3\epsilon w$ and $S_c=2w\bigl(\frac23-\epsilon^2\bigr)-\epsilon^3\ell$. Since $R'(\epsilon)=-3\epsilon\ell-2w(3\epsilon^2+6\epsilon+1)/[\epsilon(\epsilon+2)]<0$, $R$ decreases monotonically from infinity at $\epsilon\to0$ to zero at $\epsilon=2/3$.} The term $-8e^{-R}$ corresponds to an attractive instanton--anti-instanton interaction. As in the general discussion above, $\lambda_\star=-\dd S_c/\dd R=-\epsilon$, so the multiplier supplies the force needed to hold the pair apart against this attraction. It vanishes only as $R\to\infty$, so the unconstrained critical configuration is approached only at infinite separation~\cite{DWBehtashEtAl}.

Let us next examine fluctuations around the constrained saddle, as in the free-particle example. Writing $x=x_\epsilon+\eta$ and $\lambda=-\epsilon+\xi$, the quadratic terms in the exponent of \eqref{eq:dw-constrained-path-integral}, $S[x]+\lambda\bigl(\mathcal C[x]-R\bigr)$, are
\begin{equation}
    I^{(2)}=\frac12\langle\eta,\mathcal M_\epsilon\eta\rangle
      +\xi\langle c,\eta\rangle,
    \qquad \mathcal M_\epsilon=-\partial_t^2+V''(x_\epsilon)-\epsilon F''(x_\epsilon),
    \quad c=F'(x_\epsilon),
    \label{eq:dw-quadratic-main}
\end{equation}
where $\langle f,g\rangle=\int\dd t\,fg$; the linear terms vanish at the saddle. The operator $\mathcal M_\epsilon$ is the second variation of $S[x]+\lambda_\star\mathcal C[x]=S[x]-\epsilon\,\mathcal C[x]$. Integration over $\xi$ along the imaginary axis imposes $\langle c,\eta\rangle=0$. The operator $\mathcal M_\epsilon$ has precisely one negative eigenvalue: its translation zero mode $\dot x_\epsilon$ has one node, so by the Sturm oscillation theorem~\cite{Courant:1953} exactly one eigenvalue lies below zero. This negative direction is removed by the constraint. To see this, we separate the fluctuation that changes the value of the constraint from those that preserve it. Differentiating the saddle equation with respect to $\epsilon$ gives
\begin{equation}
\mathcal M_\epsilon\,\partial_\epsilon x_\epsilon=c,
\qquad \langle c,\partial_\epsilon x_\epsilon\rangle=\frac{\dd}{\dd\epsilon}\,\mathcal C[x_\epsilon]=R'(\epsilon)<0,
\label{eq:dw-susceptibility-main}
\end{equation}
so $\partial_\epsilon x_\epsilon$ changes the constraint. Any fluctuation can therefore be written as
\begin{equation}
\eta=\eta_c+u\,\partial_\epsilon x_\epsilon,\qquad
u=\frac{\langle c,\eta\rangle}{R'(\epsilon)},
\label{eq:dw-fluctuation-split}
\end{equation}
where $\langle c,\eta_c\rangle=0$, so that $\eta_c$ preserves the constraint to linear order. By \eqref{eq:dw-susceptibility-main}, the cross term $\langle\eta_c,\mathcal M_\epsilon\partial_\epsilon x_\epsilon\rangle=\langle\eta_c,c\rangle$ vanishes, and the quadratic terms \eqref{eq:dw-quadratic-main} become
\begin{equation}
I^{(2)}=\frac12\langle\eta_c,\mathcal M_\epsilon\eta_c\rangle+\frac12R'(\epsilon)\,u^2+R'(\epsilon)\,\xi u.
\label{eq:dw-quadratic-split}
\end{equation}
The multiplier fluctuation $\xi$ couples only to $u$, and its integral along the imaginary axis gives $\delta(u)/|R'(\epsilon)|$, which removes the constraint-changing direction. Since $\mathcal M_\epsilon$ has exactly one negative eigenvalue and $R'(\epsilon)<0$, it is positive-definite on the constraint-preserving modes orthogonal to the translation zero mode $\dot x_\epsilon$, which itself preserves the constraint since $\int\dd t\,F'(x_\epsilon)\dot x_\epsilon=0$; the constrained saddle is stable.

After integrating over $\xi$ and over the constraint-preserving fluctuations, two collective coordinates remain, namely the constraint value $R$ and the position $t_0$ of the instanton pair in \eqref{eq:dw-constrained-pair}. Constrained saddles with different $t_0$ are related by time translations, and the corresponding zero mode $\dot x_\epsilon$ of $\mathcal M_\epsilon$ is traded for an integral over $t_0$, with the standard Jacobian $\frac{\|\dot{x}_{\epsilon}\|}{\sqrt{2\pi \hbar}}$~\cite{DWColeman}, where $\|\dot{x}_\epsilon\|^2 := \langle \dot{x}_\epsilon, \dot{x}_\epsilon\rangle$. The constrained instanton--anti-instanton pairs therefore contribute to the path integral as
\begin{equation}
    \int\dd t_0\int\dd R\;\mu(R)\,e^{-S_c(R)/\hbar},
\label{eq:dw-pair-integral}
\end{equation}
where the measure $\mu(R)$ contains the Jacobian $\frac{\|\dot{x}_{\epsilon}\|}{\sqrt{2\pi \hbar}}$ and the Gaussian integral over the constraint-preserving fluctuations, including the factor from the $\xi$ integral. For a widely separated pair, the instanton and the anti-instanton are located at $t_{1,2}=t_0\mp R/2$, so that $\dd t_0\,\dd R=\dd t_1\,\dd t_2$, and the measure should reduce to the product of two one-instanton measures. Indeed, an explicit computation\footnote{Together with the translation Jacobian, the integrals over $\xi$ and over the fluctuations orthogonal to the zero mode give $\mu(R)=\frac{\|\dot x_\epsilon\|}{\sqrt{2\pi\hbar}}\bigl[2\pi\hbar\,D(\epsilon)R'(\epsilon)\bigr]^{-1/2}$, with $D=\det'\mathcal M_\epsilon/\det(-\partial_t^2+w^2)$. The reference operator describes fluctuations around the minimum of $W_\epsilon$ at $x=1$ and tends to $\mathcal M_{\rm vac}$ as $\epsilon\to0$. Here $D<0$ and $R'<0$, so their product is positive, as it must be for stable constrained fluctuations. The Gel'fand--Yaglom formula expresses $D$ through the asymptotic behavior $\dot x_\epsilon\simeq A_\pm e^{\mp w(t-t_0)}$ of the zero mode at $t\to\pm\infty$, $D=\|\dot x_\epsilon\|^2/(2wA_+A_-)$~\cite{DWBehtashEtAl}, and \eqref{eq:dw-constrained-pair} gives $A_\pm=\pm8w^3/b$. The zero-mode norm cancels, and $\mu(R)=\frac{1}{2\pi\hbar}\bigl[128w^7/(b^2|R'(\epsilon)|)\bigr]^{1/2}\to\frac{4}{\pi\hbar}$ as $\epsilon\to0$, since $b^2\simeq2\epsilon$ and $R'\simeq-1/\epsilon$.} gives
\begin{equation}
\lim_{R\to\infty}\mu(R)\,e^{-2S_I/\hbar}=\kappa^2,\qquad
\kappa=\sqrt{\frac{S_I}{2\pi\hbar}}\left(\frac{\det'\mathcal M_I}{\det\mathcal M_{\rm vac}}\right)^{-1/2}e^{-S_I/\hbar}
=\frac{2}{\sqrt{\pi\hbar}}\,e^{-2/(3\hbar)}.
\label{eq:dw-fugacity-main}
\end{equation}
Here $\kappa$ is the one-instanton weight per unit time, or instanton fugacity~\cite{DWColeman,DWZinnJustinJentschuraI,Marino:2012zq}. The determinant ratio compares the fluctuation operator around the instanton, $\mathcal M_I=-\partial_t^2+V''(x_I)$, with its zero mode removed, to that around the vacuum, $\mathcal M_{\rm vac}=-\partial_t^2+1$; for the double well, $\det'\mathcal M_I/\det\mathcal M_{\rm vac}=1/12$.

The remaining integral over $R$ is not Gaussian. At large separation, the integrand in \eqref{eq:dw-pair-integral} behaves as $\kappa^2e^{8e^{-R}/\hbar}$, which grows as $R$ decreases, signaling that the instanton and the anti-instanton attract each other. At small $R$ the pair collapses into a small fluctuation around the minimum, which is already part of perturbation theory, so integrating the attractive interaction over all $R$ would double count perturbative configurations. The standard way to avoid this is the Bogomolny--Zinn-Justin prescription~\cite{DWBogomolny,DWZinnJustin1981}: one evaluates the integral for a repulsive interaction, i.e.\ with the sign of the interaction term reversed, and then continues the result back to the attractive case. This continuation is ambiguous: it can be performed through the upper or the lower half of the complex plane, and the two results differ by an imaginary part. As we show below, this imaginary part cancels the ambiguity in the Borel resummation of perturbation theory~\cite{DWBogomolny,DWZinnJustin1981,DWZinnJustinJentschuraI}.

To extract the energy levels, we consider the partition function $Z(T)=\operatorname{Tr}e^{-HT/\hbar}$, with $H=-\frac{\hbar^2}{2}\partial_x^2+V(x)$, given by the path integral over periodic trajectories on a Euclidean circle of circumference $T$. An instanton--anti-instanton pair divides the circle into two intervals, of lengths $R$ and $T-R$, spent near the left and the right minimum, respectively. The instanton and the anti-instanton attract each other across both intervals, and when both intervals are long, the action of the pair is
\begin{equation}
S_c(R;T)=2S_I-8e^{-R}-8e^{-(T-R)}+\cdots
\label{eq:dw-circle-action-main}
\end{equation}
This expression follows from the infinite-line result \eqref{eq:dw-pair-action} by including the interaction across the second interval; it is not an exact constrained solution on a circle of finite circumference. In terms of $z=R-T/2$ and $a_T=16e^{-T/2}/\hbar$, the $R$-dependent part of the integrand is $e^{-S_c(R;T)/\hbar}\propto e^{a_T\cosh z}$. For a repulsive interaction, $a_T<0$, the integral over $z$ converges and gives $2K_0(-a_T)$. Continuing to $a_T>0$ from above or below the real axis gives
\begin{align}
\mathcal J_\pm(a_T)
&=2K_0(-a_T\pm \i0)=2K_0(a_T)\mp2\pi \i I_0(a_T)\nonumber\\
&=T-2L_\pm+O(a_T^2|\log a_T|),
\qquad L_\pm=\log\frac8\hbar+\gamma\pm \i\pi,
\label{eq:dw-qzm-main}
\end{align}
where $K_0$ and $I_0$ are modified Bessel functions and $\gamma$ is the Euler–Mascheroni constant. Equivalently, $\mathcal J_\pm$ is the integral over the steepest-descent contour of the pair: its two arms, along the lines $\operatorname{Im}z=\pi$ and $\operatorname{Im}z=-\pi$, give $2K_0(a_T)$, while the vertical part of the contour gives the imaginary term~\cite{DWDersySchwartz}. The expansion in the second line requires $a_T\ll1$, i.e.\ $T\gg\log(1/\hbar)$, so that typical pairs are widely separated. Together with $\hbar T\ll1$, this defines the regime $\log(1/\hbar)\ll T\ll1/\hbar$ in which we extract the coefficients of the semiclassical expansion.

For $T\gg1$, the partition function is dominated by the two lowest levels,
\begin{equation}
Z_{\rm low}(T)=e^{-E_{0,+}T/\hbar}+e^{-E_{0,-}T/\hbar}
=2e^{-\bar E_0T/\hbar}\cosh\frac{\Delta E_0T}{2\hbar},
\qquad \Delta E_0=E_{0,-}-E_{0,+}.
\label{eq:dw-low-trace-main}
\end{equation}
In the semiclassical expansion, the leading term comes from trajectories that stay near one of the two minima, $Z_{\rm low}^{\rm pert}=2e^{-T\mathcal S_\pm E_0^{\rm pert}/\hbar}$. Here the factor of two counts the wells, $E_0^{\rm pert}$ is the perturbative ground-state energy of a single well, and $\mathcal S_\pm$ denotes its lateral Borel resummation. The leading correction, which we denote by $Z_{\rm low}^{(2),\pm}$, comes from trajectories with a single instanton--anti-instanton pair. They are parametrized by the collective coordinates of \eqref{eq:dw-pair-integral}: the time $R\in(0,T)$ spent in the left well and the center $t_0\in[0,T)$ of that time interval. Since the two wells have the same perturbative energy, fluctuations around them give the same factor $e^{-T\mathcal S_\pm E_0^{\rm pert}/\hbar}$ as in $Z_{\rm low}^{\rm pert}$. By \eqref{eq:dw-fugacity-main} and \eqref{eq:dw-circle-action-main}, the pair itself contributes $\mu(R)e^{-S_c(R;T)/\hbar}\simeq\kappa^2e^{a_T\cosh z}$, and since it does not depend on $t_0$, the $t_0$ integral gives a factor of $T$, while the $R$ integral gives $\mathcal J_\pm$. Hence $Z_{\rm low}^{(2),\pm}(T)=\kappa^2T\,\mathcal J_\pm\,e^{-T\mathcal S_\pm E_0^{\rm pert}/\hbar}$, and, relative to the perturbative contribution,
\begin{equation}
\frac{Z_{\rm low}^{(2),\pm}(T)}{Z_{\rm low}^{\rm pert}(T)}
=\frac{\kappa^2T}{2}\mathcal J_\pm
=\frac{\kappa^2T^2}{2}-\kappa^2TL_\pm+\cdots.
\label{eq:dw-trace-main}
\end{equation}
The term proportional to $T^2$ describes two independent tunneling events: it matches the expansion of $\cosh(\Delta E_0T/2\hbar)$ in \eqref{eq:dw-low-trace-main}, with the one-instanton splitting $\Delta E_0=2\hbar\kappa$. The connected term, linear in $T$, instead shifts the common energy $\bar E_0$ of the two levels by
\begin{equation}
\delta E_0^{(2),\pm}
=\hbar\kappa^2L_\pm
=\frac4\pi e^{-4/(3\hbar)}
\left[\log\frac8\hbar+\gamma\pm \i\pi
+O\!\left(\hbar\log\hbar\right)\right].
\label{eq:dw-neutral-energy}
\end{equation}

The imaginary part of \eqref{eq:dw-neutral-energy} has a natural interpretation. The perturbative series for $E_0^{\rm pert}$ is not Borel summable: its Borel transform has a singularity on the positive real axis, and the lateral Borel sums $\mathcal S_\pm E_0^{\rm pert}$, obtained by integrating above or below this singularity, differ by an imaginary part of order $e^{-2S_I/\hbar}$~\cite{DWZinnJustinJentschuraI,Marino:2012zq}. For the same choice of sign in $\mathcal S_\pm$ and $\mathcal J_\pm$, the imaginary part of \eqref{eq:dw-neutral-energy} cancels that of $\mathcal S_\pm E_0^{\rm pert}$, and
\begin{equation}
\bar E_0=\operatorname{Re}\mathcal S_\pm E_0^{\rm pert}+\operatorname{Re}\delta E_0^{(2),\pm}+\cdots
\end{equation}
is real and independent of the choice of sign. Equation~\eqref{eq:dw-neutral-energy} agrees with the two-instanton contribution to the energies of the lowest doublet found by Zinn-Justin and Jentschura~\cite{DWZinnJustinJentschuraI,DWZinnJustinJentschuraII}.\footnote{Their Hamiltonian is $H_{\rm ZJJ}=-\frac g2\partial_q^2+\frac{1}{2g}q^2(1-q)^2$; with $q=(x+1)/2$ and $g=\hbar/4$, one has $H=\hbar H_{\rm ZJJ}$.} Our treatment of the pair on the circle is also consistent with the exact periodic saddle of~\cite{DWDersySchwartz}, whose action $2S_I-16e^{-T/2}+\cdots$ coincides with \eqref{eq:dw-circle-action-main} at $R=T/2$. In that saddle, the separation of the pair is a negative mode of the fluctuation operator, whereas here it is the collective coordinate $R$. Finally, the cancellation of imaginary parts fixes $\operatorname{Im}\mathcal S_+E_0^{\rm pert}=-4e^{-4/(3\hbar)}$ and determines the large-order behavior of the perturbative coefficients: for $E_0^{\rm pert}=\hbar\sum_{K\geq0}a_K\hbar^K$,
\begin{equation}
a_K\sim-\frac{3^{K+1}}{4^K\pi}K!,
\label{eq:dw-large-order-main}
\end{equation}
in agreement with~\cite{DWZinnJustinJentschuraII}. The constrained-instanton calculation thus reproduces the leading instanton--anti-instanton contribution to the spectrum from exact constrained saddles and an explicit integral over their separation. Higher-loop corrections and observables at arbitrary $T$ would require further calculations.

\subsection{Massive \texorpdfstring{$\phi^4$}{phi4} theory in two dimensions}\label{sec:scalar-ci}
As a last example, we consider massive $\phi^4$ theory in two dimensions in its symmetric phase. It has no nontrivial real saddle at positive coupling. Nevertheless, as in the double-well example, we can find exact constrained saddles. Here, a suitable constraint fixes the integral over the interaction vertex $\int\dd^2y\,\phi^4$, and the resulting saddles determine the large-order terms in the perturbative expansion of the vacuum energy. To our knowledge, the constrained-saddle calculation below has not appeared in the literature before, even though constrained instantons have a long history in scalar field theory.\footnote{Affleck considered massive four-dimensional $\phi^4$ theory with a negative quartic interaction, where the mass term obstructs the scale-invariant instanton, and fixed its size to estimate large-order perturbative behavior \cite{Affleck:1980mp}. Earlier, Lipatov formulated the large-order saddle as a joint extremization over fields and coupling, explicitly interpreting it as a variational problem at fixed interaction integral \cite{CILipatov}. Nielsen and Nielsen constructed massive four-dimensional scalar constrained instantons through a small-mass expansion and examined the conditions for regularity and exponential decay \cite{CINielsen}; Glerfoss and Nielsen subsequently studied their one-loop renormalization, including renormalization of the constraint \cite{CIGlerfoss}. More recently, Elder, Gawrych and Rajantie formulated the constrained vacuum-decay problem in the same unstable massive theory and identified distinct solution branches using their constrained negative modes \cite{CIElder}. Aoki, Ibe and Shirai revisited the matching of the small- and large-distance expansions in four-dimensional scalar and Yang--Mills--Higgs theories, finding regular solutions for constraints previously thought to be problematic \cite{CIAoki}. These constructions are also closely related to the valley method \cite{CIAoyama}.} For brevity, we discuss here only the main results of the calculation and refer to Appendix~\ref{app:scalar-ci} for the detailed derivation.

By rescaling the Euclidean coordinates, which also rescales the quartic coupling, we can set the mass-squared parameter to one, so that the $\phi^4$-theory action reads
\begin{equation}
S_g[\phi]=\int\dd^2y\left[\frac12(\partial\phi)^2+\frac12\phi^2+g\,{:}\phi^4{:}\right],
\label{eq:scalar-model}
\end{equation}
with non-negative quartic coupling $g\geq0$. Normal ordering of the quartic interaction, ${:}\phi^4{:}$, with respect to the free propagator of unit mass amounts to the addition of mass and vacuum-energy counterterms, which are the only counterterms needed in two dimensions. To see that there is no nontrivial real saddle, multiply the classical field equation $(-\partial^2+1)\phi+4g\phi^3=0$ by $\phi$ and integrate by parts: any real finite-action solution satisfies $\int\dd^2y\,[(\partial\phi)^2+\phi^2+4g\phi^4]=0$. For $g>0$ the integrand is positive-definite, so the integral vanishes only if $\phi=0$ everywhere.

We want to calculate the vacuum energy density
\begin{equation}
f(g)=-\lim_{V\to\infty}\frac{1}{V}\log\frac{Z_g}{Z_0}=\sum_{k\geq2}f_kg^k,
\label{eq:scalar-vacuum-energy}
\end{equation}
where $Z_g=\int\mathcal D\phi\,e^{-S_g[\phi]}$ and $V$ is the Euclidean space volume. We constrain the path integral $Z_g$ by inserting $1=\int_{\mathbb R}\dd\zeta\,\delta(\mathcal C[\phi]-\zeta)$ with $\mathcal C[\phi]=\int\dd^2y\,{:}\phi^4{:}$. In the classical approximation normal ordering does not change the constraint, while at the quantum level it supplies the mass and vacuum-energy counterterms. Since the term $g \int\dd^2y \,{:}\phi^4{:}$ equals $g\zeta$ on the support of the delta function, we obtain the exact identity
\begin{equation}
\frac{Z_g}{Z_0}=\int_{\mathbb R}\dd\zeta\,e^{-g\zeta}\rho_0(\zeta),\qquad
\rho_0(\zeta)=\big\langle\delta(\mathcal C[\phi]-\zeta)\big\rangle_0 ,
\label{eq:scalar-density-identity}
\end{equation}
where $\rho_0$ is the probability density of $\mathcal C[\phi]$ in the free theory. The coefficients $f_k$ at large $k$ are therefore controlled by the tail $\zeta\to+\infty$ of $\rho_0$, which is dominated by constrained saddles.

Since the interaction term is fixed by the constraint, the constrained saddles extremize the free action $S_0=S_{g=0}$ at fixed $\int\dd^2y\,\phi^4=\zeta$, and do not depend on $g$:
\begin{equation}
\phi_\zeta(y)=\pm\left(\frac{\zeta}{I_4}\right)^{1/4}Q(y-y_0),\qquad
S_c(\zeta)=g\zeta+\frac12\sqrt{I_4\zeta}.
\label{eq:scalar-family}
\end{equation}
Here $Q$ is the positive, radially symmetric solution of $(-\partial^2+1)Q=Q^3$ that decays at infinity, and $I_n=\int\dd^2y\,Q^n$. The coupling only shifts the Lagrange multiplier, $\lambda_\star=-g-\frac14\sqrt{I_4/\zeta}$. In Appendix~\ref{app:scalar-ci}, we show that these saddles are stable and evaluate the Gaussian integral over fluctuations around them, which gives the tail
\begin{equation}
\frac{\rho_0(\zeta)}{ V}\simeq\frac{\rho_0^{\mathrm{CI}}(\zeta)}{ V}=\frac{I_4^{3/4}}{16\pi^{3/2}\sqrt{|\Delta_R|}}\,\zeta^{-1/4}e^{-\frac12\sqrt{I_4\zeta}},
\label{eq:scalar-density-tail}
\end{equation}
where $\Delta_R<0$ is the renormalized determinant of the fluctuation operator around $\phi_\zeta$ with the translation zero modes removed. Through \eqref{eq:scalar-density-identity}, this gives the leading large-order behavior of the vacuum energy,
\begin{equation}
f_k\simeq\frac{(-1)^{k+1}\,k!}{2\pi^2\sqrt{|\Delta_R|}}\Bigl(\frac{16}{I_4}\Bigr)^{k}.
\label{eq:scalar-large-order}
\end{equation}
This agrees with the leading large-order behavior found in~\cite{CIMPR}, where the result was obtained from Gaussian fluctuations around the instanton at negative coupling and a dispersion relation in $g$.

\section{Gravitational constrained saddles}\label{sec:grav-constr-saddles}
In this section, we use the constrained-instanton formalism to construct charged Euclidean wormholes with two or more asymptotic boundaries. Our construction is closely related to that of Cotler and Jensen, who obtained two-boundary wormholes by constraining the proper length of the wormhole \cite{Cotler:2020lxj,Cotler:2021cqa,Cotler:2022rud}. We impose a slightly different constraint, which is better suited for constructing multi-boundary wormholes. Rather than fixing a length between two wormhole boundaries, we fix the area of a codimension-one surface. The associated Lagrange multiplier acts as the tension of a brane localized on this surface, supporting the wormhole neck. Since this tension is nonzero, the wormholes are not saddles of the Einstein--Maxwell path integral itself. As in Section~\ref{sec:constr-instantons}, the constraint makes them saddles at each fixed value of the area, and the integral over the area is kept until the end.

We first formulate the GPI with surface area constraints. We then construct multi-boundary wormholes by gluing an arbitrary number of Reissner--Nordstr\"om--AdS (RN--AdS) exteriors to an $n$-boundary core of constant curvature. Finally, we construct two-boundary wormholes by gluing two RN--AdS exteriors directly across a common constant-radius surface. We also use a constraint on the area of a codimension-two surface to construct single-boundary saddles with a conical singularity at the horizon, whose area is fixed independently of the boundary temperature.

\subsection{Constraining surface areas}
Consider spacetime $M$ with $n$ boundaries, $\partial M = \Sigma_1 \sqcup \Sigma_2\sqcup \cdots \sqcup \Sigma_n$, and take a $(p+1)$-dimensional surface $\cS$, embedded in $M$ via $X:\cS\hookrightarrow M$. Let $g_{\mu\nu}$ be the metric on $M$ and $h_{ab}^{(i)}$ be the induced metric on the $i$-th boundary, $\Sigma_i$ (with $\iota_i:\Sigma_i\hookrightarrow M$, we have $h_{ab}^{(i)} = (\iota_i^*g)_{ab}$). We will denote the metric induced on $\cS$ as $q_{AB} = (X^*g)_{AB}$. The gravitational path integral in a sector with fixed topology of $M$, with Dirichlet boundary conditions $h^{(i)}$, can be formally written as:
\begin{equation}
    Z_M[h^{(i)}] = \int_{\iota_i^*g_{ab} = h^{(i)}_{ab}}Dg\,e^{-I[g]}.
\end{equation}
We can insert the following resolution of the identity in the above path integral:
\begin{equation}
    1 = \int \dd\zeta\,\delta(\text{vol}(\cS) - \zeta) = \int_{c+\i\mathbb{R}}\frac{\dd \lambda}{2\pi\i}\int \dd\zeta\,\exp[-\lambda(\text{vol}(\cS) - \zeta)],
\end{equation}
where $\text{vol}(\cS)$ is the volume of the surface $\cS$,
\begin{equation}
    \text{vol}(\cS) = \int_{\cS}\dd^{p+1}\sigma\sqrt{\det(q_{AB}(\sigma))}.
\end{equation}
Clearly this identity does not depend on our choice of the surface $\cS$, and thus, on its embedding $X$, even though $\text{vol}(\cS)$ does depend on $q = X^*g$. The GPI becomes:
\begin{equation}
    Z_M[h^{(i)}] = \int_{c+\i\mathbb{R}}\frac{\dd\lambda}{2\pi\i}\int \dd\zeta\,e^{\lambda \zeta}\,Z_M[h^{(i)};X,\lambda],\label{eq:gpi-area}
\end{equation}
where
\begin{equation}
    Z_M[h^{(i)};X,\lambda] = \int_{\iota_i^*g = h^{(i)}}Dg\,\exp[-I[g]- \lambda\int_{\cS}\dd^{p+1}\sigma\,\sqrt{\det(X^*g)}]\label{eq:gpi-constrained}
\end{equation}
is the path integral of a gravity theory coupled to a non-dynamical $p$-brane, with tension $\lambda$ and fixed shape given by $X$. We will refer to this brane as a ``constraint brane.''\footnote{This technique was implicitly used in~\cite{deBoer:2025tmh} to construct Lorentzian wormholes with conical singularities. In our language, these singularities are codimension-two constraint branes.} For $\lambda=0$, \eqref{eq:gpi-constrained} is the original path integral. For $\lambda\neq0$, its saddles include geometries that solve the original field equations only away from $\cS$. Integrating over $\lambda$ and $\zeta$ in \eqref{eq:gpi-area} then restores the original path integral, without assuming that it has a saddle. Of course, one can impose such constraints in any gravity theory coupled to matter.

\subsection{Multi-boundary wormholes}\label{sec:multi-bdry-saddles}

\begin{figure}[H]
\centering
\includegraphics[width=0.5\textwidth]{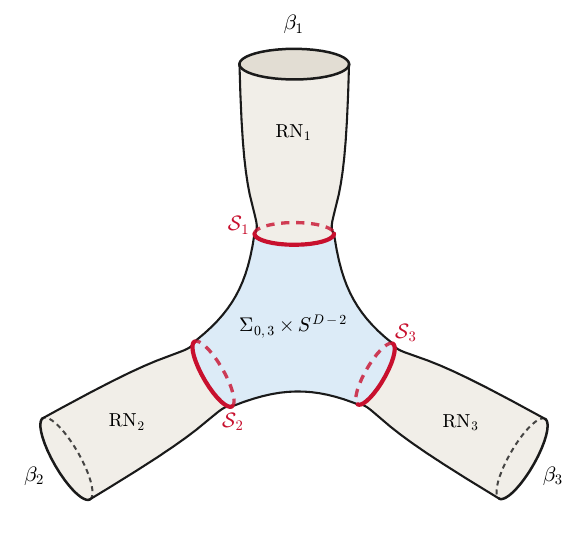}
\caption{Constrained three-boundary wormhole. A genus-zero hyperbolic core $\Sigma_{0,3}\times S^{D-2}$ is glued along three constraint branes $\mathcal{S}_i = \{r= R_0\}$ (red) to Reissner--Nordstr\"om exteriors RN${}_i$ with common flux $Q$ and mass parameters $M_i$, which end on asymptotic boundaries of inverse temperature $\beta_i$. Only the two-dimensional part of the geometry is drawn; the $S^{D-2}$ is suppressed, and the circles shown are Euclidean time circles. The branes have negative tension $\lambda_i$, given by \eqref{eq:lambda_nbdry_wormhole_Ddim}, which supports the jump in extrinsic curvature.}
\label{fig:3bdry-wormhole}
\end{figure}

Using the above-discussed constraints we can construct GPI saddles in Einstein--Maxwell theory with negative or zero cosmological constant by gluing spacetimes across surfaces with discontinuous extrinsic curvatures. The discontinuity is supported by the tension of the corresponding  constraint brane. In this way, we glue $n$ Reissner--Nordstr\"om exteriors across constant-radius surfaces to an $n$-boundary wormhole core of the form $M_{\rm core} = \Sigma_{g,n}\times S^{D-2}$, where $\Sigma_{g,n}$ is a hyperbolic surface of genus $g$ with $n$ boundaries, and $S^{D-2}$ has constant radius $R_0$ (see Fig.~\ref{fig:3bdry-wormhole}).

Geometry of the $i$-th exterior is
\begin{equation}
    \dd s_{(i)}^2 = \beta_i^2 f_i(r)\dd\tau^2 + \frac{\dd r^2}{f_i(r)} + r^2 \dd \Omega_{D-2}^2,\label{eq:RN-metric}
\end{equation}
where
\begin{equation}
    f_i(r) = \frac{r^2}{\ell^2} + 1 - \frac{2M_i}{r^{D-3}} + \frac{\mathcal{Q}_i^2}{r^{2(D-3)}}.\label{eq:RN-lapse-function}
\end{equation}
Here $\Omega_{D-2} = \frac{2 \pi^{(D-1)/2}}{\Gamma(\frac{D-1}{2})}$ is the volume of the unit $(D-2)$-dimensional sphere, $\ell$ is the AdS length, $\Lambda = -\frac{(D-1)(D-2)}{2\ell^2}$, $\mathcal{Q}_i = \sqrt{\frac{8\pi G_D}{(D-2)(D-3)}}\,\frac{Q_i}{\Omega_{D-2}}$, and $Q_i$ is the electric flux through the $S^{D-2}$. We take the coordinate $\tau$ to be periodic, with $\tau \sim \tau + 1$. Then, circumference of the Euclidean time circle at the asymptotic boundary of the $i$-th exterior is equal to $\beta_i$.\footnote{In asymptotically flat space, $f_i(r)\to1$, and the proper circumference of the time circle on a large cutoff surface approaches $\beta_i$. In asymptotically AdS space, the proper circumference grows as $\beta_i r/\ell$, and $\beta_i$ is its renormalized value, i.e., the inverse temperature of the boundary theory.} Electric field strength tensor in the $i$-th exterior is
\begin{equation}
    \frac{1}{2}F^{(i)}_{\mu\nu}\dd x^\mu \wedge \dd x^\nu = -\frac{\i Q_i}{\Omega_{D-2}\,r^{D-2}}\,\beta_i\dd \tau\wedge \dd r.
\end{equation}

We glue these solutions to an $n$-boundary core along constant-radius surfaces $\cS_i = \{r = r_0^{(i)} = \text{const.}\}$. We take the wormhole core metric to be
\begin{equation}
    \dd s_{\rm core}^2 = \hat{g}_{ab}\dd x^a \dd x^b + R_0^2\dd\Omega_{D-2}^2,\label{eq:core_metric}
\end{equation}
where $\hat{g}_{ab}$ is a metric on the hyperbolic surface $\Sigma_{g,n}$. The electric field strength two-form is given by
\begin{equation}
    \frac{1}{2}F^{(\rm core)}_{\mu\nu} \dd x^\mu \wedge \dd x^\nu = -\frac{\i Q}{\Omega_{D-2} R^{D-2}_0}\sqrt{\hat{g}}\,\dd^2 x,\label{eq:core_field_strength}
\end{equation}
where $\sqrt{\hat{g}}\,\dd^2 x$ is the volume form of $\Sigma_{g,n}$. Metric \eqref{eq:core_metric} satisfies the Einstein--Maxwell equations provided that $R_0$ and $Q$ obey
\begin{equation}
    \mathcal{Q}^2 = R_0^{2D-6} + \frac{D-1}{D-3}\frac{R_0^{2D-4}}{\ell^2},\label{eq:R(Q)_ext}
\end{equation}
where we defined $\mathcal{Q} = \sqrt{\frac{8\pi G_D}{(D-2)(D-3)}}\,\frac{Q}{\Omega_{D-2}}.$ Thus, $R_0$ is the horizon radius of an extremal RN--AdS black hole with charge $Q$. The Ricci tensor of $\Sigma_{g,n}$ is
\begin{equation}
    R_{ab} = -a^2 \hat{g}_{ab},\qquad a^2 = (D-2)(D-3)\frac{\mathcal{Q}^2}{ R_0^{2D-4}} - \frac{D-3}{R_0^2}  = \frac{(D-3)^2}{R_0^2} + \frac{(D-1)(D-2)}{\ell^2}.\label{eq:2d_curvature}
\end{equation}

Now, we need to analyze the junction conditions. Let the $i$-th boundary of $\Sigma_{g,n}$ have circumference $b_i$ and constant geodesic curvature $\hat k_i$. The normal vector used to define $\hat k_i$ points outward from the core toward the attached exterior region. The induced metric on the $(D-1)$-dimensional gluing surface $\cS_i$ is:
\begin{equation}
    \dd s_{\cS_i}^2 = b_i^2\dd\tau^2 + R_0^2\,\dd\Omega_{D-2}^2,\quad \tau \sim \tau + 1.
\end{equation}
Continuity of the sphere metric at junctions implies that all radii $r_0^{(i)}$ are the same and equal to $R_0$. Continuity of the $\tau\tau$-component of the metric fixes the boundary lengths $b_i$,
\begin{equation}
    b_i = \beta_i \sqrt{f_i(R_0)},\label{eq:bis}
\end{equation}
for every exterior region. Moreover, continuity of the electric field implies that all fluxes $Q_i$ are equal to $Q$,
\begin{equation}
    Q_1 = \cdots = Q_n = Q.
\end{equation}

Extrinsic curvature of $\cS_i$, evaluated from the core side, is:
\begin{equation}
    K^-{}^{\tau}{}_{\tau} = \hat k_i,\quad K^-{}^{m}{}_n = 0,\label{eq:interior_ext_curvs}
\end{equation}
where indices $m,n$ label the $S^{D-2}$-directions. From the exterior side, using the normal vector pointing toward increasing $r$, the extrinsic curvature is:
\begin{equation}
    K^{+\tau}{}_{\tau} = \frac{f_i'(R_0)}{2\sqrt{f_i(R_0)}},\quad K^{+m}{}_{n} = \frac{\sqrt{f_i(R_0)}}{R_0}\delta^m_n.\label{eq:exterior_ext_curvs}
\end{equation}
Note that, since $\frac{\mathcal{Q}^2}{R^{2D-6}_0} = 1 + \frac{D-1}{D-3}\frac{R_0^2}{\ell^2},$ we have $f_i'(R_0) = -\frac{D-3}{R_0}f_i(R_0)$, and therefore, $K^{+\tau}{}_\tau = -\frac{D-3}{2R_0}\sqrt{f_i(R_0)}$. The Israel junction conditions relate the discontinuity of extrinsic curvature to the stress tensor $S^a{}_b$ on the junction surface:
\begin{equation}
    (K^{+A}{}_B - \delta^A_B K^+) - (K^{-A}{}_B - \delta^A_B K^-) = -8\pi G_D\, S^A{}_B.\label{eq:israel_junction}
\end{equation}
A brane with constant tension $\lambda_i$ wrapped around the junction surface gives us stress tensor $S^A{}_B = -\lambda_i \delta^A_B.$ The jump in the extrinsic curvatures must therefore be isotropic,
\begin{equation}
    K^{+\tau}{}_{\tau} - K^{-\tau}{}_{\tau} = K^{+m}{}_m - K^{-m}{}_{m}\quad \text{(with no sum over $m$)}.
\end{equation}
Substituting \eqref{eq:interior_ext_curvs} and \eqref{eq:exterior_ext_curvs}, we obtain
\begin{equation}
    \hat k_i = -\frac{D -1}{2R_0}\sqrt{f_i(R_0)}.\label{eq:ncorek_Ddim}
\end{equation}
The remaining junction equation from \eqref{eq:israel_junction} determines the saddle-point value of the corresponding Lagrange multiplier,
\begin{equation}
    \lambda_i = -\frac{D-2}{8\pi G_D R_0}\sqrt{f_i(R_0)}.\label{eq:lambda_nbdry_wormhole_Ddim}
\end{equation}
The existence of a saddle therefore reduces to the existence of a hyperbolic core whose boundary components have the prescribed pairs $(b_i,\hat k_i)$ given by \eqref{eq:bis} and \eqref{eq:ncorek_Ddim}. In Appendix~\ref{app:three-boundary-core}, we explicitly construct the genus-zero three-boundary core and demonstrate that it exists for a finite range of parameters, including, in particular, the near-extremal exteriors.

The tension \eqref{eq:lambda_nbdry_wormhole_Ddim} of the constraint branes is negative. For a dynamical brane this would signal an instability, since growing the brane would lower its action \cite{Pilo:2000et,Kogan:2001qx,Marolf:2001ne}. A constraint brane, however, has no dynamics of its own: its tension is a Lagrange multiplier, and growing the brane changes the constrained volume, which is fixed at the saddle and integrated over afterwards. Accordingly, in Section~\ref{sec:multi-boundary-stability} we will find no negative modes associated with the core or the junctions.

The resulting wormhole has arbitrary boundary inverse temperatures $\beta_1,\ldots,\beta_n$, fixed by Dirichlet boundary conditions; a common electric flux $Q$, fixed by the gauge-field boundary conditions; and unfixed mass-parameter moduli $M_1,\ldots,M_n$. At fixed $\beta_i$ and $Q$, the mass parameters determine the boundary circumferences $b_i$ through \eqref{eq:bis}. The circumferences $b_i$ are analogous to the boundary-length variables used when gluing trumpet geometries to hyperbolic cores in the calculation of JT-gravity path integrals \cite{Saad:2019lba}. In the present construction, however, the gluing curves generally have nonzero geodesic curvature and are therefore not themselves geodesic boundaries.

It is important to note that mass parameter values $M_i$ corresponding to real saddles are restricted to the super-extremal branch, 
\begin{equation}
    M_i \leq M_{\rm ext},
\end{equation}
where $M_{\rm ext}$ is the mass parameter of the extremal RN--AdS black hole of charge $Q$,
\begin{equation}
    2M_{\rm ext} = R_0^{D-3}\left(\frac{R_0^2}{\ell^2} + 1  + \frac{\mathcal{Q}^2}{R_0^{2(D-3)}}\right).\label{eq:extr-RN--AdS-mass}
\end{equation}
To see this, observe that for a real saddle $f_i(R_0)$ must be greater than zero. Then, from the definition of $f_i(r)$, it follows that
\begin{equation}
    2M_i = R_0^{D-3}\left(\frac{R_0^2}{\ell^2} + 1  + \frac{\mathcal{Q}^2}{R_0^{2(D-3)}}\right) - R_0^{D-3}f_i(R_0) \leq 2M_{\rm ext}.
\end{equation}

\subsection{Two-boundary wormholes}\label{sec:2bdry-saddles}

It turns out that the hyperbolic-core construction described above does not apply to a two-boundary wormhole. For a genus-zero, 2-boundary Riemann surface $\Sigma_{0,2}$, the Gauss--Bonnet theorem implies that:
\begin{equation}
    -a^2 A + \hat k_1 b_1 + \hat k_2 b_2 = 2\pi \chi(\Sigma_{0,2}) = 0,
\end{equation}
where $A$ is the area of $\Sigma_{0,2}$, while $b_i$ and $\hat k_i$ are the circumference and geodesic curvature of its $i$-th boundary. Since $A > 0$, it follows that $\hat k_1 b_1 + \hat k_2 b_2 > 0,$ so at least one geodesic curvature must be positive. This is incompatible with gluing both boundaries to RN--AdS exteriors, since Eq.~\eqref{eq:ncorek_Ddim} requires the geodesic curvature at every such junction to be negative.

\begin{figure}[H]
\centering
\includegraphics[width=0.5\textwidth]{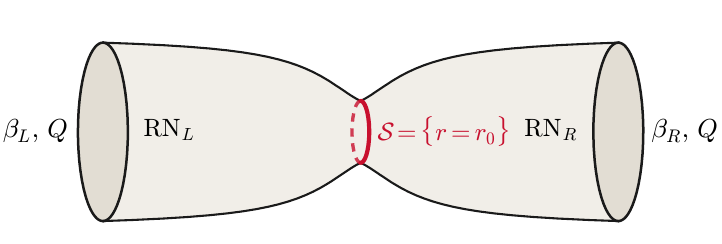}
\caption{Constrained two-boundary wormhole constructed by gluing two Reissner--Nordstr\"om exteriors RN${}_L$ and RN$_{R}$ with boundary data $(\beta_L,Q)$ and $(\beta_R,Q)$ along the surface $r=r_0$, where the constraint brane $\mathcal{S}$ (red) supports the jump in extrinsic curvature. Only the $(\tau,r)$ part of the geometry is drawn; the $S^{D-2}$ is suppressed.}
\label{fig:2bdry-wormhole}
\end{figure}

A two-boundary constrained saddle can instead be constructed by gluing two RN--AdS exteriors directly across a surface of constant radius $r=r_0$ (see Fig.~\ref{fig:2bdry-wormhole}). Continuity of the spherical part of the induced metric requires the gluing radius $r_0$ to be the same on both sides. Unlike in the multi-boundary construction, $r_0$ is not necessarily equal to the extremal horizon radius.

We label the two halves of the wormhole geometry by indices $L$ (left) and $R$ (right). Left (right) part of the wormhole has inverse temperature $\beta_L$ ($\beta_R$), charge $Q_L$ ($Q_R$), and mass $M_L$ ($M_R$). The corresponding lapse functions are $f_i(r) = \frac{r^2}{\ell^2} +1 -\frac{2M_i}{r^{D-3}} + \frac{\mathcal{Q}_i^2}{r^{2(D-3)}}$, where $i=L,R$. Continuity of the electric field requires that $Q_L = Q_R$, and continuity of the $\tau\tau$-component of the metric gives us:
\begin{equation}
    \beta_L\sqrt{f_L(r_0)} = \beta_R\sqrt{f_R(r_0)}.\label{eq:2bdry_tautau_continuity}
\end{equation}
At fixed $\beta_L$, $\beta_R$, and $Q$, Eq.~\eqref{eq:2bdry_tautau_continuity} relates the two mass parameters. In particular, when $\beta_L=\beta_R$, it implies $M_L=M_R$.

Let $K_L{}^A{}_B$ and $K_{R}{}^A{}_B$ be extrinsic curvatures on the gluing surface, evaluated from the left and right exterior, respectively. On each side, we choose the unit normal to point away from the junction and toward the corresponding asymptotic boundary. With this convention,
\begin{equation}
    K_i{}^\tau{}_\tau = \frac{f_i'(r_0)} {2\sqrt{f_i(r_0)}}, \qquad K_i{}^m{}_n = \frac{\sqrt{f_i(r_0)}}{r_0}\, \delta^m_n, \qquad i=L,R, \label{eq:2bdry_extrinsic_curvs}
\end{equation}
where $m,n$ label the $S^{D-2}$ directions. With constraint surface stress tensor $S^A{}_B = -\lambda \delta^A_B$, the Israel junction conditions give us:
\begin{align}
    &\frac{f'_L(r_0)}{2\sqrt{f_L(r_0)}} + \frac{f'_R(r_0)}{2\sqrt{f_R(r_0)}} = \frac{\sqrt{f_L(r_0)} + \sqrt{f_R(r_0)}}{r_0}.\label{eq:2bdry_wormhole_masses_eq}
    \\[3mm]
    &\lambda = -\frac{D-2}{8\pi G_D r_0}\left(\sqrt{f_L(r_0)} + \sqrt{f_R(r_0)}\right)\label{eq:2bdry_tension}
\end{align}
We see that the constraint-brane tension $\lambda$ is negative. Eq.~\eqref{eq:2bdry_wormhole_masses_eq}, together with \eqref{eq:2bdry_tautau_continuity}, determines masses $M_L$, $M_R$ in terms of the radius $r_0$. In particular, when $\beta_L = \beta_R$, the two masses are equal, $M_L = M_R = M$, and from \eqref{eq:2bdry_wormhole_masses_eq} it follows that:
\begin{equation}
    M(r_0) = \frac{r_0^{D-3}}{D-1}\left(1 + (D-2)\frac{\mathcal{Q}^2}{r_0^{2(D-3)}}\right).
\end{equation}
One can easily check that when $r_0$ approaches the horizon radius of extremal RN--AdS black hole, $r_0\to R_0$, mass parameter $M(r_0)$ approaches the mass parameter of the extremal black hole, $M(R_0) = M_{\rm ext}$. The function $M(r_0)$ is strictly convex for $r_0 > 0$. The equation $M(r_0) = M_{\rm ext}$ has two positive solutions. One is $r_0 = R_0$ and the second, which we denote by $R_1$, is larger, $R_1 > R_0.$ Consequently,
\begin{equation}
    M(r_0)= \begin{cases} >M_{\rm ext} \qquad \text{for } 0<r_0<R_0,\\ <M_{\rm ext} \qquad \text{for } R_0<r_0<R_1,\\ >M_{\rm ext} \qquad \text{for }r_0>R_1. \end{cases} \label{eq:massive-intervals}
\end{equation}
For $r_0 < R_0$ the solution is sub-extremal, so the wormhole neck lies behind the outer RN--AdS horizon and the Euclidean metric is not positive-definite at the junction. Therefore, there are no regular real solutions with $r_0 < R_0.$ Regular saddles are parametrized by $r_0 > R_0$. When $R_0 < r_0 < R_1$ the solution is super-extremal, although the complete wormhole geometry is smooth because the region containing the RN singularity has been excised.\footnote{For $r_0>R_1$, the mass parameters again exceed $M_{\rm ext}$, but the neck lies outside the outer horizon, $f_{L,R}(r_0)>0$, so these saddles are regular as well.}

In the extremal limit $r_0\to R_0$, both $M_L$ and $M_R$ approach $M_{\rm ext}$, the lapse functions $f_i(r_0)$ approach zero, and hence, the constraint brane tension vanishes,
\begin{equation}
    \lambda \propto - \left(\sqrt{f_L(r_0)} + \sqrt{f_R(r_0)}\right)\xrightarrow{r_0\to R_0}0.
\end{equation}
As we show in Appendix \ref{app:timelike-reduction}, upon dimensional reduction along the temporal circle, this limit recovers the axio-dilaton wormholes described in \cite{Maldacena:2026jqd}.

\subsection{Single-boundary saddles}\label{sec:1bdry-saddles}
Constraint branes also allow us to construct single-boundary saddles that are more general than smooth black holes. We constrain the area of a codimension-two surface $\mathcal S\simeq S^{D-2}$ by adding to the Einstein--Maxwell action the constraint-brane term $I_{\mathcal C}=\lambda\left( \operatorname{area}(\mathcal S)-\zeta\right).$ The single-boundary saddles of this system are Reissner--Nordstr\"om cigars with the brane located at the tip, where the Euclidean time circle shrinks. The brane supports a conical singularity at the tip, which allows the horizon area to be fixed independently of the boundary inverse temperature $\beta$.

Away from the tip, the saddles are Euclidean Reissner--Nordstr\"om--AdS solutions,
\begin{align}
    \dd s^2&=\beta^2f(r)\dd\tau^2+\frac{\dd r^2}{f(r)}
                 +r^2\dd\Omega_{D-2}^2,
    \qquad \tau\sim\tau+1,\nonumber\\
    \frac12F_{\mu\nu}\dd x^\mu\wedge\dd x^\nu &=-\frac{\i Q}{\Omega_{D-2}\,r^{D-2}}\,\beta\,\dd\tau\wedge\dd r,
    \label{eq:single-constrained-metric}
\end{align}
where $f(r)=1+\frac{r^2}{\ell^2}-\frac{2M}{r^{D-3}}+ \frac{\mathcal Q^2}{r^{2(D-3)}}$ and $\mathcal Q^2=8\pi G_DQ^2/[(D-2)(D-3)\Omega_{D-2}^2]$. The tip of the cigar is a root $r=r_+$ of $f(r)$, and the condition $f(r_+)=0$ fixes the mass parameter to
\begin{equation}
    2M=r_+^{D-3}\left(1+\frac{r_+^2}{\ell^2} +\frac{\mathcal Q^2}{r_+^{2(D-3)}}\right).
    \label{eq:single-constrained-energy}
\end{equation}
The Euclidean geometry $r\geq r_+$ is real only if $f(r)$ possesses real roots and $r_+$ is the largest of them. Near $r=r_+$, upon introducing the coordinate $\rho$ via $r-r_+=f'(r_+)\rho^2/4$, we can recast the metric \eqref{eq:single-constrained-metric} as
\begin{equation}
    \dd s^2\overset{\rho\to 0}{\sim}\dd\rho^2 + \frac{\beta^2f'(r_+)^2}{4}\rho^2\dd\tau^2 + r_+^2\dd\Omega_{D-2}^2,
\end{equation}
so the opening angle at the tip is $\beta f'(r_+)/2$. The Einstein equations at the tip equate the deficit angle to $8\pi G_D\lambda$, and the variation with respect to $\lambda$ fixes the area of $\mathcal S$,
\begin{equation}
    8\pi G_D\lambda=2\pi-\frac{\beta f'(r_+)}2,
    \qquad \zeta=\Omega_{D-2}r_+^{D-2}.
    \label{eq:single-tip-junction}
\end{equation}
Both deficit and excess angles are allowed. The brane becomes tensionless when $\beta f'(r_+)=4\pi$, in which case the saddle is the smooth Reissner--Nordstr\"om black hole.

\section{Stability against spherically symmetric perturbations}\label{sec:stability}
In this section, we analyze perturbative stability of the constrained saddles constructed in the previous section. One could expect that negative-tension branes can lead to an instability, since increasing the brane area would lower its energy \cite{Pilo:2000et,Kogan:2001qx,Marolf:2001ne}. However, as shown in Section~\ref{sec:constr-instantons}, positivity of the relevant quadratic action must be tested only on the constraint-preserving fluctuations. Fluctuations that change the constrained areas are removed by integration over the corresponding Lagrange multiplier, and should not be regarded as negative modes. We account for all allowed values of these modes when integrating over the constraint value $\zeta.$

We restrict our analysis to perturbations that are spherically symmetric on $S^{D-2}$, which we assume to be the sector most likely to contain a negative mode; we do not analyze the sectors with nonzero angular momentum explicitly. More general perturbations can be decomposed into scalar, vector, and tensor harmonics on $S^{D-2}$ \cite{Kodama:2003kk}. Modes with nonzero angular momentum $j$ receive positive angular-gradient contributions proportional to $j(j + D-3)/r^2$ where $r$ is the radius of $S^{D-2}$. These contributions tend to raise the eigenvalues of the quadratic action relative to those in the $j=0$ sector.

Within the spherically symmetric sector, keeping the Maxwell field on shell does not amount to discarding an independent propagating fluctuation, since a two-dimensional gauge field has no local degrees of freedom. With Neumann boundary conditions fixing electric flux, the spherically symmetric Maxwell fluctuation is therefore determined by the two-dimensional metric and dilaton fluctuations and can be integrated out exactly.

The spherically symmetric sector of the $D$-dimensional Einstein--Maxwell theory consequently reduces to two-dimensional dilaton gravity. The remaining fluctuations are global or boundary modes, together with the gluing data at the constraint surfaces. We can therefore solve the gravitational constraints before evaluating the quadratic action. This removes the usual conformal mode problem \cite{Gibbons:1978ac,Mottola:1995sj,Liu:2023jvm} in a controlled way.

\subsection{Dimensional reduction of Einstein--Maxwell theory on \texorpdfstring{$S^{D-2}$}{}}\label{sec:dim-reduction}
\paragraph{Unconstrained action.} To set the notation, let us first review the dimensional reduction of Einstein--Maxwell theory to two-dimensional gravity. We use the metric ansatz,
\begin{equation}
    \dd s^2 = \tilde{g}_{ab}\dd x^a \dd x^b + X(x)^{\frac{2}{D-2}}\dd\Omega_{D-2}^2,\label{eq:metric-reduction}
\end{equation}
where $a,b=1,2$, the coordinates $x^a$ parametrize a two-dimensional manifold $M_2$, and $X(x)$ is the dilaton field, related to the sphere radius by $r = X^{\frac{1}{D-2}}$. 

We work in the fixed-charge ensemble. The spherically symmetric solution of the Maxwell equation is
\begin{equation}
    \frac{1}{2}F_{\mu\nu}\dd x^\mu \wedge \dd x^\nu = -\frac{\i Q}{\Omega_{D-2}}\frac{1}{r^{D-2}}\sqrt{\tilde{g}}\,\dd^2x\label{eq:field-strength-reduction}
\end{equation}
where $Q$ is the electric flux through the $S^{D-2}$ and $\sqrt{\tilde{g}} = \sqrt{\det (\tilde g_{ab})}$. Substituting \eqref{eq:metric-reduction} and \eqref{eq:field-strength-reduction} into the Einstein--Maxwell action, including the Gibbons--Hawking--York term and the Maxwell boundary term which fixes the electric flux, gives the following two-dimensional action,
\begin{equation}
    I_2 = -\frac{1}{16\pi G_2}\int_{M_2} \dd^2x\sqrt{\tilde g}\,\left\{X R(\tilde{g}) + \frac{D-3}{D-2}\frac{1}{X}(\tilde\nabla X)^2 + \tilde{V}(X)\right\} - \frac{1}{8\pi G_2}\int_{\partial M_2}\dd\tilde s\,X\tilde k.\label{eq:2d-action1}
\end{equation}
Here $R(\tilde{g})$ is the Ricci scalar associated to the two-dimensional metric $\tilde{g}_{ab}$, $\tilde k$ is the geodesic curvature of the boundary $\partial M_2$, $\dd\tilde s$ is the line element of $\partial M_2$, and $G_2 = G_D/\Omega_{D-2}$ is the two-dimensional Newton's constant. The dilaton potential $\tilde{V}(X)$ is
\begin{equation}
    \tilde{V}(X) = -2\Lambda X - (D-2)(D-3)\frac{\mathcal{Q}^2}{X} + (D-2)(D-3)X^{\frac{D-4}{D-2}}.
\end{equation}

It is convenient to remove the dilaton kinetic term by performing a Weyl transformation, $\tilde g_{ab} = e^{2\omega(X)}\hat g_{ab}$. To this end, we choose conformal factor
\begin{equation}
    \omega(X) = -\frac{D-3}{2(D-2)}\log(\frac{X}{R^{D-2}_0}),\label{eq:weyl-rescaling}
\end{equation}
where $R_0$ is a fixed distance scale. We take $R_0$ to be the horizon radius of the $D$-dimensional extremal RN--AdS black hole. With this choice, the contribution generated by the Weyl transformation of $X R(\tilde g)$ cancels the kinetic term in \eqref{eq:2d-action1}. The corresponding boundary contribution cancels against the transformation of the GHY term, and the action \eqref{eq:2d-action1} becomes
\begin{equation}
    I_2 = -\frac{1}{16\pi G_2}\int_{M_2} \dd^2 x\sqrt{\hat g}\,\{X R(\hat g) + V(X)\} - \frac{1}{8\pi G_2} \int_{\partial M_2}\dd\hat s\,X\,\hat k,\label{eq:2d-action2}
\end{equation}
where
\begin{equation}
    V(X) = R_0^{D-3}\left[-2\Lambda X^{\frac{1}{D-2}} - (D-2)(D-3)\mathcal{Q}^2X^{-\frac{2D-5}{D-2}} + (D-2)(D-3)X^{-\frac{1}{D-2}}\right].
\end{equation}
The corresponding equations of motion for the metric are
\begin{equation}
    E_{ab} = \hat\nabla_a \hat\nabla_b X - \hat g_{ab}\hat\nabla^2 X + \frac{1}{2}V(X)\hat g_{ab} = 0.\label{eq:2d-metric-eom}
\end{equation}

\paragraph{Constrained surface action.} It remains to reduce the constraint-brane action. We assume that the constrained surface is spherically symmetric and has the product form $\cS = \hat \cS \times S^{D-2}$, where $\hat\cS$ is a closed loop in $M_2$. Its higher-dimensional area is $\Omega_{D-2}\int_{\hat\cS}\dd\tilde s\,X(x)$, and hence, the constraint action before the Weyl transformation is 
\begin{equation}
    I_{\mathcal{C}} = \lambda \left(\Omega_{D-2}\int_{\hat \cS}\dd\tilde s\, X(x) - \zeta\right).
\end{equation}
In the frame used in \eqref{eq:2d-action2}, the constraint action becomes
\begin{equation}
    I_{\mathcal{C}} = \lambda \left(R_0^{\frac{D-3}{2}}\Omega_{D-2}\int_{\hat \cS} \dd\hat s\,X^{\frac{D-1}{2(D-2)}} - \zeta\right).\label{eq:2d-Iconstr}
\end{equation}

Importantly, the GHY term must also be included on the constraint surface. This is not apparent from the insertion of the constraint action alone. At the constrained saddle, however, the localized brane source produces discontinuities in the normal derivatives of the metric and dilaton, so that the curvature tensors contain a Dirac delta function supported on the junction. Consequently, there are two equivalent ways to describe this geometry. One may regard the metric as a single, continuous metric with distributional curvature, in which case the bulk Einstein--Hilbert term includes the contribution localized at the junction. Alternatively, one may divide the geometry into two regions with regular curvature and treat their common interface as a boundary of each region. In the latter description, one must include a GHY term from each side of the junction.

To see this equivalence, let us split the two-dimensional spacetime $M_2$ into two parts, $M_+$ and $M_-$ glued along the curve $\hat\cS$, so that the induced metric and the dilaton are continuous across $\hat\cS$, while their normal derivatives might be discontinuous . Denote by $n_{\pm}$ the outward-pointing unit normal to $\hat\cS$ as viewed from $M_{\pm}$, and by $\hat k_{\pm}$ the corresponding geodesic curvatures. The singular part of the Ricci tensor is $R_{\rm sing} = 2(\hat k_+ + \hat k_-)\delta_{\hat\cS}$, where $\delta_{\hat\cS}$ is the Dirac delta function localized on $\hat\cS$, normalized so that $\int_{M_2}\dd^2 x\sqrt{\hat g}\,F(x)\delta_{\hat\cS}  = \int_{\hat{\cS}}\dd \hat{s}\,F(x)$. It follows that the singular contribution to the bulk action is
\begin{equation}
    -\frac{1}{16\pi G_2}\int \dd^2x\sqrt{\hat g}\,XR_{\rm sing} = -\frac{1}{8\pi G_2}\int_{\hat\cS}\dd\hat s\,X(\hat k_+ + \hat k_-),\label{eq:ghy-gluing-action}
\end{equation}
which is precisely the sum of the two GHY terms.\footnote{We can verify this with a direct calculation. Consider a two-dimensional metric in conformal gauge,
\begin{equation}
    \dd\hat s^2 = e^{2\varphi(\rho,\tau)} (\dd\rho^2 + \dd\tau^2).
\end{equation}
The corresponding Ricci scalar is
\begin{equation}
    R(\hat g) = -2e^{-2\varphi}(\partial_\rho^2 + \partial_\tau^2)\varphi.
\end{equation}
Define the constraint curve $\hat\cS$ as $\hat\cS = \{\rho = 0\}$. Divide spacetime into two regions, $M_- = \{\rho < 0\}$ and $M_+ = \{\rho > 0\}$. Let $\hat k_{\pm}$ be the geodesic curvature of $M_{\pm}$, and $\dd\hat s = \dd \tau\sqrt{\hat h} = e^{\varphi(\rho = 0,\tau)}\dd\tau$ be the volume form on $\partial M_\pm$. Explicitly,
\begin{equation}
    \hat k_{\pm} = \mp \lim_{\rho\to 0^{\pm}}e^{-\varphi}\partial_\rho \varphi.
\end{equation}
If there is a jump in the extrinsic curvatures across the $\rho=0$ surface, the term $-2e^{-2\varphi}\partial_\rho^2 \varphi$ in the Ricci scalar is singular at $\rho = 0$. Integrating $-\frac{1}{16\pi G_2}\sqrt{\hat g}\, XR(\hat g)$ over a small interval $\rho\in (-\epsilon,\epsilon)$ around $\rho = 0$ (with continuous function $X$) gives:
\begin{align}
    -\frac{1}{16\pi G_2}\int\dd\tau \int_{-\epsilon}^{\epsilon}\dd\rho\,X(-2\partial_\rho^2 \varphi) &= -\frac{1}{8\pi G_2}\int \dd\tau \,X|_{\rho = 0}(-\lim_{\rho\to 0^+}\partial_\rho \varphi + \lim_{\rho\to 0^-}\partial_\rho\varphi)\nonumber 
    \\
    &= -\frac{1}{8\pi G_2}\int \dd\tau \sqrt{\hat h}\,X|_{\rho = 0}(\hat k_+ + \hat k_-)\nonumber 
    \\
    &= -\frac{1}{8\pi G_2}\int_{\partial M_-}\dd\hat s\,X\hat k - \frac{1}{8\pi G_2}\int_{\partial M_+}\dd\hat s\,X\hat k.
\end{align}
We recovered the action \eqref{eq:ghy-gluing-action}.} Therefore, including these terms does not add a new physical contribution; it simply represents the distributional curvature in the description in which the two sides of the junction are treated as smooth manifolds with a boundary. For this reason, when calculating the dilaton gravity action, we shall evaluate the bulk curvature separately in each smooth region and include the GHY term on every boundary component.

Variation of the full action, given by the sum of \eqref{eq:2d-action2}, \eqref{eq:2d-Iconstr} and \eqref{eq:ghy-gluing-action} over the dilaton field $X$ and metric $q_{AB}$ induced on $\hat\cS$ yields two-dimensional Israel junction conditions,
\begin{align}
    \hat\nabla_{n_+}X+\hat\nabla_{n_-}X &= 8\pi G_2\lambda\,\Omega_{D-2} R_0^{\frac{D-3}{2}} X^{\frac{D-1}{2(D-2)}}, \label{eq:2d-junction-dilaton}
    \\
    \hat k_++\hat k_- &= 8\pi G_2\lambda\,\Omega_{D-2} R_0^{\frac{D-3}{2}}\, \frac{D-1}{2(D-2)}\,X^{-\frac{D-3}{2(D-2)}}. \label{eq:2d-junction-metric}
\end{align}
Note however that the dilaton and the metric induced on $\hat\cS$ are dynamical fields, and therefore, we cannot impose equations of motion \eqref{eq:2d-junction-dilaton} and \eqref{eq:2d-junction-metric} while analyzing stability of the saddles.

\paragraph{On-shell configurations.} Let us now describe the on-shell backgrounds relevant to the stability analysis. Their two-dimensional form follows directly from the dimensional reduction of the $D$-dimensional solutions constructed in the previous section. After the Weyl rescaling defined by \eqref{eq:weyl-rescaling}, each exterior region is described by
\begin{equation}
    \dd\hat s^2 = \left(\frac{r}{R_0}\right)^{D-3} \left( \beta_i^2f_i(r)\,\dd\tau^2 +\frac{\dd r^2}{f_i(r)} \right), \qquad X=r^{D-2},
\end{equation}
where $f_i(r)$ is the lapse function of the corresponding exterior and $\tau\sim\tau+1$, as in \eqref{eq:RN-metric}. In the wormhole core, the dilaton is constant and the Weyl factor equals one, so
\begin{equation}
    \dd\hat s^2=\dd s_{\Sigma}^2, \qquad X=R_0^{D-2},
\end{equation}
where $\dd s_{\Sigma}^2$ is the hyperbolic metric appearing in the higher-dimensional construction. The two-boundary saddles instead consist of two exterior regions joined directly at $r=r_0$, without an intervening hyperbolic core.

\subsection{Stability of the two-boundary wormhole}
\label{sec:two-boundary-stability}

In this subsection, we analyze the stability of the two-boundary wormhole constructed in the previous section against spherically symmetric perturbations. To identify the independent path-integration variables, we first examine the phase space of two-dimensional dilaton gravity with the constraint branes. The phase space of a general two-dimensional dilaton gravity without constraints has been studied extensively: in the absence of propagating matter, the only physical degrees of freedom are the mass parameter and its canonical conjugate \cite{Louis-Martinez:1993bge,Cavaglia:1998xj}, and exact methods for integrating out the bulk fields have been developed in \cite{Kummer:1996hy,Bergamin:2004pn}. The boundary degrees of freedom have been analyzed in \cite{Kummer:1996si,Bergamin:2007aw}; see \cite{Grumiller:2002nm} for a review and \cite{Ruzziconi:2020wrb} for a more modern treatment of the boundary phase space.

Our problem additionally requires treating fluctuations at the constraint surface. In \cite{Kummer:1996si,Bergamin:2007aw}, the boundary dilaton and induced metric are prescribed and held fixed under variation. At our internal junction, we instead integrate over their common values, imposing continuity across the junction and fixing only the brane area in the $D$-dimensional theory. We therefore give a detailed derivation adapted to the relevant junction conditions.

\paragraph{Action and boundary conditions.}The two-dimensional geometry consists of two regions, $M_L$ and $M_R$, joined along the constrained line $\hat \cS$. Their boundaries are $\partial M_i=B_i\cup\hat \cS$, with $i=L,R$, where $B_i$ is an asymptotic cutoff curve with large, constant dilaton,
\begin{equation}
 B_i=\{X=X_c=\mathrm{const.}\}.
\end{equation}
We denote the two-dimensional metric in the frame without a dilaton kinetic term by $\hat g_{ab}$ and use the two-dimensional Newton's constant $G_2=G_D/\Omega_{D-2}$. The full action is
\begin{equation}
    I=I_L+I_R+I_{\mathcal C}, \label{eq:full-action}
\end{equation}
with $I_{\cC}$ given by \eqref{eq:2d-Iconstr} and
\begin{equation}
 I_i=I_i^{\mathrm{EH}}+I_i^{\hat \cS}+I_i^B+I_i^{\mathrm{ct}}.
 \label{eq:half-action-decomposition}
\end{equation}
Here $I_i^{\mathrm{EH}}$ denotes the dilaton-gravity bulk action,
\begin{equation}
 I_i^{\mathrm{EH}}=-\frac{1}{16\pi G_2}
 \int_{M_i}\dd^2x\sqrt{\hat g}\,
 \bigl(XR(\hat g)+V(X)\bigr),
 \label{eq:dilaton-bulk-action}
\end{equation}
and $I_i^{\hat \cS}$ and $I_i^B$ are the GHY terms at the junction and the asymptotic boundary:
\begin{equation}
 I_i^\gamma=-\frac{1}{8\pi G_2}\int_\gamma\dd\hat s\,X\hat k_i,
 \qquad \gamma=\hat \cS,B_i.
 \label{eq:GHY-definition}
\end{equation}
The geodesic curvature $\hat k_i$ is defined using the outward normal of $M_i$. Continuity across $\hat \cS$ identifies the induced line element and dilaton on the two sides, while their geodesic curvatures can differ. Including the GHY term on each side of the junction $\hat \cS$ accounts for the distributional curvature of the glued geometry when the bulk integrals are evaluated separately on the two smooth regions.

On the constant-dilaton cutoff curve, the Hamilton--Jacobi counterterm is \cite{Grumiller:2007ju}
\begin{equation}
    I^{\rm ct}_{i} = \frac{1}{8\pi G_2}\int_{B_i}\dd\hat s\,\sqrt{w(X)}.\label{eq:2d-asymptotic-counterterm}
\end{equation}
For the spherical reduction of asymptotically flat Einstein--Maxwell theory, the prescription \eqref{eq:2d-asymptotic-counterterm} agrees with the dimensional reduction of the standard flat-space background-subtraction method in the infinite cutoff limit $X_c \to \infty$. For asymptotically AdS solutions, the finite part of this action might not coincide with the dimensional reduction of the usual higher-dimensional holographic counterterms, as the latter assigns a nonzero vacuum energy to the global AdS \cite{Balasubramanian:1999re}. More general boundary dilaton profiles require tangential-derivative terms in the counterterm \cite{McNees:2023tus}. In what follows, we shall use the prescription \eqref{eq:2d-asymptotic-counterterm}.

\paragraph{Gauge constraints.} Now, we need to fix the bulk diffeomorphisms on each side of the junction surface $\hat\cS$. Wherever $\dd X \neq 0$, we can identify one coordinate with the dilaton and write the metric as
\begin{equation}
    \dd\hat s^2 = \psi(T,X)^2 F(T,X) \,\dd T^2 + \frac{1}{F(T,X)}(\dd X + N(T,X)\dd T)^2.
\end{equation}
We regard $T$ as Euclidean time, so that $N$ is the shift and $\psi\sqrt F$ is the lapse. Without redefining $X$, the remaining coordinate freedom $T\mapsto\widetilde T(T,X)$ can locally set $N=0$. The residual transformations are $T\mapsto\widetilde T(T)$. 

The equation of motion obtained by varying the shift must be imposed as a gauge constraint. It is proportional to $E_{TX}$ given by \eqref{eq:2d-metric-eom}, and at $N=0$, it simplifies to $\partial_T F = 0$, implying that $F = F(X)$ is independent of $T$. The metric therefore takes the form
\begin{equation}
    \dd\hat s^2 = \psi(T,X)^2 F(X)\,\dd T^2 + \frac{\dd X^2}{F(X)}.\label{eq:gauge-fixed-2d-metric}
\end{equation}
Substituting this expression into the bulk part of the action \eqref{eq:2d-action2}, we obtain
\begin{align}
    I_i^{\rm EH} = -\frac{1}{16\pi G_2}\int\dd T\dd X\left[\psi_i V(X) - \psi_i X \partial_X^2 F_i - 3X (\partial_X F_i) (\partial_X\psi_i) - 2X F_i \partial_{X}^2\psi_i\right].
\end{align}
This action does not contain the time-derivatives of $\psi$; the field $\psi$ is not dynamical, and hence, we can integrate it out. This gives us the constraint $F'_i(X) = V(X)$, removing all of the bulk degrees of freedom except for the zero mode $C_i$ of $F_i(X)$,
\begin{equation}
    F_i(X) = w(X) + C_i,\qquad w'(X) = V(X).\label{eq:constrained-f(X)}
\end{equation}
We choose the anti-derivative $w(X)$, so that it does not have a constant term.

As we show below, combining the bulk and GHY contributions removes the explicit dependence on $\psi_i$, allowing the reduced action to be expressed in terms of $F_i$, the boundary dilaton, and the induced line element. Thus, without loss of generality, we can set $\psi_i = {\rm const.}$ Then, the $D$-dimensional uplifted metric
\begin{equation}
 g_{\mu\nu}\dd x^\mu\dd x^\nu
 =\left(\frac{R_0}{r}\right)^{D-3}
 \hat g_{ab}\dd x^a\dd x^b+r^2\dd\Omega_{D-2}^2,
 \qquad X=r^{D-2},
 \label{eq:uplift}
\end{equation}
is equal to the RN--AdS metric \eqref{eq:RN-metric}, with Euclidean time coordinates identified as
\begin{equation}
    \beta_i\tau=(D-2)R_0^{D-3}\psi_iT. \label{eq:clock-dictionary}
\end{equation}
Comparison of the radial component of \eqref{eq:RN-metric} with the RN metric yields
\begin{equation}
    F_i(X)=(D-2)^2R_0^{D-3}r^{D-3}f_i(r), \label{eq:lapse-dictionary}
\end{equation}
where $f_i$ is the standard RN lapse function \eqref{eq:RN-lapse-function}, with mass parameter proportional to the mode $C_i$,
\begin{equation}
    C_i=-2(D-2)^2R_0^{D-3}M_i.\label{eq:mass-dictionary}
\end{equation}

\paragraph{Action for boundary modes.} From the previous paragraph it follows that the only physical degrees of freedom are the dilaton fluctuations on the surfaces $\hat \cS$, $B_i$, and the constant modes $C_i$. Now we evaluate the action for these modes. We take $B_i$ and $\hat \cS$ to be smooth closed curves, parametrized by $\sigma\sim\sigma+1$. Their induced two-dimensional line element is $\dd\hat s=E\,\dd\sigma$, with
\begin{equation}
    E^2=\psi_i(T,X)^2F_i(X)\dot T^2 + \frac{\dot X^2}{F_i(X)}, \qquad \dot T=\frac{\dd T}{\dd\sigma},\quad \dot X=\frac{\dd X}{\dd\sigma}. \label{eq:2d-einbein}
\end{equation}
At the junction $\hat{\cS}$, continuity gives the same $E$ and $X$ for $i=L$ and $i=R$. It is convenient to define
\begin{equation}
    \alpha_i=\arcsin\left(\frac{\dot X}{E\sqrt{F_i(X)}}\right).
 \label{eq:tangent-angle}
\end{equation}

By substituting \eqref{eq:gauge-fixed-2d-metric} into \eqref{eq:dilaton-bulk-action}, we obtain
\begin{equation}
    I_i^{\mathrm{EH}}=-\sum_{\gamma=\hat\cS,B_i}\frac{\varepsilon_\gamma}{16\pi G_2} \int_\gamma\dd\sigma\,\dot T \left(2\psi_iF_i-X\psi_i\partial_XF_i - 2XF_i\partial_X\psi_i\right), \label{eq:EH-boundary-expression}
\end{equation}
where $\varepsilon_{\hat\cS}=-1$ and $\varepsilon_{B_i}=+1$. The geodesic curvature of the curve $\gamma = \hat \cS, B_i$ is
\begin{equation}
    \hat k_i=\frac{\varepsilon_\gamma}{E}\left[-\dot\alpha_i+ \left(\frac{\psi_iF_i'}{2}+F_i\partial_X\psi_i\right)\dot T\right]. \label{eq:geodesic-curvature}
\end{equation}
The GHY contribution \eqref{eq:GHY-definition} is consequently
\begin{equation}
    I_i^\gamma=-\frac{\varepsilon_\gamma}{8\pi G_2} \int_\gamma\dd\sigma\,X \left[-\dot\alpha_i+ \left(\frac{\psi_i\partial_XF_i}{2}+F_i\partial_X\psi_i\right)\dot T\right]. \label{eq:GHY-boundary-expression}
\end{equation}
Adding the bulk and the two GHY terms cancels the terms proportional to $\partial_XF_i$ and $\partial_X\psi_i$. Integrating $X\dot\alpha_i$ by parts on each closed curve gives
\begin{align}
 I_i^{\mathrm{EH}}+I_i^{\hat\cS}+I_i^B
 &=-\frac{1}{8\pi G_2}\sum_{\gamma=\hat\cS,B_i}\varepsilon_\gamma
 \int_\gamma\dd\sigma\,
 \left(\psi_iF_i\dot T+\alpha_i\dot X\right)
 \nonumber\\
 &=-\frac{1}{8\pi G_2}\sum_{\gamma=\hat\cS,B_i}\varepsilon_\gamma
 \int_\gamma\dd\sigma\,
 \left[
 \sqrt{E^2F_i(X)-\dot X^2}
 +\dot X\arcsin\left(\frac{\dot X}{E\sqrt{F_i(X)}}\right)
 \right].
 \label{eq:combined-boundary-action}
\end{align}
This action is clearly invariant under reparametrizations $\sigma \to \tilde{\sigma}(\sigma)$. As required, the action \eqref{eq:combined-boundary-action} does not depend on the integrated-out field $\psi_i(T,X)$. Therefore, we can set $\psi_i$ to a constant, and use the relations \eqref{eq:clock-dictionary}, $\eqref{eq:lapse-dictionary}$ and \eqref{eq:mass-dictionary} to rewrite it in terms of $D$-dimensional variables.

At the asymptotic boundaries, $X= X_c$ is fixed, so that $\dot{X} = 0$. Denoting the proper length of $B_i$ as $L_{B,i}$, and including the counterterm \eqref{eq:2d-asymptotic-counterterm}, the contribution of $B_i$ to the action becomes
\begin{equation}
    I_{i}^{B} + I_{i}^{\rm ct} = \frac{L_{B,i}}{8\pi G_2} \left[ \sqrt{w(X_c)} -\sqrt{w(X_c)+C_i} \right].\label{eq:2d-bdry-action}
\end{equation}
At $X_c \to \infty$, we have $w(X_c)\to \infty$ and $L_{B,i}/\sqrt{w(X_c)} \to \beta_i/[(D-2)R_0^{D-3}]$. Consequently, \eqref{eq:2d-bdry-action} reduces to
\begin{equation}
    I_i^B + I_i^{\rm ct} = -\frac{\beta_i C_i}{16\pi G_2(D-2)R_0^{D-3}} = \beta_i E_i^{\rm ADM},
\end{equation}
where $E_i^{\rm ADM} = \frac{(D-2)\Omega_{D-2}}{8\pi G_D}M_i$ is the ADM energy of the $D$-dimensional RN--AdS black hole.

The constraint action becomes
\begin{equation}
    I_{\mathcal C}=\lambda\left( \Omega_{D-2}\int_0^1\dd\sigma\,E\left(\frac{R_0}{r}\right)^{\frac{D-3}{2}}r^{D-2}-\zeta\right),
\end{equation}
and, as \eqref{eq:combined-boundary-action}, it is reparametrization-invariant. We choose the parametrization $\sigma$ so that $E\,(R_0/r)^{(D-3)/2}=L$ is constant. Since $E\,\dd\sigma$ is the line element of $\hat\cS$ in the Weyl-rescaled metric $\hat g$, $E\,(R_0/r)^{(D-3)/2}\,\dd\sigma$ is the line element induced by the $D$-dimensional metric, and $L$ is the length of $\hat\cS$ in that metric. Integrating over $\lambda$ imposes the constraint, determining $L$ as a functional of $r$,
\begin{equation}
    L[r]=\frac{\zeta}{\Omega_{D-2}\displaystyle\int_0^1\dd\sigma\,r^{D-2}}. \label{eq:length-functional}
\end{equation}
Mass parameters $M_i$ are determined by the periods $\beta_i$ of the $D$-dimensional Euclidean time through the condition,
\begin{equation}
    \beta_i= \int_0^1\dd\sigma\, \beta_i\dot{\tau}(\sigma) = \int_0^1\dd\sigma\, \frac{\sqrt{L[r]^2f_i(r)-\dot r^2}}{f_i(r)}. \label{eq:period-conditions}
\end{equation}
Here, in the second step we used $L[r]^2=\beta_i^2f_i(r)\dot{\tau}^2+\dot{r}^2/f_i(r)$, which follows from Eq.~\eqref{eq:RN-metric}.

Summarizing, path integral for spherically symmetric fluctuations of Einstein--Maxwell theory around the constrained two-boundary wormhole background takes the form
\begin{equation}
    Z\sim\int\dd\zeta\int\mathcal D r(\sigma)\,\exp\bigl[-I(r;\beta_i,\zeta)\bigr], \label{eq:reduced-integral}
\end{equation}
where real positive-area configurations correspond to $\zeta>0$, and
\begin{align}
    I(r;\beta_i,\zeta)=\frac{D-2}{8\pi G_2} \sum_{i=L,R}\Bigg\{\int_0^1\dd\sigma\,r^{D-3}
 \left[
 \dot r\arcsin\left(\frac{\dot r}{L[r]\sqrt{f_i(r)}}\right)
 +\sqrt{L[r]^2f_i(r)-\dot r^2}
 \right] +\beta_i M_i \Bigg\}.
 \label{eq:full-reduced-action}
\end{align}
Here, masses $M_i$ are determined via \eqref{eq:period-conditions} and $L[r]$ is given by \eqref{eq:length-functional}.

\paragraph{Quadratic action and stability.}The saddle point of \eqref{eq:reduced-integral} corresponds to a constant background $r=r_0$. In this case the constraints \eqref{eq:period-conditions} and \eqref{eq:length-functional} reduce to
\begin{equation}
    L[r_0] = \frac{\zeta}{\Omega_{D-2}r_0^{D-2}} = \beta_i \sqrt{f_i(r_0)}.\label{eq:r0-zeta-relation}
\end{equation}
This is precisely the continuity condition \eqref{eq:2bdry_tautau_continuity}. Action \eqref{eq:full-reduced-action} simplifies to
\begin{equation}
    I(r_0;\beta_i,\zeta) = \frac{D-2}{8\pi G_2} \sum_{i=L,R}\left(r_0^{D-3}L[r_0] \sqrt{f_i(r_0)} + \beta_i M_i\right).\label{eq:rconst-action}
\end{equation}
By varying $I(r_0;\beta_i,\zeta)$ over $r_0$, we can determine the classical value of $r_0$. We now need to expand the action \eqref{eq:full-reduced-action} in perturbations around the $r_0$ extremizing \eqref{eq:rconst-action},
\begin{equation}
 r(\sigma)=r_0+\eta(\sigma),
 \qquad
 \eta(\sigma)=\eta_0+\widetilde\eta(\sigma),
 \qquad
 \widetilde\eta(\sigma)=\frac{1}{\sqrt2}
 \sum_{n\neq0}\eta_n e^{2\pi\i n\sigma}, 
 \label{eq:Fourier-expansion}
\end{equation}
with $\eta_{-n}=\eta_n^*$ imposed by reality condition $\eta(\sigma)\in \mathbb R$. Constant $r_0$ is determined by $\zeta$ via \eqref{eq:r0-zeta-relation}.

One can show that the quadratic part of \eqref{eq:full-reduced-action} in the $\eta$-variables is:
\begin{equation}
    I^{(2)}=\frac{D-2}{16\pi G_2}(\beta_L+\beta_R)r_0^{D-5} \left[a_0\eta_0^2+\sum_{n=1}^{\infty}a_n|\eta_n|^2\right],
 \label{eq:quadratic-action}
\end{equation}
where
\begin{align}
 a_0&=(D-2)(D-3)+(D-1)^2\frac{r_0^2}{\ell^2}
 -(D-3)\frac{\mathcal Q^2}{r_0^{2(D-3)}},
 \label{eq:zero-mode-coefficient}\\
 a_n&=\widetilde m^2+(2\pi n)^2\mu,
 \label{eq:nonzero-mode-coefficient}\\
 \widetilde m^2
 &=(D-3)\left((D-2)\frac{\mathcal Q^2}{r_0^{2(D-3)}}-1\right)
 \nonumber\\[-1mm]
 &\quad-\frac{(D-1)^2}{4}
 \left(\frac{\beta_L}{\beta_R}+\frac{\beta_R}{\beta_L}-2\right)
 \left[\frac{r_0^2}{\ell^2}+\frac{D-3}{D-1}
 \left(1-\frac{\mathcal Q^2}{r_0^{2(D-3)}}\right)\right],
 \label{eq:nonzero-mass}
 \\
 \mu&=\frac{r_0^2}{\beta_L\beta_R}
 \left[\frac{r_0^2}{\ell^2}+\frac{D-3}{D-1}
 \left(1-\frac{\mathcal Q^2}{r_0^{2(D-3)}}\right)\right]^{-1}.
 \label{eq:kinetic-coefficient}
\end{align}

The constant mode is strictly stable. To see this, rewrite
the radial derivative of the lapse function as
\begin{equation}
    r_0f_i'(r_0) = (D-1)\frac{r_0^2}{\ell^2} +(D-3)\left[ 1-f_i(r_0)-\frac{\mathcal Q^2}{r_0^{2(D-3)}} \right].
\end{equation}
Substituting this expression into the junction condition \eqref{eq:2bdry_wormhole_masses_eq},
and using $f_L(r_0),f_R(r_0)>0$ gives
\begin{equation}
    (D-1)\frac{r_0^2}{\ell^2} +(D-3)\left[1-\frac{\mathcal Q^2}{r_0^{2(D-3)}}\right] =(D-1)\sqrt{f_L(r_0)f_R(r_0)}>0.\label{eq:2bdry-junction-ineq}
\end{equation}
Consequently, $a_0 > (D-3)^2+(D-1)(D-2)r_0^2/\ell^2 > 0.$ Moreover, comparing \eqref{eq:kinetic-coefficient} and \eqref{eq:2bdry-junction-ineq}, we immediately see that the kinetic energy coefficient $\mu$ is also strictly positive, $\mu > 0.$ Therefore, the least stable mode is the one with lowest nonzero kinetic energy, that is, $\eta_1$.

We can see that the saddles are stable close to extremality, when $r_0\to R_0$, independently of the fixed boundary temperatures. Since the extremal radius satisfies $\frac{\mathcal Q^2}{R_0^{2(D-3)}} = 1+\frac{D-1}{D-3}\frac{R_0^2}{\ell^2}$, the factor multiplying the negative temperature-dependent term in \eqref{eq:nonzero-mass} tends to zero, and $\lim_{r_0\to R_0}\widetilde m^2=(D-3)^2+(D-1)(D-2)\frac{R_0^2}{\ell^2}>0.$ Moreover, the circumference $L[r_0]$ of the wormhole neck vanishes in this limit, so that $\mu \to +\infty$ at fixed $\beta_L,\beta_R$. Thus all nonzero Fourier coefficients $a_n=\widetilde m^2+(2\pi n)^2\mu$ are positive sufficiently close to extremality. This conclusion applies to both asymptotically flat and asymptotically AdS cases.

In the asymptotically flat case, $\ell^{-2}=0$, positivity of the lapse functions and the background junction condition imply $0<\mathcal Q^2/r_0^{2(D-3)}<1$ and
$r_0>|\mathcal Q|^{1/(D-3)}$. The negative contribution to $\tilde{m}^2$ from
unequal temperatures, that is, the term proportional to $\left(\frac{\beta_L}{\beta_R}+\frac{\beta_R}{\beta_L}-2\right)$ in \eqref{eq:nonzero-mass}, is therefore bounded, whereas the kinetic contribution grows with the neck radius. More precisely,
\begin{equation}
    a_1> \frac{4\pi^2(D-1)}{(D-3)\beta_L\beta_R} |\mathcal Q|^{\frac{2}{D-3}} -(D-3) -\frac{(D-3)(D-1)}{4} \left(\frac{\beta_L}{\beta_R} + \frac{\beta_R}{\beta_L} - 2\right).
\end{equation}
Clearly, for any finite $\beta_L$ and $\beta_R$, this lower bound becomes positive for sufficiently large charge. Consequently, configurations with neck radii $r_0$ larger than the extremal horizon radius $R_0 = |\mathcal{Q}|^{\frac{1}{D-3}}$ are stable for sufficiently large $\mathcal{Q}$. Nevertheless, at finite charge, negative modes can occur at intermediate radii.

In asymptotically AdS space, at equal inverse temperatures $\beta_L=\beta_R=\beta$, all admissible saddles are stable at sufficiently large charge provided that $\beta<2\pi\ell/\sqrt{D-3}$. The restriction on $\beta$ is necessary since, at any fixed charge, $\lim_{r_0\to\infty}a_1 =\frac{4\pi^2\ell^2}{\beta^2}-(D-3).$  If $\beta>2\pi\ell/\sqrt{D-3}$, the saddle therefore becomes unstable when $r_0$ is sufficiently large.  For unequal temperatures, the negative term in \eqref{eq:nonzero-mass} proportional to $\left(\frac{\beta_L}{\beta_R} + \frac{\beta_R}{\beta_L} - 2\right)$ grows quadratically with $r_0$, whereas the kinetic contribution approaches a finite value:
\begin{equation}
    a_1 \overset{r_0\to \infty}{\sim} -\frac{(D-1)^2}{4} \left(\frac{\beta_L}{\beta_R} + \frac{\beta_R}{\beta_L} - 2\right) \frac{r_0^2}{\ell^2} + \O(1).
\end{equation}
Consequently, for every charge and fixed unequal temperatures, there are saddles with sufficiently large $r_0$ that are unstable.

We stress that the Gaussian approximation applied in this section breaks down even when $a_1$ is positive but very small, and it diverges when we sum over constrained saddles with $a_1\to 0$. This divergence arises from the integral over $\zeta$, upon accounting for the one-loop correction associated with the mode $\eta_1$. Using $\zeta=\Omega_{D-2}r_0^{D-2}L[r_0]$, we can replace this integral by an integral over $r_0$. Suppose that $a_1$ has a simple zero at $r_0=r_0^*$, with stable saddles on the side $r_0<r_0^*$. The first Fourier harmonic contains two real fluctuation amplitudes, so its Gaussian integral contributes a factor $1/a_1$. Since $\dd\zeta/\dd r_0$ is finite and nonzero at $r_0^*$, even restricting the integration to real stable saddles gives, up to smooth nonvanishing factors,
\begin{equation}
\int^{r_0^*}\dd r_0\,
\frac{\dd\zeta}{\dd r_0}\frac{1}{a_1(r_0)}
\sim
\int^{r_0^*}\frac{\dd r_0}{r_0^*-r_0},
\end{equation}
which diverges logarithmically. This does not establish a divergence of the full path integral: the Gaussian approximation fails when the quadratic coefficient vanishes, and higher-order terms in these fluctuation amplitudes must then be included near the critical radius $r_0 = r_0^*$. And, for the asymptotically flat saddles with large electric charges that we will use in the next section, $a_1$ is positive at every admissible radius, so this particular obstruction does not arise.

\subsection{Stability of multi-boundary wormholes}
\label{sec:multi-boundary-stability}

We now analyze dilaton fluctuations in the multi-boundary wormhole core and show that the linearized lapse and shift constraints eliminate every smooth, single-valued dilaton deformation. Continuity then removes the radial fluctuations at the constraint surfaces. 

The core background satisfies
\begin{equation}
    \bar X=X_0,\qquad V(X_0)=0,\qquad
    V'(X_0)=2a^2,\qquad \hat R=-2a^2,
    \qquad a^2>0.
    \label{eq:constant-dilaton-background}
\end{equation}
We write $X=X_0+\mathcal{X}$ with small $\mathcal{X}$ and denote the background Weyl-frame metric by $\hat g_{ab}$. To fix the metric gauge, we use local Gaussian normal coordinates. In particular, we should not use the gauge from the previous subsection, as the dilaton field is constant on the wormhole core background. Starting from
\begin{equation}
    \dd\hat s^2=L^2\dd\rho^2
    +h(\dd\tau+N\dd\rho)^2,
\end{equation}
we set $L=1$ and $N=0$. Here $L$ denotes the ADM lapse. This gauge is regular on the constant-dilaton background and can be implemented in a neighborhood of a constraint brane; a single Gaussian chart covering the entire core is unnecessary. On the background, we write
\begin{equation}
    \dd\hat s^2=\dd\rho^2+\gamma(\rho,\tau)^2\dd\tau^2,
    \qquad
    n=\partial_\rho,\qquad
    s=\gamma^{-1}\partial_\tau,
    \label{eq:Gaussian-core-metric}
\end{equation}
where $n$ and $s$ are the unit normal and tangent to the $\rho=\mathrm{const.}$ slices.

The equations obtained by varying the lapse and shift before gauge fixing must still be imposed as constraints. To linearize them, consider the metric equation
\begin{equation}
    E_{ab}
    =\hat\nabla_a\hat\nabla_bX-\hat g_{ab}\hat\nabla^2X
    +\frac12V(X)\hat g_{ab}.
\end{equation}
Around \eqref{eq:constant-dilaton-background},
\begin{equation}
    \delta E_{ab}
    =\hat\nabla_a\hat\nabla_b\mathcal{X}
    -\hat g_{ab}\hat\nabla^2\mathcal{X}
    +a^2\hat g_{ab}\mathcal{X}.
    \label{eq:linearized-metric-equation}
\end{equation}
Metric perturbations drop out because $\hat\nabla_aX_0=0$ and $V(X_0)=0$; we have not set them to zero. The linearized lapse and shift constraints are
\begin{equation}
    \delta E_{nn}=0,\qquad \delta E_{ns}=0.
    \label{eq:linearized-core-constraints}
\end{equation}

Off-shell we cannot independently impose the remaining equation
$\delta E_{ss}=0$ on the fluctuations. Instead, this equation follows from the constraints. For arbitrary $\mathcal{X}$,
the tensor \eqref{eq:linearized-metric-equation} obeys
\begin{equation}
    \hat\nabla^a\delta E_{ab}
    =\left(\hat R_b{}^a+a^2\delta_b^a\right)
      \hat\nabla_a\mathcal{X}
    =0,
    \label{eq:linearized-diffeomorphism-identity}
\end{equation}
where the last equality uses
$\hat R_{ab}=-a^2\hat g_{ab}$. This is an identity, not an
additional equation imposed on $\mathcal{X}$. After imposing
\eqref{eq:linearized-core-constraints}, we can write
$\delta E_{ab}=S s_as_b$, with $S=\delta E_{ss}$.
Since $\hat\nabla_s s=-(\partial_\rho\log\gamma)\,n$, where
$K=\partial_\rho\log\gamma$, the normal projection of
\eqref{eq:linearized-diffeomorphism-identity} gives
\begin{equation}
    0=n^b\hat\nabla^a(Ss_as_b)=-(\partial_\rho\log\gamma)S.
    \label{eq:KS-constraint}
\end{equation}
The term $\partial_\rho\log(\gamma)$ cannot vanish on a non-empty open set of a constant negative-curvature surface, and therefore, $S=0.$ It follows that $\delta E_{ss} = 0$, and thus, all linearized metric equations of motion are satisfied off-shell, $\delta E_{ab}=0$.

Taking the trace of $\delta E_{ab}$ and substituting it back into
\eqref{eq:linearized-metric-equation}, we obtain
\begin{equation}
    \hat\nabla^2\mathcal{X}=2a^2\mathcal{X},
    \qquad
    \hat\nabla_a\hat\nabla_b\mathcal{X}
    =a^2\mathcal{X}\,\hat g_{ab}.
    \label{eq:dilaton-Hessian-equation}
\end{equation}
It follows that
\begin{equation}
    \xi_a=\epsilon_a{}^b\hat\nabla_b\mathcal{X},
    \qquad
    \hat\nabla_{(a}\xi_{b)}=0,
    \label{eq:induced-Killing-vector}
\end{equation}
where $\epsilon_{ab}$ is the covariantly constant volume form.
Every smooth, single-valued dilaton fluctuation satisfying the
constraints therefore defines a global Killing vector on the core. However, the negative-curvature cores used in our construction admit no nonzero globally defined Killing vectors (see Sect. 3.4.2 of \cite{Saad:2019lba}). Consequently, there are no dilaton fluctuations, $\xi=0$.

In the previous subsection we argued that the only dynamical modes in a wormhole exterior are the dilaton fluctuations on the constrained surface. However, now we see that, upon imposing continuity of the fields across the constraint brane, these fluctuations are set to zero since there are no smooth dilaton fluctuations in the core. Moreover, constraint surface reparametrization modes do not acquire an action from the wormhole core. Using $X=X_0 = {\rm const.}$, $V(X_0)=0$ and including the GHY terms on all constraint surfaces, Gauss--Bonnet gives
\begin{align}
    I^{\mathrm{EH}}+\sum_i I_i^{\hat \cS} =-\frac{X_0}{16\pi G_2}
 \left[
 \int_{\Sigma_{g,n}}\dd^2x\sqrt{\hat g}\,R(\hat g)
 +2\sum_i\int_{\hat \cS_i}\dd\hat s\,\hat k_i
 \right] = -\frac{X_0}{4G_2}\chi(\Sigma_{g,n}),
\end{align}
where $\chi(\Sigma_{g,n})$ is the Euler characteristic of the core. Thus, although the individual GHY terms can change when the junctions move, their variation is exactly canceled by the bulk contribution.

Consequently, the constrained quadratic action has no negative modes associated with the core or the junctions: the dilaton and radial modes are eliminated by the constraints, while the remaining metric moduli and junction deformations have vanishing classical quadratic action. Neither their integration measure nor the treatment of the asymptotic boundary reparametrizations is determined by this argument.

\section{Positivity implies the Weak Gravity Conjecture}\label{sec:wgc-proof}

In this section we formulate an argument for electric WGC following the proof strategy of \cite{DiUbaldo:2026rly} for axionic WGC. The key input of that argument is that the action of two-boundary axion wormholes with fixed flux is linear in the flux. In the ensemble with fixed boundary values of the axion, the partition function is a sum of partition functions in the fixed-flux ensemble, and, as a result, the sum over fluxes of the two-boundary wormhole contributions diverges when the difference between the boundary values of the axion reaches a critical value. This divergence is in conflict with the positivity of the Marolf--Maxfield inner product~\cite{Marolf:2020xie} unless the wormholes are destabilized by instantons. Requiring that instantons destabilize the two-boundary wormholes before the boundary axion value reaches the critical value leads to the axionic weak gravity conjecture.

More specifically, the GPI with one open asymptotic boundary and boundary condition $\mathcal{J}$ prepares a state $|\mathcal{J}\rangle$, with inner product defined as $\langle\mathcal{J}_1|\mathcal{J}_2\rangle = Z(\mathcal{J}_1^\Theta\cup\mathcal{J}_2)$, where $Z(\mathcal J)$ is the GPI sum over geometries satisfying the boundary condition $\mathcal J$, and $\Theta$ denotes CPT conjugation. Positivity of the inner product implies that the GPI with $n$ asymptotic boundaries, each with boundary condition $\mathcal J$, which we denote as $\overline{Z(\mathcal{J})^n}$, can be interpreted as the $n$-th moment of a random variable $Z(\mathcal{J})$ with respect to an average or a pseudo-average \cite{DiUbaldo:2026tkx}. It follows that the corresponding cumulants $\kappa_n = \overline{Z(\mathcal{J})^n}_{\rm connected}$ obey \cite{DiUbaldo:2026rly}
\begin{equation}
    \kappa_1\ge0,\qquad \kappa_2\ge0,\qquad \kappa_1\kappa_3 \ge\kappa_2\bigl(\kappa_2-\kappa_1^2\bigr). \label{eq:wgc-positivity}
\end{equation}
The $n$-th cumulant is determined by connected geometries with $n$ asymptotic boundaries. We want to show that for certain boundary conditions $\mathcal{J}$ the two-boundary contribution is so large that $\kappa_2^2\gg \kappa_1\kappa_3$ and $\kappa_2\gg\kappa_1^2$, so that the third inequality in \eqref{eq:wgc-positivity} is violated. Requiring wormholes satisfying such boundary conditions to be unstable leads to the WGC.

As we will show below, similar reasoning applies to Einstein--Maxwell theory in asymptotically flat space. The role of the axion wormholes is played by our constrained two-boundary wormholes, the boundary values of the axion translate to the electric potential, and the instantons translate to electrically charged particles. There is, however, an important difference from the axion-wormhole case. The action of our two-boundary wormholes becomes linear in the electric flux only in the limit of vanishing constrained area, $\zeta\to0$. In this limit, the Euclidean time circle at the wormhole neck shrinks to zero size, quantum fields near the neck are at a very high local temperature, and thus, their backreaction invalidates the semiclassical description. We therefore introduce a UV cutoff $\zeta_c$, restricting the constrained areas to $\zeta\geq\zeta_c$. At finite cutoff, the sum over fluxes of the two-boundary contributions converges. However, above a critical value of the electric potential, it grows so fast at large inverse temperature in Planck units that the third inequality in \eqref{eq:wgc-positivity} remains violated, assuming that there are no nontrivial cancellations with the $\zeta < \zeta_c$ sector contributions.

To estimate $\kappa_1$, we analyze the single-boundary contribution in asymptotically flat space. At fixed charge there are two candidate saddles: small and large RN black holes, the latter being thermodynamically unstable. The black-hole contributions are problematic once the geometries with a conical singularity at the horizon are included, since then the single-boundary GPI becomes an integral over the horizon area \cite{Carlip:1993sa,Marolf2022thermal}, in which the Boltzmann factor $e^{-\beta E}$ competes with $e^{\frac{\rm Area}{4G_D}}$. In AdS, this integral converges, and its saddles were analyzed in~\cite{AMN2026,KM2026}. In flat space, however, the entropy grows faster than the energy at large horizon areas, and the integral diverges.\footnote{A finite cavity makes the canonical ensemble well defined~\cite{York:1986it,Braden:1990hw,LW1995}, but the divergence reappears when the cavity is removed.} To make the flat-space integral convergent, we rotate its integration contour in the complex plane of the horizon radius $r_+$. This can be thought of as the analytic continuation of the AdS contour~\cite{AMN2026,KM2026} to complex values of $\beta$. A Picard--Lefschetz analysis~\cite{Fedoryuk:1976,Pham:1983,Arnold:1988,Berry:1990,Berry:1991,Howls:1997}\footnote{See~\cite{Witten:2010cx} for a physicist-oriented review.} then shows that at fixed charge only one saddle contributes: the small black hole with positive heat capacity (or its analytic continuation in the range of parameters where the real saddle does not exist). The larger black hole, which has negative heat capacity, does not contribute. At fixed electric potential there are also two types of saddles: hot flat space\footnote{Gross, Perry and Yaffe argued that hot flat space is unstable to the nucleation of black holes~\cite{GPY1982}: the Euclidean Schwarzschild solution has a single negative mode, which gives the free energy of hot flat space an imaginary part. We do not address this instability here. In particular, we do not include saddles with several black holes, which would describe the nucleation of more than one black hole; smooth saddles of this type are not known at finite temperature.} and RN black holes. The black-hole contributions come from saddles whose electric potentials are shifted by $\frac{2\pi\i n}{\beta e}$, where $e$ is the elementary charge and $n$ is a positive integer for $\Phi>0$ (in $D>4$, only some of these saddles contribute). They dominate over hot flat space once $\Phi$ exceeds $\sin[\pi/(2(D-2))]$ times the extremal potential. We present these results in Section~\ref{sec:moments} and detail the Picard--Lefschetz analysis in Appendix~\ref{app:picard-lefschetz}.

\subsection{Calculating on-shell actions}\label{sec:onshell-actions}
First, we need to estimate how the wormhole saddles contribute to the gravitational path integral. To this end, we evaluate the Einstein--Maxwell action on the configurations constructed in Section~\ref{sec:grav-constr-saddles}. We work in the ensemble with fixed electric flux, in which the Euclidean action, including the constraint branes, reads
\begin{align}
I ={}& I_{\rm EM} + \sum_i\lambda_i\bigl(\mathrm{vol}(\cS_i)-\zeta_i\bigr), \label{eq:os-full-action}
\end{align}
where $I_{\rm EM}$ is the standard action of Einstein--Maxwell theory,
\begin{align}
    I_{\rm EM} ={}& -\frac{1}{16\pi G_D}\int_M\dd^D x\sqrt{g}\,(R-2\Lambda) + \frac{1}{4}\int_M \dd^D x\sqrt{g}\,F_{\mu\nu}F^{\mu\nu} \nonumber\\
    & - \frac{1}{8\pi G_D}\int_{\partial M}\dd^{D-1}x\sqrt{h}\,K + I_N + I_{\rm ct}.
    \label{eq:os-EM-action}
\end{align}
Here
\begin{equation}
I_N = -\int_{\partial M}\dd^{D-1}x\sqrt{h}\, n_\mu A_\nu F^{\mu\nu}\label{eq:INeumann}
\end{equation}
is the boundary term which fixes the electric flux through the asymptotic boundaries \cite{Hawking:1995ap}, and $I_{\rm ct}$ is a suitable counterterm, which, upon dimensional reduction on $S^{D-2}$, reduces to \eqref{eq:2d-asymptotic-counterterm}. As explained in Section~\ref{sec:stability}, we evaluate the curvature separately in each smooth region and include a GHY term on both sides of every constraint surface. This accounts for the distributional curvature supported on the branes; see Eq.~\eqref{eq:ghy-gluing-action}.

On a constrained saddle, $\mathrm{vol}(\cS_i) = \zeta_i$, so the term proportional to the Lagrange multiplier $\lambda_i$ in  \eqref{eq:os-full-action} vanishes, and the saddle contributes to the path integral with weight $e^{-I_{\rm EM}}$ integrated over the constrained areas,
\begin{equation}
Z \supset \int\prod_i\dd\zeta_i\,\exp\bigl[-I_{\rm EM}(\zeta)\bigr],
\label{eq:os-weight}
\end{equation}
up to loop corrections.

\paragraph{Wormhole exterior.} First, let us calculate the Einstein--Maxwell action of one half of the two-boundary wormhole constructed in Section~\ref{sec:2bdry-saddles}. To facilitate the computation, we use the canonical formalism \cite{Hawking:1995fd}. 

We consider the region $r_0\le r\le r_c$ of the exterior geometry \eqref{eq:RN-metric}, bounded by the constraint surface $\cS=\{r=r_0\}$ and the cutoff surface $r=r_c\to\infty$. We refer to this region as a wormhole exterior. It is stationary, non-rotating, and contains no horizon, since $f(r)>0$ for $r\ge r_0$. Therefore, the Euclidean time $\tau\sim\tau+1$ of \eqref{eq:RN-metric} defines a regular foliation of the wormhole exterior by the slices $\Sigma=[r_0,r_c]\times S^{D-2}$, and its boundary has the form $S^1\times\partial\Sigma$, where $\partial\Sigma$ consists of the two spheres at $r=r_0$ and $r=r_c$. The corresponding ``$3+1$'' field decomposition is
\begin{equation}
\dd s^2 = N^2\dd\tau^2 + \gamma_{ab}\,(\dd x^a + N^a\dd\tau)(\dd x^b + N^b\dd\tau),\qquad A = A_\tau\,\dd\tau + A_a\,\dd x^a,
\label{eq:os-ADM}
\end{equation}
where indices $a,b$ label the spatial directions, $N$ is the lapse, $N^a$ is the shift, and $\gamma_{ab}$ is the metric induced on $\Sigma$. The momenta conjugate to $\gamma_{ab}$ and $A_a$ are
\begin{equation}
\pi^{ab} = \frac{\sqrt{\gamma}}{16\pi G_D}\bigl(K^{ab}-K\gamma^{ab}\bigr),\qquad \mathcal{E}^a = \sqrt{\gamma}\,n_\mu F^{\mu a},
\end{equation}
where $n^\mu$ is the unit normal to $\Sigma$, and $K_{ab} = \frac{1}{2N}\bigl(\dot\gamma_{ab}-D_aN_b-D_bN_a\bigr)$ is the extrinsic curvature of $\Sigma$. Dots denote derivatives with respect to $\tau$, and $D_a$ is the covariant derivative compatible with $\gamma_{ab}$. In terms of these variables, the action \eqref{eq:os-EM-action} takes the following form
\begin{align}
I_{\rm EM} ={}& \int_0^1\dd\tau\int_\Sigma\dd^{D-1}x\,\Bigl(\pi^{ab}\dot\gamma_{ab} + \mathcal{E}^a\dot A_a + N\cH + N^a\cH_a + A_\tau\,D_a\mathcal{E}^a\Bigr)\nonumber\\
&-\frac{1}{8\pi G_D}\int_0^1\dd\tau\int_{\partial\Sigma}\dd^{D-2}y\,\sqrt{\sigma}\,\Bigl[N\,(\vartheta-\vartheta_0) + r_aN_b\bigl(K^{ab}-K\gamma^{ab}\bigr)\Bigr]\nonumber\\
&-\int_0^1\dd\tau\int_{\partial\Sigma}\dd^{D-2}y\,\sqrt{\sigma}\,N\,r_aA_bF^{ab},
\label{eq:os-canonical-action}
\end{align}
where
\begin{align}
\cH &= -\frac{16\pi G_D}{\sqrt{\gamma}}\Bigl(\pi_{ab}\pi^{ab}-\frac{\pi^2}{D-2}\Bigr) - \frac{\sqrt{\gamma}}{16\pi G_D}\bigl(\mathcal{R}[\gamma]-2\Lambda\bigr) - \frac{\mathcal{E}_a\mathcal{E}^a}{2\sqrt{\gamma}} + \frac{\sqrt{\gamma}}{4}F_{ab}F^{ab},\nonumber\\
\cH_a &= 2\gamma_{ac}D_b\pi^{bc} - F_{ab}\mathcal{E}^b,
\label{eq:os-constraints}
\end{align}
are the Hamiltonian, momentum and the Gauss constraint $D_a \mathcal{E}^a$, and $\pi = \gamma_{ab}\pi^{ab}$. In the boundary terms, $\sigma_{AB}$ is the metric induced on $\partial\Sigma$, $r^a$ is the outward unit normal to $\partial\Sigma$ in $\Sigma$, and $\vartheta = D_ar^a$ is the trace of the extrinsic curvature of $\partial\Sigma$ in $\Sigma$. The term proportional to $\vartheta_0$ comes from the counterterm $I_{\rm ct}$, and $\vartheta_0$ is the value of $\vartheta$ for a sphere of the same radius in the reference background, either flat space or pure AdS.

The action \eqref{eq:os-canonical-action} imposes Neumann boundary conditions for the flux through the boundary $\partial\Sigma$, namely $\delta\mathcal{E}^a|_{\partial \Sigma} = 0$, since the variation of the term $\int A_\tau D_a \mathcal E^a$ gives a boundary term $\int_{\partial \Sigma} \dd^{D-2}y\sqrt{\sigma}\,r_aA_\tau \delta \mathcal E^a$.

Let us now evaluate the terms in \eqref{eq:os-canonical-action} on the wormhole exterior, for which $N=\beta\sqrt{f(r)}$, $N^a=0$, $\gamma_{ab}\dd x^a\dd x^b = \dd r^2/f(r) + r^2\dd\Omega_{D-2}^2$, and, in a suitable gauge, $A = A_\tau\dd\tau$. The wormhole exterior is static, $\dot\gamma_{ab}=\dot A_a=0$, so the terms $\pi^{ab}\dot\gamma_{ab}$ and $\mathcal{E}^a\dot A_a$ vanish. Since $N^a=0$, the term $N^a\cH_a$ and the shift-dependent boundary term vanish as well. The magnetic boundary term vanishes because $A_a=0$. The remaining bulk terms, $N\cH$ and $A_\tau D_a \mathcal{E}^a$, are equal to zero upon integrating out the Lagrange multipliers $N$ and $A_\tau$. Therefore, the on-shell action of the wormhole exterior reduces to
\begin{equation}
I_{\rm EM} = -\frac{1}{8\pi G_D}\int_0^1\dd\tau\int_{\partial\Sigma}\dd^{D-2}y\,\sqrt{\sigma}\,N\,(\vartheta-\vartheta_0),
\label{eq:os-canonical-onshell}
\end{equation}
which receives contributions from the two components of $\partial\Sigma$. At the cutoff sphere, $\vartheta = (D-2)\sqrt{f(r_c)}/r_c$, while the reference value in flat space ($\ell\to\infty$) or pure AdS is $\vartheta_0 = (D-2)\sqrt{\tfrac{r_c^2}{\ell^2}+1}\,/r_c$. Hence,
\begin{align}
    &-\frac{1}{8\pi G_D}\int_0^1\dd\tau\int_{r=r_c}\dd^{D-2}y\,\sqrt{\sigma}\,N\,(\vartheta-\vartheta_0) \nonumber
    \\
    &= -\frac{(D-2)\Omega_{D-2}}{8\pi G_D}\,\beta\, r_c^{D-3}\sqrt{f(r_c)}\left(\sqrt{f(r_c)}-\sqrt{\tfrac{r_c^2}{\ell^2}+1}\right) \xrightarrow{r_c\to\infty}\beta E, \label{eq:os-asymptotic}
\end{align}
where $E = \frac{(D-2)\Omega_{D-2}}{8\pi G_D}M$ is the ADM energy. The same result follows from the reduced counterterm \eqref{eq:2d-asymptotic-counterterm}; see Section~\ref{sec:two-boundary-stability}. At the constraint surface, the outward normal points toward decreasing $r$, so $\vartheta = -(D-2)\sqrt{f(r_0)}/r_0$, and there is no contribution from the counterterm. Hence, the action of the wormhole exterior, including the GHY term on $\cS$, is
\begin{equation}
I_{\rm ext}(\beta,Q;r_0,M) = \beta E + \frac{(D-2)\Omega_{D-2}}{8\pi G_D}\,\beta\, r_0^{D-3}f(r_0).
\label{eq:os-ext-action}
\end{equation}
The second term is the contribution of the GHY term on the constrained surface.

It is instructive to separate the extremal mass in \eqref{eq:os-ext-action}:
\begin{equation}
I_{\rm ext}(\beta,Q;r_0,M) = \beta E_{\rm ext}(Q) + \frac{(D-2)\Omega_{D-2}}{8\pi G_D}\,\beta\left[\bigl(M_H(r_0)-M_{\rm ext}\bigr) + \frac{1}{2}\,r_0^{D-3}f(r_0)\right],
\label{eq:os-ext-action-MH}
\end{equation}
where $E_{\rm ext}(Q) = \frac{(D-2)\Omega_{D-2}}{8\pi G_D}M_{\rm ext}$ is the energy of the extremal RN--AdS black hole of charge $Q$, and
\begin{equation}
M_H(r_0) = \frac{r_0^{D-3}}{2}\left(1+\frac{r_0^2}{\ell^2}+\frac{\mathcal{Q}^2}{r_0^{2(D-3)}}\right).
\end{equation}
Both terms in the square brackets are non-negative. The second one is non-negative for every real wormhole exterior, $f(r_0)\ge0$. To see that the first one is non-negative, note that $\dd M_H/\dd r_0 = 2\pi r_0^{D-3}/\beta_H(r_0)$, where $\beta_H(r_0)$ is the inverse Hawking temperature of RN--AdS black hole of charge $Q$ and horizon radius $r_H=r_0$. Since $\beta_H(r_0)<0$ for $r_0<R_0$ and $\beta_H(r_0)>0$ for $r_0>R_0$, $M_H$ reaches its minimum at the extremal radius, $r_0 = R_0$, where $M_H(R_0) = M_{\rm ext}$. Therefore,
\begin{equation}
    I_{\rm ext}(r_0) \geq \beta E_{\rm ext}(Q).\label{eq:os-action-bound}
\end{equation}
The inequality is saturated at $r_0 = R_0.$

\paragraph{Two-boundary wormhole.} The two-boundary saddle of Section~\ref{sec:2bdry-saddles} is obtained by gluing two wormhole exteriors, labeled $L$ and $R$, with common charge $Q$ and radius $r_0$ along $\cS$. Hence, its action is simply the sum of the actions of the two exteriors,
\begin{equation}
    I_{\rm 2\text{-}bdry}(\beta_L,\beta_R,Q;r_0) = I_{{\rm ext},L} + I_{{\rm ext},R} = \sum_{i=L,R}\beta_i\left[E_i + \frac{(D-2)\Omega_{D-2}}{8\pi G_D}\,r_0^{D-3}f_i(r_0)\right],
\label{eq:os-2bdry-action}
\end{equation}
where the mass parameters $M_i$ are determined by $r_0$ through \eqref{eq:2bdry_tautau_continuity} and \eqref{eq:2bdry_wormhole_masses_eq}. From Eq.~\eqref{eq:os-action-bound} it follows that
\begin{equation}
    I_{\rm 2\text{-}bdry}(\beta_L,\beta_R,Q;r_0) \ge (\beta_L+\beta_R)\,E_{\rm ext}(Q),
\end{equation}
with equality attained at the extremal end of the family, $r_0\to R_0$, where the brane tension vanishes and the wormhole develops an infinitely thin, long throat.

\paragraph{Multi-boundary wormholes.} For the $n$-boundary saddles of Section~\ref{sec:multi-bdry-saddles}, the Einstein--Maxwell action decomposes into the action of the core and the actions of the $n$ exterior regions,
\begin{equation}
I_{\rm EM} = I_{\rm core} + \sum_{i=1}^n I_{{\rm ext},i}.
\end{equation}
Each exterior region is of the form considered above, with $r_0=R_0$, so \eqref{eq:os-ext-action} gives
\begin{equation}
I_{{\rm ext},i} = \beta_iE_i + \frac{(D-2)\Omega_{D-2}}{8\pi G_D}\,\beta_iR_0^{D-3}f_i(R_0) = \beta_i\left[E_{\rm ext}(Q) + \frac{(D-2)\Omega_{D-2}}{16\pi G_D}\,R_0^{D-3}f_i(R_0)\right],
\label{eq:os-exterior}
\end{equation}
where in the second step we used $f_i(R_0) = 2(M_{\rm ext}-M_i)/R_0^{D-3}$.

The core $M_{\rm core} = \Sigma_{g,n}\times S^{D-2}$ is spherically symmetric. As shown in Section~\ref{sec:dim-reduction}, for such configurations the Einstein--Maxwell action reduces to the action \eqref{eq:2d-action2} of two-dimensional dilaton gravity. On the core, the dilaton is constant, $X = X_0 \equiv R_0^{D-2}$, the Weyl factor \eqref{eq:weyl-rescaling} is trivial, $\hat g$ is the hyperbolic metric, and $V(X_0)=0$. Therefore, including the GHY terms on the core side of the junctions, the Einstein--Maxwell action of the core reduces to the Euler characteristic of $\Sigma_{g,n}$ by the Gauss--Bonnet theorem,
\begin{equation}
I_{\rm core} = -\frac{X_0}{16\pi G_2}\left[\int_{\Sigma_{g,n}}\dd^2x\sqrt{\hat g}\,R(\hat g) + 2\sum_{i=1}^n\int_{\hat\cS_i}\dd\hat s\,\hat k_i\right] = -\frac{X_0}{4G_2}\,\chi(\Sigma_{g,n}) = -\chi(\Sigma_{g,n})\,S_{\rm BH}(Q).
\label{eq:os-core-action}
\end{equation}
Here $G_2 = G_D/\Omega_{D-2}$, $\chi(\Sigma_{g,n}) = 2-2g-n$, and
\begin{equation}
    S_{\rm BH}(Q) = \frac{\Omega_{D-2}R_0^{D-2}}{4G_D}
\end{equation}
is the Bekenstein--Hawking entropy of the extremal black hole of charge $Q$.

Combining \eqref{eq:os-exterior} and \eqref{eq:os-core-action}, we obtain
\begin{equation}
I_{\rm EM} = \sum_{i=1}^n\beta_i\left[E_{\rm ext}(Q) + \frac{(D-2)\Omega_{D-2}}{16\pi G_D}\,R_0^{D-3}f_i(R_0)\right] - \chi(\Sigma_{g,n})\,S_{\rm BH}(Q).
\label{eq:os-multi-action}
\end{equation}
In terms of the constrained areas $\zeta_i = \Omega_{D-2}R_0^{D-2}\beta_i\sqrt{f_i(R_0)}$, the $f_i$-dependent terms in \eqref{eq:os-multi-action} are $\frac{(D-2)\,\zeta_i^2}{16\pi G_D\Omega_{D-2}R_0^{D-1}\beta_i}$.

\paragraph{Single-boundary saddles.}
We now evaluate the on-shell action \eqref{eq:os-full-action} of the single-boundary saddles of Section~\ref{sec:1bdry-saddles}, using the Lagrangian description. We introduce an IR cutoff at $r=r_c$ and take $r_c\to\infty$ at the end.

On shell, the trace of the Einstein equations gives $R=\frac{2D}{D-2}\,\Lambda+\frac{4\pi(D-4)}{D-2}\,G_DF_{\mu\nu}F^{\mu\nu}$, while the field strength in \eqref{eq:single-constrained-metric} gives $F_{\mu\nu}F^{\mu\nu}=-2Q^2/(\Omega_{D-2}^2r^{2(D-2)})$. Thus, integrating the bulk Lagrangian over the smooth part of the cigar, $r_+<r<r_c$, we find
\begin{align}
    I_{\rm bulk}
    &=\int\dd^Dx\sqrt{g}\left[-\frac{R-2\Lambda}{16\pi G_D}+\frac14F_{\mu\nu}F^{\mu\nu}\right] \nonumber\\
    &=\frac{\beta\Omega_{D-2}}{8\pi G_D}\left[\frac{r_c^{D-1}-r_+^{D-1}}{\ell^2}+\mathcal Q^2\left(r_c^{3-D}-r_+^{3-D}\right)\right].
\end{align}

Near the tip, the geometry is the product of a two-dimensional cone with opening angle $\beta f'(r_+)/2$ and the sphere $\mathcal S$. Hence, the scalar curvature contains the distributional term $2\left(2\pi-\beta f'(r_+)/2\right)\delta_{\mathcal S}$, where $\delta_{\mathcal S}$ is the delta function supported on the tip of the cigar. By \eqref{eq:single-tip-junction}, the corresponding contribution of the Einstein--Hilbert term cancels the term $\lambda\,{\rm area}(\mathcal S)$ in the constraint action, and we obtain
\begin{equation}
    I_{\rm tip}+I_{\mathcal C}
    =-\frac{1}{8\pi G_D}\left(2\pi-\frac{\beta f'(r_+)}{2}\right)\operatorname{area}(\mathcal S)
     +\lambda\left(\operatorname{area}(\mathcal S)-\zeta\right)
    =-\lambda\zeta,
\end{equation}
as for cosmic branes \cite{Lewkowycz:2013nqa,Dong:2016fnf}.

The rest of the calculation proceeds in the standard way. The GHY term is given by
\begin{align}
    -\frac{1}{8\pi G_D}\int_{r=r_c}\dd^{D-1}x\sqrt{h}\,K
    &=-\frac{\beta\Omega_{D-2}}{8\pi G_D}\left[(D-2)r_c^{D-3}\bigl(\tfrac{r_c^2}{\ell^2}+1\bigr)+\frac{r_c^{D-1}}{\ell^2}-(D-1)M+\mathcal Q^2r_c^{3-D}\right].
\end{align}
To evaluate the Maxwell boundary term \eqref{eq:INeumann}, we use the gauge field that is regular at the tip of the cigar, $A=\frac{\i\beta Q}{(D-3)\Omega_{D-2}}\left(r_+^{3-D}-r^{3-D}\right)\dd\tau.$ This gives
\begin{equation}
    I_N=\frac{(D-2)\beta\Omega_{D-2}}{8\pi G_D}\,\mathcal Q^2\left(r_+^{3-D}-r_c^{3-D}\right).
\end{equation}
Finally, the counterterm subtracts the action of the reference background, the thermal AdS (or flat space in the $\ell\to \infty$ limit), with inverse temperature $\beta_0$ fixed by requiring that the reference background induces the same metric on the cutoff surface as the single-boundary saddle of interest. Thus, the thermal circles at $r=r_c$ must have the same proper length, $\beta_0\sqrt{\tfrac{r_c^2}{\ell^2}+1}=\beta\sqrt{f(r_c)}$ \cite{Gibbons:1976ue,Hawking:1982dh}. Evaluating the bulk and GHY terms on the reference background, we find
\begin{align}
    I_{\rm ct}
    &
    =\frac{(D-2)\beta_0\Omega_{D-2}}{8\pi G_D}\,r_c^{D-3}\bigl(\tfrac{r_c^2}{\ell^2}+1\bigr) \nonumber\\
    &=\frac{\beta\Omega_{D-2}}{8\pi G_D}\left[(D-2)r_c^{D-3}\bigl(\tfrac{r_c^2}{\ell^2}+1\bigr)-(D-2)M\right]+O\bigl(r_c^{3-D}\bigr).
\end{align}

Adding all contributions, the terms that grow with $r_c$ cancel, and in the limit $r_c\to\infty$ we obtain
\begin{align}
    I_{\rm 1\text{-}bdry}(\beta,Q;\zeta)
    &=\frac{\beta\Omega_{D-2}}{8\pi G_D}\left[M-\frac{r_+^{D-1}}{\ell^2}+(D-3)\,\mathcal Q^2r_+^{3-D}\right]-\lambda\zeta \nonumber\\
    &
    =\beta E(\zeta,Q)-\frac{\zeta}{4G_D}.
    \label{eq:single-tip-reduced-action}
\end{align}
To write the second line, we used $\frac12r_+^{D-2}f'(r_+)=r_+^{D-1}/\ell^2+(D-3)\bigl(M-\mathcal Q^2r_+^{3-D}\bigr)$ and \eqref{eq:single-tip-junction}, and we introduced the ADM energy $E(\zeta,Q)=(D-2)\Omega_{D-2}M/(8\pi G_D)$, with $M$ given by \eqref{eq:single-constrained-energy}. The ADM energy is determined by $\zeta$ and $Q$ (and the AdS length) since $r_+$ is given by $\zeta=\Omega_{D-2}r_+^{D-2}$.

\subsection{From actions to moments}\label{sec:moments}

The actions of the two- and multi-boundary wormholes computed in Section~\ref{sec:onshell-actions} are minimized when the constrained areas vanish. By \eqref{eq:os-action-bound}, the action of the two-boundary wormholes decreases to $(\beta_L+\beta_R)E_{\rm ext}(Q)$ as $\zeta\to0$, and the $f_i$-dependent terms in \eqref{eq:os-multi-action}, which are proportional to $\zeta_i^2$, vanish as $\zeta_i\to0$. One may thus be tempted to estimate the integrals over the constrained areas in \eqref{eq:os-weight} by the contributions of their lower endpoints, i.e., of the extremal wormholes. However, this estimate is not reliable. As the constrained areas go to zero, the proper length of the Euclidean time circle goes to zero at the neck of the two-boundary wormholes, $b=\beta_L\sqrt{f_L(r_0)}=\beta_R\sqrt{f_R(r_0)}$, and at the junctions of the multi-boundary wormholes, $b_i=\beta_i\sqrt{f_i(R_0)}$. Quantum fields near the neck are then at a very high local temperature, even though the temperature at the asymptotic boundaries is fixed, and their backreaction on the geometry becomes large. 

To see this more explicitly, consider a free massless real scalar field $\phi$, periodic around the Euclidean time circle, on a static background of the form \eqref{eq:os-ADM} with $N^a=0$,
\begin{equation}
\dd s^2=N^2\dd\tau^2+\gamma_{ab}\,\dd x^a\dd x^b,\qquad \tau\sim\tau+1.
\end{equation}
The lapse $N$ is the proper length of the Euclidean time circle through a given point of the spatial slice $\Sigma$, so $1/N$ is the local temperature. In terms of the optical metric $\bar\gamma_{ab}=N^{-2}\gamma_{ab}$ and the field $\psi=N^{(D-2)/2}\phi$, the action of the scalar takes the form
\begin{equation}
I[\psi]=\frac12\int_0^1\dd\tau\int_\Sigma\dd^{D-1}x\sqrt{\bar\gamma}\,\psi\bigl(-\partial_\tau^2+H\bigr)\psi,\qquad H=-\bar\nabla^2+U,
\label{eq:optical-action}
\end{equation}
where $\bar\nabla$ is the covariant derivative compatible with $\bar\gamma_{ab}$, and
\begin{equation}
U=\frac{D-2}{2}\,\bar\nabla^2\log N+\frac{(D-2)^2}{4}\,\bigl|\bar\nabla\log N\bigr|^2.
\end{equation}
Since $H$ does not depend on $\tau$, the eigenfunctions of $-\partial_\tau^2$ are $e^{2\pi\i n\tau}$ with $n\in\mathbb{Z}$, and Poisson resummation gives
\begin{equation}
\mathrm{Tr}\,e^{-s(-\partial_\tau^2+H)}=\sum_{n\in\mathbb{Z}}e^{-4\pi^2n^2s}\,\mathrm{Tr}\,e^{-sH}=\frac{1}{\sqrt{4\pi s}}\sum_{w\in\mathbb{Z}}e^{-w^2/(4s)}\,\mathrm{Tr}\,e^{-sH},
\end{equation}
where $w$ is the winding number of paths around the time circle. Upon removing the $w=0$ vacuum contribution, which renormalizes the couplings of the local effective action, effective action for the scalar field is
\begin{equation}
\log Z_m=\frac{1}{\sqrt{4\pi}}\sum_{w=1}^\infty\int_0^\infty\frac{\dd s}{s^{3/2}}\,e^{-w^2/(4s)}\,\mathrm{Tr}\,e^{-sH}.
\label{eq:thermal-proper-time}
\end{equation}
To estimate contribution from the UV, short distance regime (with small proper length parameter~$s$), we use the heat-kernel expansion \cite{Vassilevich:2003xt}
\begin{equation}
\mathrm{Tr}\,e^{-sH}=\frac{1}{(4\pi s)^{(D-1)/2}}\int_\Sigma\dd^{D-1}x\sqrt{\bar\gamma}\left[1+s\left(\frac{\bar R}{6}-U\right)+\mathcal{O}(s^2)\right],
\label{eq:heat-kernel}
\end{equation}
where $\bar R$ is the scalar curvature of $\bar\gamma_{ab}$. Substituting the leading term into \eqref{eq:thermal-proper-time} and using $\int_0^\infty\dd s\,s^{-1-D/2}e^{-w^2/(4s)}=\Gamma(D/2)(4/w^2)^{D/2}$, we find
\begin{equation}
\log Z_m=\frac{\Gamma(D/2)\,\zeta_{\rm R}(D)}{\pi^{D/2}}\int_\Sigma\dd^{D-1}x\sqrt{\bar\gamma}=\frac{\Gamma(D/2)\,\zeta_{\rm R}(D)}{\pi^{D/2}}\int_\Sigma\dd^{D-1}x\,\frac{\sqrt{\gamma}}{N^{D-1}},
\label{eq:thermal-logZ}
\end{equation}
where $\zeta_{\rm R}$ is the Riemann zeta function. This is the free energy of a gas of massless particles at the local temperature $1/N(x)$.

Near the neck of a two-boundary wormhole, the lapse function $N$ is proportional to $b$, $N^{-(D-1)}\sim b^{-(D-1)}$. Thus, the neck contributes to \eqref{eq:thermal-logZ} a term of order $b^{-(D-1)}$, which diverges as $b\to0$, invalidating the constrained saddle-point approximation. The energy density of the radiation is obtained by varying $-\log Z_m$ with respect to the lapse at fixed $\gamma_{ab}$. At the wormhole neck, it equals
\begin{equation}
\rho_m=\frac{(D-1)\,\Gamma(D/2)\,\zeta_{\rm R}(D)}{\pi^{D/2}\,b^D},
\label{eq:rho-neck}
\end{equation}
which is the energy density of thermal radiation at temperature $1/b$. Photons and gravitons give contributions of the same order, with different numerical coefficients. The radiation backreacts strongly on the saddle when its energy density, multiplied by $8\pi G_D$, is comparable to the curvature of the saddle. Near the neck, the curvature is of order $1/R_0^2$, so the backreaction is small only if $8\pi G_D\rho_m\ll1/R_0^2$, i.e., $b^D\gg G_DR_0^2$. Since $\zeta=\Omega_{D-2}r_0^{D-2}b\simeq\Omega_{D-2}R_0^{D-2}b$, this condition reads
\begin{equation}
\zeta\geq\zeta_c(Q),\qquad \zeta_c(Q)=K_c\,G_D^{1/D}R_0^{D-2+2/D},
\label{eq:zeta-cutoff}
\end{equation}
with the dimensionless constant $K_c$ determining the allowed backreaction. If we choose $K_c\gg 1$, we can neglect backreaction.

The same analysis applies at the junctions of the multi-boundary wormholes. By \eqref{eq:bis}, the circumference of the time circle at the $i$-th junction is $b_i=\beta_i\sqrt{f_i(R_0)}=\zeta_i/(\Omega_{D-2}R_0^{D-2})$, so it goes to zero as $\zeta_i\to0$. Near a small junction surface, the core is a thin collar around a short closed geodesic, see \eqref{eq:core-geodesic-metric}, in which the time circle has circumference $L_i\cosh(a\rho_i)$ with $L_i=b_i\sqrt{1-\hat k_i^2/a^2}\simeq b_i$. On the exterior side, $f_i(r)$ changes only on the scale $R_0$, since $R_0f_i'(R_0)=-(D-3)f_i(R_0)$. Therefore, the size of the time circle changes slowly over distances of order $b_i$, and the energy density of the radiation near the junction is given by \eqref{eq:rho-neck} with $b$ replaced by $b_i$. We thus impose the cutoff \eqref{eq:zeta-cutoff} on each constrained area, $\zeta_i\geq\zeta_c(Q)$.

\subsubsection{Cumulants in the grand canonical ensemble} 

We can now estimate the cumulants $\kappa_1,\kappa_2,\kappa_3$. We work in the grand canonical ensemble, in which the inverse temperature $\beta$ and the electric potential $\Phi$ are fixed at every asymptotic boundary. The grand canonical partition function is obtained from the fixed-flux one by summing over the quantized charges $Q$,
\begin{equation}
\mathcal Z(\beta,\Phi) = \sum_{Q\in e\mathbb{Z}} e^{\beta\Phi Q}\,Z(\beta,Q),
\label{eq:wgc-grand-canonical}
\end{equation}
where $e$ is the elementary charge. By charge conjugation, we can take $\Phi\geq0$ without loss of generality. By \eqref{eq:os-weight}, the first three cumulants read
\begin{align}
\kappa_1 &= 1+\mathcal Z_{\rm 1\text{-}bdry}(\beta,\Phi),\nonumber\\
\kappa_2 &= \sum_{Q\in e\mathbb{Z}} e^{2\beta\Phi Q}\int_{\zeta_c(Q)}^\infty\dd\zeta\,e^{-I_{\rm 2\text{-}bdry}(\beta,\beta,Q;r_0(\zeta))}+\kappa_2^{\rm UV},\label{eq:wgc-cumulants}\\
\kappa_3 &= \sum_{Q\in e\mathbb{Z}} e^{3\beta\Phi Q}\int_{\cD_c(Q)}\dd\zeta_1\,\dd\zeta_2\,\dd\zeta_3\,e^{-I_{\rm 3\text{-}bdry}(\zeta_1,\zeta_2,\zeta_3)}+\begin{pmatrix}\text{wormholes with}\\ \text{unequal charges}\end{pmatrix}+\kappa_3^{\rm UV},\nonumber
\end{align}
up to one-loop factors. The terms in $\kappa_1$ as well as the first terms in $\kappa_2$ and $\kappa_3$ account for the IR contributions of the GPI to each moment for which the semi-classical approximation is under control. The terms $\kappa_2^{\rm UV}$ and $\kappa_3^{\rm UV}$ stand for the contributions of the configurations with $\zeta < \zeta_c(Q)$, which cannot be evaluated semiclassically. To distinguish the terms that are under semi-classical control from these corrections we will use the notation $\kappa_2^{\rm IR}$ and $\kappa_3^{\rm IR}$ for the IR contributions to these two moments.

\subsubsection{IR contributions}
Above, the first term in $\kappa_1$ is the contribution of thermal flat space (thermal AdS in the AdS case), in which the sphere shrinks smoothly at the center. It has $Q=0$ and vanishing action for any $\Phi$. The second term, $\mathcal Z_{\rm 1\text{-}bdry}(\beta,\Phi)$, is the contribution of the single-boundary saddles of Section~\ref{sec:1bdry-saddles} at fixed electric potential, which we evaluate below. The action $I_{\rm 2\text{-}bdry}$ is given by \eqref{eq:os-2bdry-action}, and $I_{\rm 3\text{-}bdry}$ is the action \eqref{eq:os-multi-action} of the three-boundary wormhole with the core $\Sigma_{0,3}$, for which $\chi(\Sigma_{0,3})=-1$. The integrals over the constrained areas are restricted to the region above the cutoff \eqref{eq:zeta-cutoff}. In particular, the $\zeta_i$-integrals in $\kappa_3^{\rm IR}$ run over the region $\cD_c(Q)$ in which $\zeta_i\geq\zeta_c(Q)$ for $i=1,2,3$ and the three-boundary wormholes exist (see Appendix~\ref{app:three-boundary-core} for the detailed analysis of the existence conditions of genus-zero three-boundary wormholes). 

With the notation clarified, let us now calculate the sums and integrals in \eqref{eq:wgc-cumulants} explicitly. In order to stress-test the inequalities between moments in \eqref{eq:wgc-positivity}, we work in the limit $\beta/\lpl\to\infty$, where $\lpl = (G_D)^{1/(D-2)}$ is the Planck length, and take $\Phi-\Phi_{\rm ext}$ of order $G_D^{-1/2}$, where
\begin{equation}
\Phi_{\rm ext}=\sqrt{\frac{1}{8\pi G_D}\frac{D-2}{D-3}}.
\label{eq:wgc-Phi-ext}
\end{equation}
Then $\Phi/\Phi_{\rm ext}$ is independent of $G_D$ at the leading order in the small $\lpl$-expansion. As we shall see, in this limit the IR contribution $\kappa_2^{\rm IR}$ is very large, leading to a potential violation of the third inequality in \eqref{eq:wgc-positivity}.

\paragraph{Single-boundary contribution.} Real saddles exist for $\zeta$ larger than the horizon area of the extremal black hole, $\zeta_{\rm ext}=\Omega_{D-2}R_0^{D-2}$. For $\zeta<\zeta_{\rm ext}$, the constrained surface is the inner horizon of the Reissner--Nordstr\"om solution, $r_+<R_0$; the corresponding constrained saddles are complex, with $r$-coordinate contour avoiding the outer horizon \cite{KM2026}. Following \cite{AMN2026,KM2026}, we include them, so that the integral over $\zeta$ starts at $\zeta=0$. This gives us
\begin{equation}
    Z_{\rm 1\text{-}bdry}(\beta,Q)\simeq \int_0^\infty\dd\zeta\, \mu(\beta,Q;\zeta)\exp\left[-\beta E(\zeta,Q) +\frac{\zeta}{4G_D}\right].
    \label{eq:single-tip-area-integral}
\end{equation}
The measure factor $\mu$ depends on one-loop determinants. The integration contour in \eqref{eq:single-tip-area-integral} agrees with the one derived from the Lorentzian path integral in \cite{Marolf2022thermal,KM2026}. In AdS, this integral is well defined.\footnote{At finite $\ell$, $\beta E$ grows as $r_+^{D-1}/\ell^2$ at large $\zeta$, faster than the entropy term $\zeta/(4G_D)\propto r_+^{D-2}$, so the integral converges.} In asymptotically flat space, however, the integral \eqref{eq:single-tip-area-integral} diverges at large $\zeta$, since the term $\beta E$ grows slower than the entropy term at $\zeta \to \infty$. We define the flat-space integral as the $\ell\to\infty$ limit of the AdS integral as follows. Changing the integration variable to the horizon radius, we write the latter as
\begin{equation}
    Z_{\rm 1\text{-}bdry}(\beta,Q)\simeq\int_0^\infty\dd r_+\,\widetilde\mu(\beta,Q;r_+)\exp\left[-\beta E+\frac{\Omega_{D-2}r_+^{D-2}}{4G_D}\right],
    \label{eq:1bdry-radius-integral}
\end{equation}
where $\widetilde\mu=\mu\,\dd\zeta/\dd r_+$. At real $\beta$, this integral is dominated by the large AdS black hole, with $r_+\simeq4\pi\ell^2/[(D-1)\beta]$, whose weight grows without bound as $\ell\to\infty$. Hence, we first continue to complex $\beta=|\beta|e^{\i\theta}$ and rotate the tail of the integration contour to $\arg r_+=-\psi$, with
\begin{equation}
    \frac{\pi}{2(D-2)}<\theta<\frac{\pi}{D-2},\qquad
    \frac{\pi}{2(D-2)}<\psi<\frac{\pi/2+\theta}{D-1}.
    \label{eq:1bdry-beta-continuation}
\end{equation}
Along this tail, $\operatorname{Re}(\beta r_+^{D-1})>0$ and $\operatorname{Re}(r_+^{D-2})<0$, so the integrand decays at large $|r_+|$ both at finite $\ell$ and in the limit $\ell\to\infty$. After taking this limit, we rotate the tail to $\arg r_+=-\pi/(D-2)$, where the factor $e^{\Omega_{D-2}r_+^{D-2}/(4G_D)}$ still decays, and continue the inverse temperature back to real values. Neglecting the measure factor, we obtain
\begin{equation}
    Z_{\rm 1\text{-}bdry}(\beta,Q)\simeq\int_\Gamma\dd r_+\exp\left[-\beta E(Q,r_+)+\frac{\Omega_{D-2}r_+^{D-2}}{4G_D}\right],
    \label{eq:1bdry-flat-integral}
\end{equation}
where the contour $\Gamma$ starts at $r_+=0$ along the positive real axis and approaches $\infty\cdot e^{-\i\pi/(D-2)}$.

The stationary points of the integrand in \eqref{eq:1bdry-flat-integral} are given by
\begin{equation}
    \beta=\frac{4\pi r_+}{D-3}\left(1-\mathcal Q^2/r_+^{2(D-3)}\right)^{-1}.
    \label{eq:os-1bdry-beta}
\end{equation}
At these points, the flat space limit $\ell \to \infty$ of the action \eqref{eq:single-tip-reduced-action} becomes
\begin{equation}
    I_{\rm 1\text{-}bdry}(\beta,Q) = \frac{\beta\Omega_{D-2}}{16\pi G_D} r_+^{D-3}\left[1+(2D-5)\frac{\mathcal Q^2}{r_+^{2(D-3)}}\right].
    \label{eq:os-1bdry-action}
\end{equation}
Equation \eqref{eq:os-1bdry-beta} has two positive solutions when
\begin{equation}
    \beta>\beta_{\min}(Q)
    =\frac{2\pi(2D-5)}{(D-3)^2}r_\star,
    \qquad r_\star^{2(D-3)}=(2D-5)\mathcal Q^2.
    \label{eq:os-beta-min}
\end{equation}
The two real solutions, $r_s$, $r_{\ell}$, to which we will refer as small and large RN black holes, satisfy $r_s < r_\star < r_{\ell}$.\footnote{They are the $\ell \to \infty$ limit of the small and intermediate AdS-RN black hole. The large AdS black hole ceases to exist in the flat space limit.} The small black hole saddle has positive heat capacity at fixed charge, while the large flat space black hole has negative heat capacity. They merge at $r_+=r_\star$ when $\beta=\beta_{\min}(Q)$. Since $\beta_{\min}(Q)\propto|Q|^{1/(D-3)}$, at fixed $\beta$ there are no real saddles for sufficiently large $|Q|$. For such charges, the two real saddles become a pair of complex-conjugate saddles. 

In Appendix~\ref{app:picard-lefschetz}, using Picard--Lefschetz analysis, we show that in the canonical ensemble only one saddle contributes to \eqref{eq:1bdry-flat-integral}: the small real black hole for $\beta>\beta_{\min}(Q)$, and its continuation into the lower half of the complex $r_+$-plane for $\beta<\beta_{\min}(Q)$. Thus, we can approximate the first cumulant as
\begin{equation}
\kappa_1=\sum_{Q\in e\mathbb{Z}}e^{\beta\Phi Q - I_{\text{1-bdry}}(\beta,Q)},
\label{eq:1bdry-charge-sum}
\end{equation}
with action $I_{\text{1-bdry}}(\beta,Q)$ evaluated on the small RN black hole saddles. In Appendix~\ref{app:picard-lefschetz} we evaluate this sum in the saddle-point approximation. It turns out that for $\Phi>\Phi_{\rm ext}\sin[\pi/(2(D-2))]$, the sum is dominated by complex saddles, and $\log\kappa_1\sim(\beta/\lpl)^{D-2}$. For smaller $\Phi$, the charged saddles are exponentially suppressed, and $\kappa_1$ is of order one and dominated by thermal flat space.

We can also Poisson-resum the sum \eqref{eq:1bdry-charge-sum},
\begin{equation}
\mathcal Z_{\rm 1\text{-}bdry}(\beta,\Phi)=\sum_{n\in\mathbb Z}\int\dd Q\,e^{\beta\Phi_nQ -I_{\rm 1\text{-}bdry}(\beta,Q)},
\qquad
\Phi_n=\Phi+\frac{2\pi\i n}{\beta e}.
\label{eq:1bdry-poisson}
\end{equation}
The saddle point of the $n$-th term in the sum is given by
\begin{equation}
r_+=\tilde r_n\equiv\frac{(D-3)\beta}{4\pi}\left(1-\frac{\Phi_n^2}{\Phi_{\rm ext}^2}\right),
\qquad
Q=Q_n\equiv(D-3)\Omega_{D-2}\bigl(\tilde r_n\bigr)^{D-3}\Phi_n.
\label{eq:1bdry-stationary}
\end{equation}
In Appendix~\ref{app:picard-lefschetz}, we show that for $\Phi>0$ only the saddles with $n>0$ contribute: in $D=4$ all of them, and in $D>4$ all but a finite number. The saddle point of the sum over the potential shifts $n$ coincides with the saddle point of the charge sum \eqref{eq:1bdry-charge-sum}. Hence, the saddle-point approximations before and after the Poisson resummation agree.

\paragraph{Two-boundary contribution.} Since the brane tension \eqref{eq:2bdry_tension} is negative, the action of the two-boundary wormholes increases with the constrained area, $\partial_\zeta I_{\rm 2\text{-}bdry}=-\lambda>0$. Hence, the $\zeta$-integral in $\kappa_2^{\rm IR}$ is dominated by its lower endpoint. Assuming that $\kappa_2^{\rm UV}$ does not cancel the contribution of the region above the cutoff, so that $\kappa_2^{\rm IR}$ sets the size of $\kappa_2$, we estimate $\kappa_2^{\rm IR}$ by the contribution of the constrained saddle with $\zeta=\zeta_c(Q)$,
\begin{equation}
\kappa_2^{\rm IR}\simeq\sum_{Q\in e\mathbb{Z}}P_2(Q)e^{2\beta\Phi Q-I_c(Q)},\qquad I_c(Q)=I_{\rm 2\text{-}bdry}\bigl(\beta,\beta,Q;r_0(\zeta_c(Q))\bigr),
\label{eq:kappa2-cutoff}
\end{equation}
where $P_2(Q)$ represents the one-loop factors and the subleading terms of the $\zeta$-integral. To evaluate $I_c(Q)$, note that for $\beta_L=\beta_R$ we have $f_L(r_0) = f_R(r_0) =\frac{D-3}{D-1}\bigl(1-x^{-2(D-3)}\bigr)$, where $x=r_0/R_0$. Hence, at the cutoff,
\begin{equation}
\frac{D-3}{D-1}\left(x^{2D-4}-x^2\right)=\frac{\zeta_c(Q)^2}{\Omega_{D-2}^2\beta^2R_0^{2D-4}}=\frac{K_c^2\,(G_DR_0^2)^{2/D}}{\Omega_{D-2}^2\beta^2},
\label{eq:cutoff-x}
\end{equation}
and \eqref{eq:os-2bdry-action} gives
\begin{equation}
I_c(Q)=2\beta\Phi_{\rm ext}|Q|\,\frac{(D-2)\,x^{D-3}+x^{-(D-3)}}{D-1}.
\label{eq:Ic}
\end{equation}
At large $|Q|$, \eqref{eq:cutoff-x} gives $x^{D-3}\propto|Q|^{2/(D(D-2))}$, so that $I_c(Q)\propto|Q|^{1+2/(D(D-2))}$ grows faster than linearly. Therefore, at finite cutoff, the sum over charges in \eqref{eq:kappa2-cutoff} converges for every $\Phi$.

Since $I_c(Q)\geq2\beta\Phi_{\rm ext}|Q|$, for $\Phi<\Phi_{\rm ext}$ each term in \eqref{eq:kappa2-cutoff} with $Q\neq0$ is suppressed at least by $e^{-2\beta(\Phi_{\rm ext}-\Phi)|Q|}$. With $\Phi - \Phi_{\rm ext}\propto G_D^{-1/2}$, the two-boundary wormholes are therefore exponentially suppressed as $G_D\to0$.

For $\Phi>\Phi_{\rm ext}$, the argument of the exponential, $2\beta\Phi Q-I_c(Q)$, is positive for a range of positive charges and has a maximum where $\dd I_c/\dd Q=2\beta\Phi$. Taking into account that $x$ depends on $Q$ through \eqref{eq:cutoff-x}, we find, for $Q>0$,
\begin{equation}
\frac{1}{2\beta\Phi_{\rm ext}}\frac{\dd I_c}{\dd Q}=\frac{(D-2+2/D)\,x^{D-3}+(1-2/D)\,x^{-(D-3)}}{D-1},
\end{equation}
which increases monotonically from one at $x=1$ to infinity. Hence, the maximum is unique, and it is located at $x=x_*$, where
\begin{equation}
x_*^{D-3}=\frac{(D-1)\Phi/\Phi_{\rm ext}+\sqrt{(D-1)^2\Phi^2/\Phi_{\rm ext}^2-4(D-2+2/D)(1-2/D)}}{2(D-2+2/D)}.
\label{eq:xstar}
\end{equation}
By \eqref{eq:cutoff-x}, the corresponding extremal radius is
\begin{equation}
R_{0,*}=\frac{1}{\sqrt{G_D}}\left(\frac{\Omega_{D-2}\beta}{K_c}\right)^{D/2}\left[\frac{D-3}{D-1}\left(x_*^{2D-4}-x_*^2\right)\right]^{D/4}.
\label{eq:R0star}
\end{equation}
The sum \eqref{eq:kappa2-cutoff} is dominated by the charges close to $Q_*$, and we obtain
\begin{align}
\log\kappa_2^{\rm IR}\simeq2\beta\Phi Q_*-I_c(Q_*)=C_2\left(\frac{\beta}{\lpl}\right)^{\frac{(D-1)(D-2)}{2}},
\label{eq:logkappa2}
\end{align}
where
\begin{equation}
C_2=\frac{(D-2)\,\Omega_{D-2}}{2\pi D(D-1)}\left(\frac{\Omega_{D-2}}{K_c}\right)^{\frac{D(D-3)}{2}}\left[\frac{D-3}{D-1}\left(x_*^{2D-4}-x_*^2\right)\right]^{\frac{D(D-3)}{4}}\left(x_*^{D-3}-x_*^{-(D-3)}\right)
\end{equation}
is positive and depends only on $D$, $K_c$ and $\Phi/\Phi_{\rm ext}$. In terms of Newton's constant, $\kappa_2^{\rm IR}\sim\exp\bigl[C_2\,\beta^{(D-1)(D-2)/2}/G_D^{(D-1)/2}\bigr]$, so that $\log\kappa_2^{\rm IR}$ grows faster than $1/G_D$.\footnote{The one-loop factor $P_2(Q)$ does not change this behavior for sufficiently large $K_c$. At the maximum, the circumference of the time circle at the neck, $b=\beta\sqrt{f(r_0)}$, is of order $\beta$, while the region around the neck in which $f(r)$ is of order one has a volume of order $R_{0,*}^{D-1}$. By \eqref{eq:thermal-logZ}, the thermal contribution of matter fields to $\log P_2$ is then of order $(R_{0,*}/\beta)^{D-1}$. It scales with $\beta/\lpl$ in the same way as \eqref{eq:logkappa2}, but it is suppressed relative to \eqref{eq:logkappa2} by a factor of order $(\Omega_{D-2}/K_c)^D$.} Thus, in flat space, the IR contribution $\kappa_2^{\rm IR}$ grows faster than any saddle-point contribution.\footnote{Note that with a similar cutoff, the constrained wormhole saddles of Cotler and Jensen~\cite{Cotler:2022rud}, reviewed in Appendix~\ref{app:cotler-jensen}, give the same scaling of $\log\kappa_2^{\rm IR}$ with $\beta/\lpl$.}

In AdS, the extremal energy grows faster than linearly with the charge, so the sum over charges in $\kappa_2^{\rm IR}$ converges for every $\Phi$, even without the cutoff, and it is dominated by charges with $R_0/\beta$ of order one. Hence, in AdS, $\log\kappa_2^{\rm IR}$ is at most of order $(\beta/\lpl)^{D-2}$.

\paragraph{Three-boundary contribution.} Consider first the equal-charge contribution in asymptotically flat space. For $Q>0$, we have $\Phi_{\rm ext}Q=(D-2)\Omega_{D-2}R_0^{D-3}/(8\pi G_D)$, and by \eqref{eq:os-multi-action} the exponent of the equal-charge contribution to $\kappa_3^{\rm IR}$ reads
\begin{equation}
3\beta\Phi Q-I_{\rm 3\text{-}bdry}=\frac{\Omega_{D-2}R_0^{D-3}}{4G_D}\left[\frac{3(D-2)\beta}{2\pi}\left(\frac{\Phi}{\Phi_{\rm ext}}-1\right)-R_0\right]-\frac{D-2}{16\pi G_D\Omega_{D-2}\beta R_0^{D-1}}\sum_{i=1}^3\zeta_i^2.
\label{eq:3bdry-exponent}
\end{equation}
The action of the three-boundary wormholes increases with the constrained areas, so, as in the two-boundary case, the $\zeta_i$-integrals are dominated by the smallest allowed areas, $\zeta_i=\zeta_c(Q)$. There, the second term in \eqref{eq:3bdry-exponent} is a higher-order correction in $G_D$, of relative order $(\lpl/\beta)^{2(D-2)/D}$, and we drop it. Thus, up to the one-loop factors, the equal-charge contribution to $\kappa_3^{\rm IR}$ is
\begin{equation}
\kappa_3^{\rm IR,\,eq} = \sum_{Q\in e\mathbb{Z}}P_3(Q)\,e^{3\beta\Phi Q-I_{\rm 3\text{-}bdry}(Q)},
\label{eq:kappa3-eq}
\end{equation}
where, for $Q>0$, the exponent is given by the first term in \eqref{eq:3bdry-exponent}, and $P_3(Q)$ represents the one-loop factors and the subleading terms of the $\zeta_i$-integrals. For $\Phi<\Phi_{\rm ext}$, each term with $Q\neq0$ is suppressed at least by $e^{-3\beta(\Phi_{\rm ext}-\Phi)|Q|}$, so the three-boundary wormholes are exponentially suppressed as $G_D\to0$. For $\Phi>\Phi_{\rm ext}$, the exponent is maximal at $R_{0,*}=3(D-3)\beta(\Phi/\Phi_{\rm ext}-1)/(2\pi)$, so the sum \eqref{eq:kappa3-eq} is dominated by the charges close to the corresponding charge $Q_*$,
\begin{equation}
\log\kappa_3^{\rm IR,\,eq} \simeq 3\beta\Phi Q_*-I_{\rm 3\text{-}bdry}(Q_*)=\frac{\Omega_{D-2}}{4(D-3)G_D}\left[\frac{3(D-3)\beta}{2\pi}\left(\frac{\Phi}{\Phi_{\rm ext}}-1\right)\right]^{D-2}.
\label{eq:3bdry-bound}
\end{equation}
It grows as $(\beta/\lpl)^{D-2}$, slower than \eqref{eq:logkappa2} for $D\geq4$.\footnote{In AdS, the sum over charges is again dominated by charges with $R_0/\beta$ of order one, and $\log\kappa_3^{\rm IR,\,eq}$ is also at most of order $(\beta/\lpl)^{D-2}$.}

Thus, this result suggests that $\kappa_3^{\rm IR}$ might be dominated by wormholes with unequal charges, which break the symmetry under permutations of the three boundaries; the theory may be in a replica-symmetry-breaking phase \cite{Engelhardt:2020qpv,Chandrasekaran:2022asa}. This is accounted for by the middle term in $\kappa_3$, which stands for the contributions of the three-boundary wormholes with unequal charges $Q_1, Q_2, Q_3$ at the three boundaries. Such wormholes are not part of the spherically symmetric family constructed in Section~\ref{sec:multi-bdry-saddles}, in which Gauss's law forces the same flux through all three boundaries. However, we cannot exclude their contributions since non-spherically symmetric saddles with unequal boundary fluxes might exist.\footnote{With different boundary conditions there are known wormhole geometries where the electric flux splits between several boundaries; e.g. the Brill instanton \cite{Brill:1992ydn,Maldacena:1998uz}.}  We expect the largest contribution to come from wormholes in which one charge is much smaller than the other two. Their geometry is likely similar to that of two-boundary wormholes, so we expect this contribution to be at most as divergent as $\kappa_2^{\rm IR}$.

\subsection{Electric Weak Gravity Conjecture}\label{sec:electric-wgc}

We can now put all the results for the moments together to give a positivity argument for the WGC. Following the argument for the axionic WGC \cite{DiUbaldo:2026rly}, we now want to test the third inequality in \eqref{eq:wgc-positivity}, which can be recast as
\begin{equation}
\frac{\kappa_1^2}{\kappa_2}+\frac{\kappa_1\kappa_3}{\kappa_2^2}\geq1,
\label{eq:wgc-third-inequality-ratio}
\end{equation}
which should hold for any chemical potential $\Phi$.

Let us first analyze the regime  $\Phi/\Phi_{\rm ext} > 1$ in flat space. Above, we argued that, for $\beta /\lpl \to \infty$, the IR contributions to $\log\kappa_2$ grow as $(\beta/\lpl)^{(D-1)(D-2)/2}$. In contrast,  $\log(\kappa_1) \sim (\beta/\lpl)^{D-2}$. Finally, we have argued that $\kappa_3^{\rm IR}$ grows at most as quickly as $\kappa_2^{\rm IR}$, so that $(\kappa_2^{\rm IR})^2 \gg \kappa_3^{\rm IR}$, even though the logarithm of $\kappa_3^{\rm IR}$ may grow with the same power of $\beta/\lpl$ as that of $\kappa_2^{\rm IR}$. Consequently, both terms on the left-hand side of the inequality \eqref{eq:wgc-third-inequality-ratio} approach zero as $\beta/\lpl \to \infty$, so if we were to include only the IR contribution to $\kappa_2$ and $\kappa_3$, the inequality \eqref{eq:wgc-third-inequality-ratio} would be violated for every $\Phi > \Phi_{\rm ext}$ with $\Phi - \Phi_{\rm ext} = \O(G_D^{-1/2})$.\footnote{ $\Phi - \Phi_{\rm ext} = \O(G_D^{-1/2})$ is the natural scaling for chemical potentials with $G_D$, so there is no fine-tuning necessary to see a violation of positivity. Furthermore, the exact choice of the cutoff $K_c$ affects only the coefficient $C_2$ in \eqref{eq:logkappa2}, not the power of $\beta/\lpl$, so this conclusion does not rely on the one-loop determinants. As argued in Section~\ref{sec:two-boundary-stability}, the two-boundary saddles are perturbatively stable, so the saddle-point description is reliable. } For $\kappa_2^\text{UV}$ and  $\kappa_3^\text{UV}$ to restore positivity, we would need them to grow faster than any saddle point contribution; since we expect UV effects to regulate the GPI contributions in this regime rather than make them more divergent, we do not expect this to be possible. Thus, under this assumption, the inequality \eqref{eq:wgc-third-inequality-ratio} is violated for $\Phi/\Phi_{\rm ext} > 1$ in flat space. For $\Phi\leq\Phi_{\rm ext}$, neither $\kappa_2$ nor $\kappa_3$ is enhanced, and there is no conflict with positivity in flat space. Nevertheless, the fact that there exists some value of $\Phi$ for which positivity is violated indicates that pure Einstein–Maxwell gravity with zero cosmological constant and without any matter fields is not, at least at the semi-classical level, a consistent theory of quantum gravity.\footnote{ In AdS, the logarithms of all three cumulants are at most of order $(\beta/\lpl)^{D-2}$, so this argument does not apply. }

To restore positivity, the semiclassical saddles must become unstable for every $\Phi>\Phi_{\rm ext}$, just as the axion wormholes of \cite{DiUbaldo:2026rly} must be destabilized by the proliferation of instantons. In Einstein--Maxwell theory, the saddles can be destabilized by charged particles wrapping the Euclidean time circle. Here, by an instability we mean that the saddles of pure Einstein--Maxwell theory no longer dominate the gravitational path integral, because there are saddles with nontrivial matter field configurations, such as the wrapped worldlines, that have a lower on-shell action. This instability differs from the one considered in the original arguments for the WGC \cite{Gibbons:1975kk,Arkani-Hamed:2006emk,Harlow:2022ich,Gonzalo:2020kke,Lin:2024jug} in two ways. First, we need to destabilize the two-boundary wormholes rather than single black holes. Second, the instability comes from worldlines wrapping the Euclidean time circle, which is non-contractible in the wormhole, rather than from the small contractible worldline loops near the horizon that describe Schwinger pair production.

Consider a particle of mass $m$ and charge $q$. We can take $Q,q,\Phi>0$ without loss of generality. With the asymptotic boundary condition $A_\tau\to\i\beta\Phi$, the gauge field in one half of the two-boundary wormhole is
\begin{equation}
A_\tau(r)=-\i\beta\bigl[\Phi(r)-\Phi\bigr],\qquad \Phi(r)=\frac{Q}{(D-3)\Omega_{D-2}\,r^{D-3}}=\Phi_{\rm ext}\left(\frac{R_0}{r}\right)^{D-3}.
\label{eq:wgc-gauge-field}
\end{equation}
The contribution to the free energy from a worldline wrapping the Euclidean time circle at radius $r$ in one of the exteriors is then
\begin{equation}
I_p(r)=\beta m\sqrt{f(r)}+\i q A_\tau(r)=\beta\bigl[m\sqrt{f(r)}+q\,\Phi(r)-q\,\Phi\bigr].
\label{eq:wgc-worldline}
\end{equation}
The wormhole saddle is unstable against the particle production at radius $r$ if and only if\footnote{The particle trajectories along the Euclidean time circle are the most unstable. To see this, consider an arbitrary closed worldline $\gamma$. Since the gauge field has only a temporal component and $\dd s\geq\beta\sqrt{f(r)}\,|\dd\tau|$, its action satisfies
\begin{equation}
I_p[\gamma]\geq\beta\int_\gamma|\dd\tau|\,
\left[m\sqrt{f(r)}-\left|q\bigl(\Phi(r)-\Phi\bigr)\right|\right].
\end{equation}
If $I_p[\gamma]<0$, the expression in brackets must be negative at some radius. The temporal circle at that radius, with the appropriate orientation, saturates this bound and therefore also has negative action. Thus radial or angular motion cannot trigger a negative-action instability before a temporal circle does.} $I_p(r) < 0$.

At the asymptotic boundary, $I_p\to\beta(m-q\Phi)$. Hence, a particle with $q/m\geq1/\Phi_{\rm ext}$ destabilizes the wormholes for every $\Phi>\Phi_{\rm ext}$, while a particle with $q/m<1/\Phi_{\rm ext}$ has $I_p(r)>0$ at every radius.\footnote{Consider a particle satisfying $q\Phi_{\rm ext}/m\leq1-\eta$ for some $\eta>0$, and write $\Phi=(1+\delta)\Phi_{\rm ext}$. The two-boundary wormholes that dominate $\kappa_2$ have the neck at $x=x_*$ given by \eqref{eq:xstar}. Their mass parameter is
\begin{equation}
\frac{M}{M_{\rm ext}}=\frac{x_*^{D-3}+(D-2)\,x_*^{-(D-3)}}{D-1}=1-\frac{D(D-3)}{D(D-3)+4}\,\delta+\mathcal{O}(\delta^2).
\end{equation}
In terms of $y=(R_0/r)^{D-3}$, the RN function reads $f=(y-M/M_{\rm ext})^2+1-M^2/M_{\rm ext}^2$. Minimizing $m\sqrt{f}+q\Phi_{\rm ext}y$ over $y$, we find that at every radius
\begin{equation}
\frac{I_p(r)}{\beta}\geq q\left(\frac{M}{M_{\rm ext}}\,\Phi_{\rm ext}-\Phi\right)+\sqrt{\bigl(m^2-q^2\Phi_{\rm ext}^2\bigr)\left(1-\frac{M^2}{M_{\rm ext}^2}\right)}.
\label{eq:wgc-Ip-bound}
\end{equation}
The first term is negative and of order $q\Phi_{\rm ext}\delta$, while the second one is positive and of order $m\sqrt{\eta\delta}$. Therefore, for sufficiently small $\delta$, $I_p(r)>0$ at every radius. For $q\Phi_{\rm ext}\geq m$ and $M\leq M_{\rm ext}$, instead, $m\sqrt{f}+q\Phi_{\rm ext}y$ increases with $y$, so $I_p$ is minimized at the asymptotic boundary.} Consequently, $m-q\Phi$ must be negative for $\Phi$ arbitrarily close to $\Phi_{\rm ext}$, which requires
\begin{equation}
\frac{q}{m}\ \ge\ \frac{1}{\Phi_{\rm ext}} = \sqrt{\frac{8\pi(D-3)G_D}{D-2}},
\label{eq:wgc-bound}
\end{equation}
for at least one particle in the theory. The right-hand side is the charge-to-energy ratio $|Q|/E_{\rm ext}(Q)$ of extremal Reissner--Nordstr\"om black holes, so \eqref{eq:wgc-bound} is precisely the electric weak gravity conjecture bound \cite{Arkani-Hamed:2006emk,Harlow:2022ich}.

\subsection{Generalizations to magnetic and multi-charge cases}\label{sec:wgc-generalizations}
The argument of Section~\ref{sec:electric-wgc} can be easily extended to derive the magnetic version of the weak gravity conjecture, as well as its multi-charge generalization, the convex hull condition \cite{Cheung:2014vva}.

\paragraph{Dyonic wormholes and the magnetic WGC.} In $D=4$, the RN exteriors can carry both an electric charge $Q_e\in e\mathbb{Z}$ and a magnetic charge $Q_m\in g\mathbb{Z}$, where $g$ is a unit magnetic charge; the Dirac quantization condition implies $eg=2\pi$. Since the Maxwell stress tensor is invariant under electric--magnetic duality, the metric depends on the charges only through $|\vec Q| = \sqrt{Q_e^2+Q_m^2}$, where $\vec Q = (Q_e,Q_m)$, and the constrained wormholes of Section~\ref{sec:grav-constr-saddles} exist for every $\vec Q$. Both the on-shell action of each exterior, \eqref{eq:os-canonical-action}, and the free energy of the thermal radiation near the neck are fully determined by the metric, so we can impose the cutoff \eqref{eq:zeta-cutoff} with $R_0=\sqrt{G_4/(4\pi)}\,|\vec Q|$.

We work in the grand canonical ensemble with fixed electric and magnetic potentials $\vec\Phi = (\Phi_e,\Phi_m)$. The corresponding partition function is $\mathcal Z(\beta,\vec\Phi) = \sum_{\vec Q\in\Gamma}e^{\beta\vec\Phi\cdot\vec Q}Z(\beta,\vec Q)$, with the charge lattice $\Gamma = e\mathbb{Z}\times g\mathbb{Z}$, and the contribution of the two-boundary wormholes generalizes \eqref{eq:kappa2-cutoff} to
\begin{equation}
\kappa_2 \simeq \sum_{\vec Q\in\Gamma} P_2(\vec Q)\,e^{2\beta\vec\Phi\cdot\vec Q-I_c(|\vec Q|)},
\label{eq:wgc-dyonic-kappa2}
\end{equation}
where $I_c$ is given by \eqref{eq:Ic} with $D=4$. In polar coordinates in the charge plane, $\vec Q = |\vec Q|(\cos\theta,\sin\theta)$ and $\vec\Phi = |\vec\Phi|(\cos\varphi,\sin\varphi)$, the exponent in \eqref{eq:wgc-dyonic-kappa2} becomes $2\beta|\vec Q||\vec\Phi|\cos(\theta-\varphi)-I_c(|\vec Q|)$. If $|\vec\Phi|\leq\Phi_{\rm ext}$, it is negative for every $\vec Q\neq0$, since $I_c(|\vec Q|)\geq2\beta\Phi_{\rm ext}|\vec Q|$. If $|\vec\Phi|>\Phi_{\rm ext}$, it is maximized in the direction of $\vec\Phi$, $\theta=\varphi$, where it coincides with the exponent in \eqref{eq:kappa2-cutoff} with $\Phi$ replaced by $|\vec\Phi|$. Since $Q_*\to\infty$ as $\beta/\lpl\to\infty $, the lattice $\Gamma$ contains charges whose distance from $Q_*\vec\Phi/|\vec\Phi|$ is negligible compared to $Q_*$, even when no lattice direction is aligned with $\vec\Phi$. Hence, the sum \eqref{eq:wgc-dyonic-kappa2} behaves as \eqref{eq:logkappa2} with $\Phi$ replaced by $|\vec\Phi|$, which for $D=4$ gives
\begin{equation}
\log\kappa_2\simeq\frac{16\pi^2}{9K_c^2}\left(\frac{\beta}{\lpl}\right)^3x_*\bigl(x_*^2-1\bigr)^2,\qquad x_*=\frac{3|\vec\Phi|/\Phi_{\rm ext}+\sqrt{9|\vec\Phi|^2/\Phi_{\rm ext}^2-5}}{5}.
\end{equation}
Therefore, $\kappa_2$ is enhanced for
\begin{equation}
\Phi_e^2+\Phi_m^2>\Phi_{\rm ext}^2,
\label{eq:wgc-dyonic-condition}
\end{equation}
while $\kappa_1$ and $\kappa_3$ remain small compared to $\kappa_2^2$ for the same reasons as in Section~\ref{sec:electric-wgc}, and consequently, the positivity constraint \eqref{eq:wgc-third-inequality-ratio} is violated.

The violation is avoided if the saddles are destabilized by dyons wrapping the Euclidean time circle. Consider a dyon of mass $m$ and charges $\vec q = (q_e,q_m)$, coupled to the electromagnetic field as in \cite{Zwanziger:1970hk,Deser:1997se}. The charges of the dominant wormholes point in the direction $\hat n=\vec\Phi/|\vec\Phi|$, up to corrections that vanish as $\beta/\lpl\to\infty$. On these wormholes, the action of the dyon wrapping the time circle at radius $r$ is given by \eqref{eq:wgc-worldline} with $q$ replaced by $\hat n\cdot\vec q$ and $\Phi$ by $|\vec\Phi|$, so the bound \eqref{eq:wgc-Ip-bound} holds with the same replacements. Therefore, as in Section~\ref{sec:electric-wgc}, the wormholes with $|\vec\Phi|$ close to $\Phi_{\rm ext}$ can only be destabilized by dyons with $\hat n\cdot\vec q/m\geq1/\Phi_{\rm ext}$. The action of such a dyon is minimized at $r\to\infty$, where it equals $\beta(m-\vec q\cdot\vec\Phi)$. Requiring that, for every direction $\hat n$ of $\vec\Phi$, the wormholes are destabilized for all $|\vec\Phi|>\Phi_{\rm ext}$, we find that for every unit vector $\hat n$, there must be a particle with
\begin{equation}
\vec z\cdot\hat n\ge1,\qquad \vec z\equiv\frac{\Phi_{\rm ext}\,\vec q}{m} = \frac{\vec q}{\sqrt{4\pi G_4}\,m}.
\label{eq:wgc-chc}
\end{equation}
Equivalently, the unit disk must be contained in the convex hull of the charge-to-mass vectors $\vec z$ of all particles and antiparticles, which is the convex hull condition of \cite{Jones:2019nev} for dyonic states. For $\hat n=(1,0)$, \eqref{eq:wgc-chc} reduces to \eqref{eq:wgc-bound}, while for $\hat n = (0,1)$ it requires a magnetically charged state with $|q_m|\ge\sqrt{4\pi G_4}\,m$, which is the standard magnetic weak gravity conjecture \cite{Arkani-Hamed:2006emk,Harlow:2022ich}.

\paragraph{Multiple gauge fields.} Consider now $N$ gauge fields $A^a$, $a=1,\dots,N$, with canonically normalized kinetic terms. The RN exteriors with charges $\vec Q=(Q^1,\dots,Q^N)$ depend only on $|\vec Q|$, and in the ensemble with fixed potentials $\vec\Phi=(\Phi^1,\dots,\Phi^N)$ we find
\begin{equation}
\kappa_2 \simeq \sum_{\vec Q\in\Gamma} P_2(\vec Q)\,e^{2\beta\vec\Phi\cdot\vec Q-I_c(|\vec Q|)},\qquad \Gamma\subset\mathbb{R}^N,
\label{eq:wgc-multi-kappa2}
\end{equation}
where $\Gamma \simeq \mathbb{Z}^{N}$ is the charge lattice, and $I_c$ is given by \eqref{eq:Ic} with $|Q|$ replaced by $|\vec{Q}|$. One can follow the same argument as in the dyonic case, replacing the polar coordinates in the charge plane with spherical coordinates in $\mathbb{R}^N$, $\vec Q = |\vec Q|\,\hat n$ with $\hat n\in S^{N-1}$. For $|\vec\Phi|>\Phi_{\rm ext}$, $\log\kappa_2$ is given by \eqref{eq:logkappa2} with $\Phi$ replaced by $|\vec\Phi|$. Positivity then requires that the unit ball in $\mathbb{R}^N$ is contained in the closed convex hull of the vectors $\pm\vec z_i$, where $\vec z_i = \Phi_{\rm ext}\,\vec q_i/m_i$, in agreement with \cite{Cheung:2014vva,Jones:2019nev,Harlow:2022ich}. In $D=4$, magnetic charges can be included in the same way, extending the charge lattice to $\Gamma\simeq \mathbb{Z}^{2N}$.

\section{Discussion}\label{sec:discussion}

In this paper, we used a new family of constrained saddles to compute the first three cumulants of the black hole partition function in asymptotically flat and AdS spacetimes. By constraining the area of a codimension-one surface, we constructed two-boundary wormholes, obtained by joining two Reissner--Nordstr\"om exteriors across a constraint brane, and multi-boundary wormholes, obtained by gluing Reissner--Nordstr\"om exteriors to an $n$-boundary core. Against constraint-preserving, spherically symmetric perturbations, the two-boundary wormholes are stable in asymptotically flat space at sufficiently large charge (Section~\ref{sec:two-boundary-stability}). This regime includes the wormholes that dominate $\kappa_2$ in Section~\ref{sec:moments}. The multi-boundary wormholes have no negative modes associated with the core or the junctions (Section~\ref{sec:multi-boundary-stability}). We also performed a minisuperspace analysis in Appendix~\ref{app:minisuperspace} which independently confirms the stability of the static modes of the asymptotically flat two-boundary wormholes. 

In asymptotically flat space and in the absence of charged particles, the resulting cumulants are inconsistent with a positive-semidefinite inner product for closed and open universes. Since the Euclidean time circle shrinks at the neck of the wormholes as the constrained area goes to zero, we restricted the constrained areas to $\zeta\geq\zeta_c(Q)$, below which the backreaction of thermal radiation invalidates the semiclassical description. For $|\Phi|>\Phi_{\rm ext}$, the remaining two-boundary contribution is dominated by large charges, and in the limit $\beta/\lpl\to\infty$ it grows as $\log\kappa_2^{\rm IR}\sim(\beta/\lpl)^{(D-1)(D-2)/2}$, see \eqref{eq:logkappa2}. This is parametrically faster than the single-boundary contribution, $\log\kappa_1\lesssim(\beta/\lpl)^{D-2}$, and than the equal-charge three-boundary contribution, $\log\kappa_3^{\rm IR,\,eq}\sim(\beta/\lpl)^{D-2}$. Under the assumptions that we state explicitly below, both terms on the left-hand side of \eqref{eq:wgc-third-inequality-ratio} then vanish, and the inequality $\kappa_1\kappa_3\geq\kappa_2(\kappa_2-\kappa_1^2)$ is violated for every $|\Phi|>\Phi_{\rm ext}$ with $|\Phi|-\Phi_{\rm ext}=\O(G_D^{-1/2})$. For $|\Phi|\leq\Phi_{\rm ext}$, neither $\kappa_2$ nor $\kappa_3$ is enhanced, and there is no conflict with positivity.

Once a charged particle species that satisfies the WGC \eqref{eq:intro-wgc} couples to gravity and to the electromagnetic field, the wormholes responsible for this violation become unstable for every $|\Phi|>\Phi_{\rm ext}$. Worldlines of such particles wrapping the Euclidean time circle near the asymptotic boundaries have negative action, so it becomes favorable to add a large number of charged particles, which backreact on the geometry and can drastically change the cumulants $\kappa_n$. We have not shown that positivity is restored once this backreaction is accounted for. Conversely, if no charged particle satisfies \eqref{eq:intro-wgc}, then for $|\Phi|$ above but sufficiently close to $\Phi_{\rm ext}$, the wrapped worldlines have positive action at every radius, see \eqref{eq:wgc-Ip-bound}. The saddles of pure Einstein--Maxwell theory then continue to dominate the GPI, and there is no mechanism that could restore positivity. Consequently, under the assumptions listed below, a positive-semidefinite inner product for closed and open universes requires at least one particle species that satisfies the WGC. The same argument applied to dyonic wormholes requires the convex hull condition for dyonic states \eqref{eq:wgc-chc}, which contains the magnetic WGC, and in theories with several $\mathrm{U}(1)$ gauge fields it yields the convex hull condition.

  Our argument relies on the following assumptions: 
\begin{enumerate}
\item \emph{Constrained instantons compute the wormhole contributions.} Since there are no saddles with the required boundary conditions, we computed $\kappa_2$ and $\kappa_3$ using constrained instantons, keeping the integral over the constrained areas until the end. The constrained wormholes of Cotler and Jensen \cite{Cotler:2022rud}, obtained by fixing the proper length of the wormhole rather than the area of a surface, give the same scaling of $\log\kappa_2$ with $\beta/\lpl$ (see Appendix~\ref{app:cotler-jensen}). This suggests that the large-charge behavior of $\kappa_2$ is not an artifact of a particular choice of constraint.

\item \emph{UV cutoff on the constrained areas.} We neglected the contribution from constrained wormholes with $\zeta<\zeta_c(Q)$ whose necks are at local temperatures at which the backreaction of thermal radiation cannot be controlled. We assumed that the contributions $\kappa_2^{\rm UV}$ and $\kappa_3^{\rm UV}$ of configurations below the cutoff, which cannot be evaluated semiclassically, cannot cancel $\kappa_2^{\rm IR}$, which grows faster than any ordinary saddle-point contribution (i.e., $\sim 1/G_D$) for $\Phi > \Phi_\text{ext}$. We expect UV effects to regulate rather than enhance the contributions of $\kappa_2^{\rm UV}$ and $\kappa_3^{\rm UV}$ so we believe this assumption is sensible.

\item \emph{Three-boundary wormholes with unequal charges.} Our estimate of $\kappa_3$ includes only the replica-symmetric wormholes of Section~\ref{sec:multi-bdry-saddles}, in which the same electric flux threads all three asymptotic boundaries. Wormholes whose boundaries carry different charges, which are not spherically symmetric, might dominate $\kappa_3$, signaling a replica-symmetry-breaking phase. We assumed that their contribution grows at most as fast as $\kappa_2^{\rm IR}$. We expect the largest such contribution to come from wormholes in which one charge is much smaller than the other two, whose geometry is likely similar to that of two-boundary wormholes. Restoring positivity would instead require $\kappa_3\gtrsim\kappa_2^2/\kappa_1$, i.e., $\log\kappa_3$ at least twice as large as $\log\kappa_2^{\rm IR}$. Constructing the corresponding constrained saddles would test this assumption and give a more accurate estimate of $\kappa_3$, as well as of the higher cumulants.

\item \emph{Stability against non-spherical perturbations.} We established stability only against spherically symmetric perturbations. Modes with nonzero angular momentum on $S^{D-2}$ receive positive angular-gradient contributions, so we do not expect them to destabilize the saddles, but a complete analysis of non-spherical perturbations is left for future work.

\end{enumerate}

In asymptotically AdS spacetimes, the argument does not apply. At large charge, the extremal energy grows as $|Q|^{\frac{D-1}{D-2}}$, faster than the linear term $\beta\Phi Q$, so the sum over charges converges for every value of the electric potential, even without a cutoff. The logarithms of all three cumulants are then at most of order $(\beta/\lpl)^{D-2}$, and the inequalities between $\kappa_1$, $\kappa_2$, $\kappa_3$ are not violated at the semiclassical level. Therefore, at least at the level of the first three cumulants and of the saddles considered here, positivity of the GPI does not constrain the spectrum of charged particles in AdS. This does not imply that the WGC fails in AdS, only that a different mechanism is needed to derive it. For black holes much smaller than the AdS radius $\ell$, the extremal energy is approximately linear in $|Q|$, and $\kappa_2$ approaches its flat-space form as $\ell\to\infty$. It would be interesting to understand whether, at large but finite $\ell$, the growth of $\kappa_2$ for $|\Phi|>\Phi_{\rm ext}$ is still sufficient to violate positivity, which would lead to an AdS version of the bound.

\paragraph{Relation to axion wormholes and previous work.} Our derivation closely parallels the derivation of the axionic WGC from positivity of the GPI presented in \cite{DiUbaldo:2026rly}. The electric potential plays the role of the difference between the boundary values of the axion, the fixed electric flux plays the role of the axion flux, and charged particles wrapping the Euclidean time circle play the role of the instantons that destabilize the axion wormholes. This analogy can be made precise near extremality. As shown in Appendix~\ref{app:timelike-reduction}, upon reducing Einstein--Maxwell theory on the Euclidean time circle, the gauge field becomes an axion and the size of the thermal circle becomes a dilaton. In the limit $r_0\to R_0$, taken together with $\beta_L,\beta_R\to\infty$, the neck of our constrained wormholes approaches the axio-dilaton wormholes of \cite{Maldacena:2026jqd}, whose uplift is an $\mathrm{AdS}_2\times S^{D-2}$ throat. We also stress that the instability underlying our argument differs from the one used in the original arguments for the WGC \cite{Arkani-Hamed:2006emk,Harlow:2022ich}, which require that extremal black holes of fixed charge be able to decay, for instance through the Schwinger production of charged particles near the horizon \cite{Gibbons:1975kk,Gonzalo:2020kke,Lin:2024jug}. Here, the bound instead follows from requiring that the saddles of pure Einstein--Maxwell theory cease to dominate the GPI for every $|\Phi|>\Phi_{\rm ext}$, where they would otherwise violate positivity. It would also be interesting to construct constrained multi-boundary axion wormholes, which would allow for an explicit computation of the higher cumulants in the setting of \cite{DiUbaldo:2026rly}.

\paragraph{Other swampland conjectures.} Positivity of the GPI has now been shown to imply two swampland conjectures \cite{Vafa:2005ui,Palti:2019pca}: the axionic WGC \cite{DiUbaldo:2026rly} and, in this paper, the WGC for electric and magnetic charges, together with its convex hull generalization. This suggests that positivity of the GPI might be harnessed to derive other swampland conjectures. A natural first step is to sharpen the bound derived here. Positivity yields the convex hull condition, which is not preserved under dimensional reduction, whereas stronger versions of the conjecture, such as the tower and sublattice WGC, require an infinite tower of superextremal states (see \cite{Harlow:2022ich} for a review). It would be interesting to understand whether applying our argument to compactified theories, or to higher cumulants, leads to these stronger versions. Similarly, in theories with massless dilatons, the charge-to-mass ratio of extremal black holes depends on the dilaton couplings, and the WGC no longer coincides with the repulsive force conjecture \cite{Harlow:2022ich}. Extending our constrained wormholes to Einstein--Maxwell--dilaton theories would determine which of the two follows from positivity. Finally, it should be straightforward to extend our analysis to higher-form gauge fields, for which positivity may imply the WGC for charged branes.


\subsection*{Acknowledgments}
We thank Jeevan Chandra, Kristan Jensen and Ben Safdi for useful discussions. 
LVI is supported in part by the Leinweber Institute for Theoretical Physics at UC Berkeley, by the Department of Energy, Office of Science, Office of High Energy Physics through the award DE-SC0025522, and by the Department of Energy through QuantISED award DE-SC0019380. GDU is supported by the Japan Science and Technology Agency (JST) as part of Adopting Sustainable Partnerships for Innovative Research Ecosystem (ASPIRE), Grant No.\ JPMJAP2318. GDU is partly supported by the Leinweber Institute for Theoretical Physics at UC Berkeley, by the Department of Energy, Office of Science, Office of High Energy Physics through the award DE-SC0025522, and by the Department of Energy through QuantISED award DE-SC0019380.

\subsection*{Statement on AI use}

The manuscript was written by the authors, who take full responsibility for any errors. AI (Claude, ChatGPT, and Gemini) helped with typographical errors, revision of the manuscript, the study of constrained saddles in non-gravitational theories, stability analysis of multi-boundary wormholes, the Picard--Lefschetz analysis of single-boundary saddles, and an initial analysis that showed that there are no spherically symmetric asymptotically flat or asymptotically AdS$_{D>2}$ wormhole saddles.

\appendix

\section{Constrained instantons in scalar field theory}\label{app:scalar-ci}
In this appendix we consider the massive $\phi^4$ theory in two dimensions in its symmetric phase.  We rescale the spacetime coordinates to set the mass to one, so that the Euclidean action reads
\begin{equation}
S_g[\phi]=\int\dd^2y\left[\frac12(\partial\phi)^2+\frac12\phi^2+g\,{:}\phi^4{:}\right],
\label{eq:app-scalar-action}
\end{equation}
where $g\geq0$ is the quartic coupling in units of the squared mass. The quartic term is normal ordered with respect to the free propagator of unit mass. The normal ordering amounts to adding a mass counterterm and a vacuum-energy counterterm,
\begin{equation}
    {:}\phi^4{:} = \phi^4 - 6\tau_\Lambda \phi^2 + \mu_\Lambda.
\label{eq:app-scalar-counterterms}
\end{equation}
By constructing constrained instanton saddles in this theory, we will calculate the large-order terms in the perturbative expansions of the vacuum energy density and of the two-point correlation function,
\begin{equation}
    f(g)=-\lim_{V\to\infty}\frac{1}{V}\log\frac{Z_g}{Z_0}=\sum_{k\geq2}f_kg^k,\qquad \mathcal G_2(x)=\langle\phi(x)\phi(0)\rangle_g=\sum_{k\geq0}\mathcal G_{2,k}(x)\,g^k.
\label{eq:app-scalar-f}
\end{equation}
Here $Z_g=\int\mathcal D\phi\,e^{-S_g[\phi]}$ is the partition function, $V$ is the Euclidean spacetime volume, and $\langle\cdot\rangle_g$ denotes the expectation value at coupling $g$. The counterterm $\mu_\Lambda$ shifts $f(g)$ by $g\mu_\Lambda$, so it affects only the free-energy coefficient $f_1$, and not the large-order behavior. We fix it by requiring $f_1=0$, so that the expansion of $f$ starts at $k=2$.

To determine the coefficients $f_k$ at large $k$, we fix the value of the interaction vertex integral $\mathcal C[\phi]=\int\dd^2y\,{:}\phi^4{:}$. We insert $1=\int_{\mathbb R}\dd\zeta\,\delta(\mathcal C[\phi]-\zeta)$ into $Z_g$ and write the delta function as an integral over a Lagrange multiplier $\lambda$, which gives
\begin{equation}
Z_g=\int_{\mathbb R}\dd\zeta\int_{c+\i\mathbb R}\frac{\dd\lambda}{2\pi\i}\int\mathcal D\phi\,e^{-S_g[\phi]-\lambda(\mathcal C[\phi]-\zeta)},
\label{eq:app-scalar-constrained}
\end{equation}
where $c$ is an arbitrary real constant. Since $S_g[\phi]=S_0[\phi]+g\,\mathcal C[\phi]$, where $S_0=S_{g=0}$ is the free action, the shift $\lambda\to\lambda-g$ removes the coupling from the exponent, up to a factor $e^{-g\zeta}$. More precisely,
\begin{equation}
\frac{Z_g}{Z_0}=\int_{\mathbb R}\dd\zeta\,e^{-g\zeta}\rho_0(\zeta),
\label{eq:app-scalar-density-identity}
\end{equation}
where $\rho_0$ is the probability density of $\mathcal{C}[\phi]$ in the free theory,
\begin{equation}
\rho_0(\zeta)=\frac{1}{Z_0}\int_{c+\i\mathbb R}\frac{\dd\lambda}{2\pi\i}\int\mathcal D\phi\,e^{-S_0[\phi]-\lambda(\mathcal C[\phi]-\zeta)}=\big\langle\delta(\mathcal C[\phi]-\zeta)\big\rangle_0.
\label{eq:scalar-rho-pi-app}
\end{equation}
Expanding $e^{-g\zeta}$ in powers of $g$ shows that the free-energy perturbative expansion coefficients $f_k$ are determined by the moments $\int\dd\zeta\,\zeta^k\rho_0(\zeta)$. At large $k$, these moments are dominated by large values of $\zeta$, so the large-order behavior is controlled by the tail of $\rho_0(\zeta)$ as $\zeta\to+\infty$. We evaluate this tail by a saddle-point approximation of \eqref{eq:scalar-rho-pi-app}.

\paragraph{Constrained saddles.}
At large $\zeta = \int \dd^2y\,{:}\phi(y)^4{:}$, the values of $\phi$ are large. The counterterms in ${:}\phi(y)^4{:}$ are of lower order in $\phi$ than the bare $\phi^4$-term, and they contribute at most at the same order as the Gaussian fluctuations. We therefore first use the classical constraint $\int\dd^2y\,\phi^4=\zeta$ to find the saddles, and include the counterterms when we evaluate the Gaussian integral. The saddle-point equations of \eqref{eq:scalar-rho-pi-app} with respect to $\phi$ and $\lambda$ are then
\begin{equation}
(-\partial^2+1)\phi+4\lambda\phi^3=0,\qquad \int\dd^2y\,\phi^4=\zeta.
\label{eq:app-scalar-eom}
\end{equation}
Multiplying the first equation by $\phi$ and integrating by parts gives $\int\dd^2y\,[(\partial\phi)^2+\phi^2]=-4\lambda\zeta$. Hence, nontrivial solutions require $\lambda<0$, i.e.\ at the saddles the multiplier acts as a negative quartic coupling, and their action is $S_0[\phi]=-2\lambda\zeta$. 

To construct solutions to \eqref{eq:app-scalar-eom}, consider a positive, radially symmetric field configuration $Q(y)$ satisfying
\begin{equation}
    (-\partial^2+1)Q-Q^3=0,\qquad Q'(0)=0,\qquad Q(\infty)=0.
\label{eq:scalar-profile}
\end{equation}
It has a width of order one and decays as $e^{-r}/\sqrt r$ at large $r=|y|$. We denote its moments by $I_n=\int\dd^2y\,Q^n$. Inserting $\phi=aQ(y-y_0)$ into \eqref{eq:app-scalar-eom} gives $(1+4\lambda a^2)aQ^3=0$ and $a^4I_4=\zeta$. Hence the constrained saddles are
\begin{equation}
\phi_\zeta(y)=\pm aQ(y-y_0),\qquad a=\left(\frac{\zeta}{I_4}\right)^{1/4},\qquad \lambda_\zeta=-\frac{1}{4a^2}=-\frac14\sqrt{\frac{I_4}{\zeta}},
\label{eq:app-scalar-saddle}
\end{equation}
where the position $y_0$ is arbitrary. The free-theory action evaluated on this configuration is $S_0[\phi_\zeta]=-2\lambda_\zeta\zeta=\frac12\sqrt{I_4\zeta}$. Including the factor $e^{-g\zeta}$ in \eqref{eq:app-scalar-density-identity}, the constrained saddles contribute to $Z_g/Z_0$ with the total action
\begin{equation}
S_c(\zeta)=g\zeta+\frac12\sqrt{I_4\zeta}.
\label{eq:app-scalar-Sc}
\end{equation}

\paragraph{Stability of the constrained saddles.}
We now show that the constrained saddles are the minima of $S_0$ on the constraint surface $\int\dd^2y\,\phi^4=\zeta$, so that no constraint-preserving fluctuation can decrease their action. For this, we need to evaluate $I_2 = \int\dd^2 y\, Q^2$ and $K_Q = \int \dd^2y\, (\partial Q)^2$. Multiplying \eqref{eq:scalar-profile} by $Q$ and integrating by parts gives $K_Q+I_2=I_4$. Moreover, $Q$ is a stationary point of the functional
\begin{equation}
E[Q]=\int\dd^2y\left[\frac12(\partial Q)^2+\frac12Q^2-\frac14Q^4\right],
\end{equation}
whose Euler--Lagrange equation is \eqref{eq:scalar-profile}. Under the rescaling $Q(y)\mapsto Q(by)$, the kinetic term is invariant in two dimensions, while the other two terms scale as $b^{-2}$. Stationarity at $b=1$ therefore gives $I_2=I_4/2$, and hence
\begin{equation}
    K_Q=I_2=\frac{I_4}{2}. \label{eq:scalar-virial}
\end{equation}
Using the sharp interpolation inequality (that is, the Gagliardo--Nirenberg inequality with its optimal constant, which is attained by $Q$~\cite{CIWeinstein}), we obtain
\begin{equation}
    \int\dd^2y\,\phi^4\leq\frac{2}{I_2}\left(\int\dd^2y\,\phi^2\right)\left(\int\dd^2y\,(\partial\phi)^2\right).
\end{equation}
Consequently, with $K=\int\dd^2y\,(\partial\phi)^2$ and $M=\int\dd^2y\,\phi^2$, every field with $\int\dd^2y\,\phi^4=\zeta$ satisfies
\begin{equation}
S_0=\frac{K+M}{2}\geq\sqrt{KM}\geq\sqrt{\frac{I_2\zeta}{2}}=\frac12\sqrt{I_4\zeta}.
\label{eq:scalar-min}
\end{equation}
The saddles $\pm aQ(y-y_0)$ saturate both inequalities, since $K=M=a^2I_2$ by \eqref{eq:scalar-virial}. They are therefore the minima of $S_0$ on the constraint surface, which proves their stability.

\paragraph{One-loop corrections.}
We now calculate the one-loop corrections around the saddle $\phi_\zeta=aQ(y-y_0)$; the saddle $-\phi_\zeta$ gives the same contribution. We choose $c=\lambda_\zeta$ in \eqref{eq:scalar-rho-pi-app}, so that the contour of $\lambda$ passes through the saddle, and expand $\phi=\phi_\zeta+\eta$ and $\lambda=\lambda_\zeta+\xi$. The terms linear in $\eta$ and $\xi$ cancel by \eqref{eq:app-scalar-eom}, and up to quadratic order in fluctuations $\eta,\xi$, we have
\begin{equation}
S_0[\phi]+\lambda\Bigl(\int\dd^2y\,\phi^4-\zeta\Bigr)=S_0[\phi_\zeta]+\frac12\langle\eta,L\eta\rangle+\xi\,\langle v_\zeta,\eta\rangle,
\label{eq:app-scalar-quadratic}
\end{equation}
where $\langle f,h\rangle=\int\dd^2y\,fh$ and
\begin{equation}
L=-\partial^2+1+12\lambda_\zeta\phi_\zeta^2=-\partial^2+1-3Q^2,\qquad v_\zeta=4\phi_\zeta^3=4a^3Q^3,
\label{eq:scalar-hessian}
\end{equation}
with $Q$ centered at $y_0$. The fluctuations with $\langle v_\zeta,\eta\rangle=0$ preserve the constraint, because $v_\zeta$ is the gradient of $\int\dd^2y\,\phi^4$ at $\phi_\zeta$. To evaluate the Gaussian integral over $\eta$, we first separate the negative and zero modes of $L$. Let $(r,\theta)$ be polar coordinates centered at $y_0$. Then, we can split the fluctuations into sectors with fixed angular momenta $\ell$, $\eta\propto e^{\i\ell\theta}$. In each sector, $L$ acts as
\begin{equation}
L_\ell=-\partial_r^2-\frac1r\partial_r+\frac{\ell^2}{r^2}+1-3Q(r)^2.
\end{equation}
Differentiating \eqref{eq:scalar-profile} shows that the translation modes $\partial_\mu Q=Q'(r)\,y_\mu/r$, $\mu=1,2$, are zero modes of $L$. They have $\ell=1$, and their radial function $Q'(r)$ has no nodes at $r>0$, since $Q$ decreases monotonically. Hence $Q'$ is the lowest eigenfunction of $L_1$, because the lowest eigenfunction has no nodes either, and two functions without nodes cannot be orthogonal. Therefore $L_1\geq0$, and since $L_\ell=L_1+(\ell^2-1)/r^2$, all sectors with $\ell\geq2$ are strictly positive.

In the sector $\ell=0$, equation \eqref{eq:scalar-profile} reduces to $L_0Q=-2Q^3$, so that $\langle Q,L_0Q\rangle=-2I_4<0$, and consequently, $L_0$ has a negative eigenvalue. The fluctuation $\eta\propto Q$ does not preserve the constraint since $\langle v_\zeta,Q\rangle=4a^3I_4\neq0$. This is the only negative mode of $L_0$. To see this, we first show that $\langle\eta,L\eta\rangle\geq0$ for all constraint-preserving fluctuations. Consider a curve $\phi(t)$ on the constraint surface with $\phi(0)=\phi_\zeta$ and $\dot\phi(0)=\eta$, so that $\langle v_\zeta,\eta\rangle=0$. Differentiating the constraint twice at $t=0$ gives $\langle v_\zeta,\ddot\phi(0)\rangle=-12\langle\eta,\phi_\zeta^2\eta\rangle$, with $\phi_\zeta$ satisfying $(-\partial^2+1)\phi_\zeta=-\lambda_\zeta v_\zeta$. Hence
\begin{equation}
\frac{\dd^2}{\dd t^2}S_0[\phi(t)]\Big|_{t=0}=\langle\eta,(-\partial^2+1)\eta\rangle+\langle(-\partial^2+1)\phi_\zeta,\ddot\phi(0)\rangle=\langle\eta,L\eta\rangle.
\end{equation}
Since $\phi_\zeta$ minimizes $S_0$ on the constraint surface, the left-hand side is nonnegative, so $\langle\eta,L\eta\rangle\geq0$ whenever $\langle v_\zeta,\eta\rangle=0$. If $L$ had two negative eigenvalues, their span would contain a nonzero $\eta$ with $\langle v_\zeta,\eta\rangle=0$, since this is a single linear condition, and this $\eta$ would have $\langle\eta,L\eta\rangle<0$. Thus, $L$ can have only one negative eigenvalue, so all other modes in the $\ell=0$ sector are positive. The sector $\ell=0$ contains no further zero modes, by the nondegeneracy of the ground-state profile~\cite{CIWeinsteinModStab,CIKwong}, so the kernel of $L$ is spanned by $\partial_1Q$ and $\partial_2Q$.

We can now write the Gaussian integral explicitly, first with the classical constraint; the counterterms are included in the next paragraph. In the quadratic action \eqref{eq:app-scalar-quadratic}, $\xi$ appears only in the term $\xi\langle v_\zeta,\eta\rangle$, so the integral over $\xi$ along the imaginary axis gives $\delta(\langle v_\zeta,\eta\rangle)$, and the two saddles contribute
\begin{equation}
\rho_0^{\mathrm{CI}}(\zeta)=2\,e^{-S_0[\phi_\zeta]}\,
\frac{\int\mathcal D\eta\,\delta(\langle v_\zeta,\eta\rangle)\,e^{-\frac12\langle\eta,L\eta\rangle}}{\int\mathcal D\eta\,e^{-S_0[\eta]}},
\label{eq:scalar-oneloop-app}
\end{equation}
where $S_0[\phi_\zeta]=\frac12\sqrt{I_4\zeta}$, and the denominator is $Z_0$. The measure $\mathcal D\eta$ is normalized so that the Gaussian integral over a single mode with eigenvalue $\omega$ gives $\omega^{-1/2}$. Following the analysis above, we split the fluctuations into the translation zero modes, the component along $v_\zeta$, which violates the constraint, and the constraint-preserving fluctuations orthogonal to the zero modes,
\begin{equation}
    \eta=\sum_{\mu=1,2}c_\mu\,\frac{\partial_\mu\phi_\zeta}{\|\partial_\mu\phi_\zeta\|}+s\,\frac{v_\zeta}{\|v_\zeta\|}+\eta_\perp,\qquad \langle\partial_\mu\phi_\zeta,\eta_\perp\rangle=\langle v_\zeta,\eta_\perp\rangle=0, \label{eq:app-scalar-split}
\end{equation}
where $\|f\|^2=\langle f,f\rangle$. The zero modes are orthogonal to $v_\zeta$, since $v_\zeta$ is radially symmetric, so the integral \eqref{eq:scalar-oneloop-app} factorizes. The integral over the zero modes is traded for the integral over the position $y_0$ of the saddle, and gives the volume $V$ times the Jacobian $J_{\mathrm{tr}}$ of the change of variables from $\dd c_\mu/\sqrt{2\pi}$ to $\dd y_0^\mu$. The integral over the constraint-violating component $s$, with measure $\dd s/\sqrt{2\pi}$, is fixed by the delta function, $\delta(\langle v_\zeta,\eta\rangle)=\delta(s)/\|v_\zeta\|$, and gives the factor $1/(\sqrt{2\pi}\,\|v_\zeta\|)$. In this way, the constraint removes the negative mode of $L$. Finally, the Gaussian integral over all other modes $\eta_\perp$ converges, since $L$ is positive on these fluctuations, and gives the factor $[\det(PLP)]^{-1/2}$. The normalization constant $1/Z_0$ contributes the free theory functional determinant $\det(-\partial^2+1)^{1/2}$. Altogether,
\begin{equation}
\rho_0^{\mathrm{CI}}(\zeta)=2\,e^{-S_0[\phi_\zeta]}\times VJ_{\mathrm{tr}}\times\frac{1}{\sqrt{2\pi}\,\|v_\zeta\|}\times\left[\frac{\det(PLP)}{\det(-\partial^2+1)}\right]^{-1/2},
\label{eq:app-scalar-factorized}
\end{equation}
where $P$ is the projector onto the fluctuations $\eta_\perp$. It remains to compute $J_{\mathrm{tr}}$ and $\det(PLP)$.

Since $\partial\phi_\zeta/\partial y_0^\mu=-\partial_\mu\phi_\zeta$, the coefficients of the zero modes change as $\dd c_\mu=-\|\partial_\mu\phi_\zeta\|\,\dd y_0^\mu$. Hence, each direction contributes the factor $\|\partial_\mu\phi_\zeta\|/\sqrt{2\pi}$ to the Jacobian. By rotational symmetry and \eqref{eq:scalar-virial}, $\|\partial_\mu\phi_\zeta\|^2=a^2\int\dd^2y\,(\partial_\mu Q)^2=a^2I_2/2=a^2I_4/4$, and hence
\begin{equation}
J_{\mathrm{tr}}=\prod_{\mu=1,2}\frac{\|\partial_\mu\phi_\zeta\|}{\sqrt{2\pi}}=\frac{a^2I_4}{8\pi}.
\end{equation}
To express $\det(PLP)$ through the determinant of $L$, we use that the diagonal element of $L^{-1}$ along the unit vector $v_\zeta/\|v_\zeta\|$ equals $\det(PLP)/\det'L$. Here $L^{-1}$ is the inverse of $L$ on the subspace orthogonal to the zero modes, and the prime denotes the determinant on this subspace. Hence
\begin{equation}
\det(PLP)=\det\!'L\,\frac{B_\zeta}{\|v_\zeta\|^2},\qquad B_\zeta\equiv\langle v_\zeta,L^{-1}v_\zeta\rangle=-8a^6I_4,
\label{eq:scalar-susceptibility}
\end{equation}
where we used $L^{-1}v_\zeta=-2a^3Q$, which follows from $LQ=-2Q^3$. The factors of $\|v_\zeta\|$ cancel, and the last two factors in \eqref{eq:app-scalar-factorized} combine into
\begin{equation}
\frac{1}{\sqrt{2\pi}}\left[\frac{\det'L}{\det(-\partial^2+1)}\,B_\zeta\right]^{-1/2}.
\label{eq:scalar-gaussian-app}
\end{equation}
Both factors in the brackets are negative, since $L$ has exactly one negative eigenvalue and $B_\zeta<0$, so their product is positive. Collecting the factors in \eqref{eq:app-scalar-factorized} and using $a=(\zeta/I_4)^{1/4}$, we obtain
\begin{equation}
\frac{\rho_0^{\mathrm{CI}}(\zeta)}{V}=\frac{2J_{\mathrm{tr}}}{\sqrt{2\pi|B_\zeta|}}\left|\frac{\det'L}{\det(-\partial^2+1)}\right|^{-1/2}e^{-\frac12\sqrt{I_4\zeta}}
=\frac{I_4^{3/4}}{16\pi^{3/2}}\left|\frac{\det'L}{\det(-\partial^2+1)}\right|^{-1/2}\zeta^{-1/4}e^{-\frac12\sqrt{I_4\zeta}}.
\label{eq:scalar-tail-bare-app}
\end{equation}

\paragraph{Renormalization.}
To define and calculate the functional determinants in \eqref{eq:scalar-tail-bare-app}, we need to introduce a UV regulator. We use a cutoff $\Lambda$ and denote by $G_{0,\Lambda}$ the regulated free propagator, $(-\partial^2+1)G_{0,\Lambda}=\delta^2(y)$, whose value at coincident points is $G_{0,\Lambda}(0)\simeq(2\pi)^{-1}\log\Lambda$. Since $L=-\partial^2+1-3Q^2$, the ratio of determinants is
\begin{equation}
\log\frac{\det L}{\det(-\partial^2+1)}=\operatorname{Tr}\log\bigl(1-3G_{0,\Lambda}Q^2\bigr)=-3\operatorname{Tr}\bigl(G_{0,\Lambda}Q^2\bigr)+\dots=-3G_{0,\Lambda}(0)\,I_2+\text{finite}.
\end{equation}
Only the first term of the expansion of the logarithm diverges in two dimensions, and removing the zero modes does not change this. Hence $|\det'L/\det(-\partial^2+1)|^{-1/2}$ contains the factor $e^{3G_{0,\Lambda}(0)I_2/2}$, which diverges when the regulator is removed. This divergence is canceled by the mass counterterm in the constraint. Normal ordering with respect to $G_{0,\Lambda}$ fixes the mass renormalization constant in \eqref{eq:app-scalar-counterterms} to $\tau_\Lambda=G_{0,\Lambda}(0)$, and the exponent of \eqref{eq:scalar-rho-pi-app} contains the mass term $6\tau_\Lambda\lambda\int\dd^2y\,\phi^2$. On the saddle, where $\lambda=\lambda_\zeta=-1/(4a^2)$ and $\int\dd^2y\,\phi_\zeta^2=a^2I_2$, this term equals $-\frac32\tau_\Lambda I_2$, and the factor $e^{-3\tau_\Lambda I_2/2}$ cancels the divergence. This cancellation requires the constraint itself to include the mass counterterm: with the bare constraint $\int\dd^2y\,\phi^4$, the divergence of the determinant would remain. There are no other divergences at the leading order in the large-$\zeta$ expansion, and the finite result is
\begin{equation}
\frac{\rho_0^{\mathrm{CI}}(\zeta)}{V}=\frac{I_4^{3/4}}{16\pi^{3/2}\sqrt{|\Delta_R|}}\,\zeta^{-1/4}e^{-\frac12\sqrt{I_4\zeta}}\Bigl[1+\O\bigl(\zeta^{-1/2}\bigr)\Bigr],\qquad
\Delta_R=\lim_{\Lambda\to\infty}e^{3\tau_\Lambda I_2}\,\frac{\det'L}{\det(-\partial^2+1)}.
\label{eq:scalar-DeltaR-app}
\end{equation}
The corrections of relative order $\zeta^{-1/2}$ come from higher orders in the fluctuations, which are suppressed by powers of the effective $\phi^4$-coupling $\lambda_\zeta\propto\zeta^{-1/2}$. The leading term is precisely the result \eqref{eq:scalar-density-tail} quoted in Section~\ref{sec:scalar-ci}.

\paragraph{Large-order behavior.}
The $k$-th moment of the tail \eqref{eq:scalar-DeltaR-app} is
\begin{align}
\frac{1}{k!}\int_0^\infty\dd\zeta\,\zeta^k\,\frac{\rho_0^{\mathrm{CI}}(\zeta)}{V}
&=\frac{4^k\,\Gamma(2k+3/2)}{(2\pi)^{3/2}\sqrt{|\Delta_R|}\,k!\,I_4^k}\nonumber\\
&=\frac{1}{2\pi^2\sqrt{|\Delta_R|}}\Bigl(\frac{16}{I_4}\Bigr)^{k}\frac{\Gamma(k+3/4)\Gamma(k+5/4)}{\Gamma(k+1)},
\label{eq:scalar-mellin-app}
\end{align}
where we used the duplication formula for the gamma function. At large $k$, the ratio of gamma functions is asymptotic to $\Gamma(k+1)=k!$. The integrand $\zeta^ke^{-\frac12\sqrt{I_4\zeta}}$ is maximal at $\zeta_k=16k^2/I_4$, so the $k$-th moment indeed probes only the tail of $\rho_0$.

The perturbative coefficients follow from these moments. Since $\log(Z_g/Z_0)=\log\langle e^{-g\mathcal C}\rangle_0$ is the generating function of the cumulants of $\mathcal C$ in the free theory, \eqref{eq:app-scalar-f} gives
\begin{equation}
f_k=\frac{(-1)^{k+1}}{k!\,V}\,\langle\mathcal C^k\rangle_{0,c},
\end{equation}
where $\langle\mathcal C^k\rangle_{0,c}$ is the $k$-th cumulant, and the limit $V\to\infty$ is understood. The density $\rho_0$ also receives contributions from configurations with several widely separated saddles. These cancel in the cumulants, which retain only the contribution of a single saddle, proportional to $V$. At large $k$, we can therefore replace $\langle\mathcal C^k\rangle_{0,c}/V$ by the moment of $\rho_0^{\mathrm{CI}}/V$, and \eqref{eq:scalar-mellin-app} gives
\begin{equation}
f_k\simeq\frac{(-1)^{k+1}\,k!}{2\pi^2\sqrt{|\Delta_R|}}\Bigl(\frac{16}{I_4}\Bigr)^{k},
\label{eq:app-scalar-fk}
\end{equation}
which is \eqref{eq:scalar-large-order}. Equivalently, the Borel transform $\mathcal Bf(t)=\sum_kf_kt^k/k!$ has the leading singularity $-\bigl[2\pi^2\sqrt{|\Delta_R|}\,(1+16t/I_4)\bigr]^{-1}$ at $t=-I_4/16$.

Inserting fields into the constrained path integral determines the large-order behavior of correlation functions in the same way. On the saddle, the insertion of $\phi(0)\phi(x)$ gives $a^2Q(-y_0)Q(x-y_0)$, which becomes $a^2\int\dd^2y\,Q(y)Q(y+x)$ after the integral over the position $y_0$, in place of the factor $V$. Since $a^2=\sqrt{\zeta/I_4}$, the corresponding moment acquires an extra factor $\sqrt{\zeta_k/I_4}=4k/I_4$. For the coefficients $\mathcal G_{2,k}(x)$ of the two-point function \eqref{eq:app-scalar-f}, at fixed separation $x$, this gives
\begin{equation}
\mathcal G_{2,k}(x)\simeq(-1)^k\frac{\Gamma(k+2)}{8\pi^2\sqrt{|\Delta_R|}}\Bigl(\frac{16}{I_4}\Bigr)^{k+1}\int\dd^2y\,Q(y)Q(y+x).
\label{eq:scalar-two-point}
\end{equation}
The profile $Q$ thus determines both the factorial growth and the dependence on the separation. In particular, the coefficients of the susceptibility $\chi=\int\dd^2x\,\mathcal G_2(x)$ behave as
\begin{equation}
\chi_k\simeq(-1)^k\frac{I_3^2\,\Gamma(k+2)}{8\pi^2\sqrt{|\Delta_R|}}\Bigl(\frac{16}{I_4}\Bigr)^{k+1},
\end{equation}
where we used $\int\dd^2y\,Q=\int\dd^2y\,Q^3=I_3$, which follows by integrating \eqref{eq:scalar-profile}.

\paragraph{Comparison with the literature.}
The same constrained saddles determine the imaginary part of the vacuum energy at negative coupling, $g=-s<0$. There, the integral over $\zeta$ in \eqref{eq:app-scalar-density-identity} diverges at large $\zeta$, and $f(g)$ is defined by analytic continuation from $g>0$, approaching $g=-s$ from above or below the real axis. The total action \eqref{eq:app-scalar-Sc}, $S_c(\zeta)=-s\zeta+\frac12\sqrt{I_4\zeta}$, now has a stationary point at $\zeta_\star=I_4/(16s^2)$, where $S_c(\zeta_\star)=I_4/(16s)$ and $S_c''(\zeta_\star)=-8s^3/I_4<0$. At this point, the multiplier of \eqref{eq:app-scalar-constrained}, $\lambda_{\zeta_\star}-g=s-\frac14\sqrt{I_4/\zeta_\star}$, vanishes, and $\phi_{\zeta_\star}=\pm Q/(2\sqrt s)$ is an ordinary saddle of $S_g$, the instanton at negative coupling. The continued contour passes through $\zeta_\star$ along the imaginary direction, along which $e^{-S_c}$ decreases, and the imaginary part is given by half of the Gaussian integral through $\zeta_\star$. For $g=-s+\i0$, this gives
\begin{equation}
\operatorname{Im}f(-s+\i0)=\frac12\,\frac{\rho_0^{\mathrm{CI}}(\zeta_\star)}{V}\,e^{s\zeta_\star}\sqrt{\frac{2\pi}{|S_c''(\zeta_\star)|}}=\frac{I_4}{32\pi\sqrt{|\Delta_R|}\,s}\,e^{-I_4/(16s)}.
\end{equation}
If $f(g)$ is analytic in the $g$-plane cut along the negative real axis, Cauchy's theorem gives the dispersion relation
\begin{equation}
f_k=\frac{1}{\pi}\int_0^\infty\frac{\dd s}{(-s)^{k+1}}\operatorname{Im}f(-s+\i0).
\end{equation}
At large $k$, the integral is dominated by $s\simeq I_4/(16k)$, where the semiclassical result applies, and it reproduces \eqref{eq:app-scalar-fk}.

Thus, at negative coupling, the constrained-saddle calculation reduces to the one of~\cite{CIMPR}, where the large-order behavior of the vacuum energy was obtained from the Gaussian fluctuations around the instanton $\phi_{\zeta_\star}$ and the dispersion relation. To compare the two results, note that their coupling multiplies $\phi^4/4$, so $g_{\mathrm{MPR}}=4g$, and their coefficients in the free energy expansion, $F_k$, are related to ours as $f_k = 4^k F_k$. In two dimensions, their Eqs.~(53) and~(56) give, for the coefficients of expansion in $g_{\mathrm{MPR}}$,
\begin{equation}
F_k=-C\Bigl(-\frac{4}{I_4}\Bigr)^{k}k!,\qquad C=\frac{I_6-I_4}{\pi^2I_4\sqrt{|\Omega_R^\perp|}}.
\label{eq:scalar-MPR-app}
\end{equation}
This agrees with \eqref{eq:app-scalar-fk} provided that $C=1/(2\pi^2\sqrt{|\Delta_R|})$. The renormalized functional determinant $\Omega_R^\perp$ of~\cite{CIMPR} is defined as the ratio $\det L/\det(-\partial^2+1)=\det(1-T)$, with $T=3G_{0,\Lambda}^{1/2}Q^2G_{0,\Lambda}^{1/2}$ and the same renormalization factor $e^{3\tau_\Lambda I_2}$, but with the two zero modes removed from $1-T$ rather than from $L$. When a zero mode $u_\mu=\partial_\mu Q$ is lifted, the small eigenvalues of $L$ and of $1-T$ differ by the factor $\langle u_\mu,u_\mu\rangle/\langle u_\mu,(-\partial^2+1)u_\mu\rangle$. Here $\langle u_\mu,u_\mu\rangle=I_4/4$, by rotational symmetry and \eqref{eq:scalar-virial}, and $\langle u_\mu,(-\partial^2+1)u_\mu\rangle=3\langle u_\mu,Q^2u_\mu\rangle=(I_6-I_4)/2$, where we used $Lu_\mu=0$ and integrated $\int\dd^2y\,Q^3(-\partial^2+1)Q=I_6$ by parts. Hence
\begin{equation}
\Delta_R=\left[\frac{I_4}{2(I_6-I_4)}\right]^2\Omega_R^\perp.
\label{eq:scalar-determinant-map-app}
\end{equation}
Substituting this relation into \eqref{eq:scalar-MPR-app} gives $C=1/(2\pi^2\sqrt{|\Delta_R|})$, so the results \eqref{eq:scalar-MPR-app} and \eqref{eq:app-scalar-fk} agree.

\section{Cotler--Jensen wormholes in Einstein--Maxwell theory}\label{app:cotler-jensen}
In \cite{Cotler:2022rud}, Cotler and Jensen constructed constrained two-boundary wormholes in Einstein--Maxwell theory with a negative cosmological constant by constraining the proper length of the wormhole. In this appendix, we review their construction and show that their contribution to $\kappa_2$ has the same form as that of our constrained wormholes, see \eqref{eq:kappa2-cutoff}. To simplify the discussion of the smoothness conditions, we work in four dimensions and with asymptotically flat boundary conditions, although the computation below straightforwardly extends to any number of dimensions.

\paragraph{Constrained saddles.}\label{app:cj-saddles}
We consider Einstein--Maxwell theory with the Euclidean action
\begin{equation}
I = \int\dd^4x\sqrt{g}\left(-\frac{R}{16\pi G_4}+\frac{1}{4}F_{\mu\nu}F^{\mu\nu}\right),
\end{equation}
supplemented with suitable boundary terms. In the gauge $\dd s^2 = \dd\rho^2+g_{ij}(\rho,x)\,\dd x^i\dd x^j$, the constraint of \cite{Cotler:2022rud} reads
\begin{equation}
\cC(\phi) \equiv \frac{1}{8\pi G_4}\int\dd^4x\,\sqrt{g_{\rho\rho}}\,\sqrt{g_{S^2}} = \zeta,
\label{eq:cotler-jensen-constraint}
\end{equation}
where $\phi$ collectively denotes the metric and the gauge field, and $\sqrt{g_{S^2}}$ is the volume density of the unit two-sphere. The constraint is imposed with a Lagrange multiplier,
\begin{equation}
Z = \int\dd\zeta\int_{c+\i\mathbb{R}}\frac{\dd\lambda}{2\pi \i}\int\mathcal{D}\phi\,\exp\Bigl[-I(\phi)-\lambda\bigl(\cC(\phi)-\zeta\bigr)\Bigr].
\end{equation}
For the static, spherically symmetric configurations considered below, it fixes the proper length between the two wormhole boundaries. The constraint does not affect the Maxwell equations, and it modifies only the $\rho\rho$ component of the Einstein equations.

The two-boundary constrained saddles with inverse temperatures $\beta_L$, $\beta_R$ and electric flux $Q$ are given by \cite{Cotler:2022rud}
\begin{equation}
\dd s^2 = \frac{\dd r^2}{g(r)}+g(r)\,\xi_i(r)^2\,\dd\tau^2+r^2\dd\Omega_2^2,\qquad A = -\i\,h_i(r)\,\dd\tau,\qquad \tau\sim\tau+1,
\end{equation}
where $i=L,R$ labels the two halves of the wormhole. Each half covers the range $r_+\le r\le r_c$, where $r_c$ is a large-radius cutoff, and the two halves are glued at the throat $r=r_+$, where $r(\rho)$ has a minimum. Here,
\begin{equation}
g(r) = 1-\frac{2M}{r}+\frac{\cQ^2}{r^2} = \frac{(r-r_+)(r-r_-)}{r^2},\qquad \cQ^2 = \frac{G_4}{4\pi}Q^2,
\end{equation}
is the RN function with mass parameter $M$, and
\begin{equation}
\xi_i(r) = \beta_i-\frac{\lambda}{2}\Bigl[\mathcal{J}(r;M,Q)-\mathcal{J}(r_c;M,Q)\Bigr],\qquad h_i(r) = \frac{Q}{4\pi}\int_{r_+}^r\dd r'\,\frac{\xi_i(r')}{r'^2},
\end{equation}
where $\mathcal{J}$ is an antiderivative of $r^{-1}g(r)^{-3/2}$,
\begin{equation}
\mathcal{J}(r;M,Q) = \log\left(\frac{\sqrt{r-r_-}+\sqrt{r-r_+}}{\sqrt{r-r_-}-\sqrt{r-r_+}}\right)+2\,\frac{r_+r_-(r_++r_-)-r(r_+^2+r_-^2)}{\sqrt{(r-r_+)(r-r_-)}\,(r_+-r_-)^2}.
\end{equation}
Thus, the constraint does not change the RN function $g(r)$, and it only deforms the lapse through $\xi_i(r)$, which is constant for $\lambda=0$. The boundary conditions $\xi_i(r_c)=\beta_i$ fix the inverse temperatures at the two cutoff surfaces.

Near the throat, $\rho = \pm\frac{2}{\sqrt{g'(r_+)}}\sqrt{r-r_+}+\mathcal{O}(r-r_+)$, with opposite signs on the two halves of the wormhole. Using
\begin{align}
\mathcal{J}(r;M,Q) &= -\frac{2r_+^2}{(r_+-r_-)^{3/2}}\,\frac{1}{\sqrt{r-r_+}}+\mathcal{O}\bigl(\sqrt{r-r_+}\bigr),\nonumber\\
\mathcal{J}(r_c;M,Q) &= \log\left(\frac{4r_c}{r_+-r_-}\right)-\frac{2(r_+^2+r_-^2)}{(r_+-r_-)^2}+\mathcal{O}\left(\frac{1}{r_c}\right),
\end{align}
we find
\begin{equation}
\xi_i(r) = \frac{A_i}{\sqrt{r-r_+}}+B_i+\mathcal{O}\bigl(\sqrt{r-r_+}\bigr),\qquad A_i = \frac{\lambda\,r_+^2}{(r_+-r_-)^{3/2}},\qquad B_i = \beta_i+\frac{\lambda}{2}\,\mathcal{J}(r_c;M,Q),
\end{equation}
so that $g\,\xi_i^2 = A_i^2\,g'(r_+)+2A_iB_i\,g'(r_+)\sqrt{r-r_+}+\mathcal{O}(r-r_+)$. Therefore, the Euclidean time circle does not shrink at the throat, and the metric is smooth at $\rho=0$ provided that
\begin{equation}\label{smoothness_condition}
A_L=A_R,\qquad B_L=-B_R.
\end{equation}
The first condition holds automatically, while the second one gives
\begin{equation}\label{smoothness2}
\beta_L+\beta_R+\lambda\,\mathcal{J}(r_c;M,Q) = 0.
\end{equation}

Finally, the proper length of the wormhole is
\begin{align}
\ell_{\rm WH}(M,Q,r_c) &= 2\int_{r_+}^{r_c}\frac{\dd r}{\sqrt{g(r)}} = 2\left[\sqrt{(r-r_+)(r-r_-)}+\frac{r_++r_-}{2}\log\left(\frac{\sqrt{r-r_-}+\sqrt{r-r_+}}{\sqrt{r-r_-}-\sqrt{r-r_+}}\right)\right]_{r=r_+}^{r=r_c}\nonumber\\
&= 2r_c+(r_++r_-)\log\left(\frac{4r_c}{r_+-r_-}\right)-(r_++r_-)+\mathcal{O}\left(\frac{1}{r_c}\right),
\end{align}
and one can check that
\begin{equation}\label{aux1}
\frac{\partial}{\partial M}\ell_{\rm WH}(M,Q,r_c) = 2\,\mathcal{J}(r_c;M,Q).
\end{equation}
The constraint \eqref{eq:cotler-jensen-constraint} sets the proper length of the wormhole to $\ell_{\rm WH}(M,Q,r_c) = L$, where $L\equiv 2G_4\zeta$, and thus determines $M$ in terms of $L$. Since the construction requires $g(r)$ to have a zero at $r=r_+$, these wormholes exist only for $M\ge\cQ$. In the extremal limit, $M\to\cQ$, the throat becomes infinitely long, $\ell_{\rm WH}\to\infty$.

\paragraph{Calculating on-shell action.}\label{app:cj-onshell}
We evaluate the on-shell action using the canonical formalism of Section~\ref{sec:onshell-actions}, in the ensemble with fixed temperature and electric flux. Since the Euclidean time circle is nowhere contractible, it defines a regular foliation of the wormhole by slices $\Sigma$ which connect the two asymptotic boundaries. The boundary $\partial\Sigma$ consists only of the two spheres at $r=r_c$.

Importantly, the constraint \eqref{eq:cotler-jensen-constraint} depends only on the induced metric $\gamma_{ab}$ on $\Sigma$, since $g_{\rho\rho} = \gamma_{\rho\rho}$ in the decomposition \eqref{eq:os-ADM}. It does not involve the lapse $N$, the shift $N^a$, or $A_\tau$. Therefore, the term $\lambda(\cC-\zeta)$ modifies only the evolution equations for $\gamma_{ab}$, while the equations of motion for $N$, $N^a$ and $A_\tau$, namely the Hamiltonian, momentum and Gauss constraints \eqref{eq:os-constraints}, are unchanged and hold on the entire wormhole. Consequently, the terms in \eqref{eq:os-canonical-action} proportional to $N$, $N^a$ and $A_{\tau}$ vanish, and the action, including the constraint term, reduces to
\begin{align}
I ={}& \int_0^1\dd\tau\int_\Sigma\dd^3x\,\bigl(\pi^{ab}\dot\gamma_{ab}+\mathcal{E}^a\dot A_a\bigr)\nonumber 
\\
&-\frac{1}{8\pi G_4}\int_0^1\dd\tau\int_{\partial\Sigma}\dd^2y\,\sqrt{\sigma}\,\Bigl[N\,(\vartheta-\vartheta_0)+r_aN_b\bigl(K^{ab}-K\gamma^{ab}\bigr)\Bigr]\nonumber\\
&-\int_0^1\dd\tau\int_{\partial\Sigma}\dd^2y\,\sqrt{\sigma}\,N\,r_aA_bF^{ab}+\lambda\bigl(\cC-\zeta\bigr).
\label{eq:cj-reduced-action}
\end{align}
The spatial metric $\gamma_{ab}\dd x^a\dd x^b = \dd\rho^2+r(\rho)^2\dd\Omega_2^2$ and the gauge field $A = -\i\,h_i(r)\,\dd\tau$ do not depend on $\tau$, and $A_a=0$. Hence, $\dot\gamma_{ab}=\dot A_a=0$, and the only nonvanishing terms in \eqref{eq:cj-reduced-action} are the boundary term proportional to $N(\vartheta-\vartheta_0)$ and the constraint term. On the side $i$, the lapse at the cutoff is $N = \sqrt{g(r_c)}\,\xi_i(r_c) = \beta_i\sqrt{g(r_c)}$, so by \eqref{eq:os-asymptotic} each asymptotic boundary contributes $\beta_iE$, where $E = M/G_4$ is the ADM energy. Hence,
\begin{equation}\label{eq:cj-onshell}
I(\beta_L,\beta_R,Q;M,\lambda,L) = (\beta_L+\beta_R)\,E+\frac{\lambda}{2G_4}\bigl(\ell_{\rm WH}(M,Q,r_c)-L\bigr).
\end{equation}
This expression holds for any $M$ and $\lambda$, even when the smoothness condition \eqref{smoothness2} is not imposed, because its derivation only uses the constraints, which are satisfied by all members of the family. Extremizing \eqref{eq:cj-onshell} with respect to $\lambda$ and $M$ gives
\begin{equation}
0 = \frac{\partial I}{\partial\lambda} = \frac{1}{2G_4}\bigl(\ell_{\rm WH}(M,Q,r_c)-L\bigr),\qquad 0 = \frac{\partial I}{\partial M} = \frac{1}{G_4}\bigl(\beta_L+\beta_R+\lambda\,\mathcal{J}(r_c;M,Q)\bigr),
\end{equation}
where we used \eqref{aux1}. The first equation is the constraint, and the second one is precisely the smoothness condition \eqref{smoothness2}. On the constrained saddles, the action is therefore $(\beta_L+\beta_R)E$.

\paragraph{Thin wormholes and the cutoff.}
On the constrained saddles, the circumference of the Euclidean time circle at the throat is
\begin{equation}
b=|A_i|\sqrt{g'(r_+)}=\frac{|\lambda|\,r_+}{r_+-r_-}.
\label{eq:cj-b}
\end{equation}
At fixed $\beta_L$, $\beta_R$ and $Q$, both $L$ and $b$ are determined by $M$, so $L$ is a function of $b$, and we can change the integration variable from $\zeta = L/2G_4$ to $b$. The contribution of these wormholes to the connected two-boundary GPI with fixed temperatures $\beta_L$, $\beta_R$ and electric flux $Q$ is then
\begin{equation}
\overline{Z(\beta_L,Q)Z(\beta_R,Q)}_{\rm connected}\simeq\int_0^\infty\dd b\,\mu_{\beta_L,\beta_R,Q}(b)\,e^{-(\beta_L+\beta_R)E(b)},
\label{eq:cj-ZZ}
\end{equation}
where $E(b)$ is the energy of the wormhole with throat circumference $b$, and the measure factor $\mu_{\beta_L,\beta_R,Q}(b)$ includes the Jacobian $|\dd L/\dd b|$ and loop corrections. The lower limit, $b\to0$, corresponds to $L\to\infty$. In this limit, the wormhole approaches extremality, $M\to\cQ$, and \eqref{smoothness2} gives $\lambda\to0$, since $\mathcal{J}(r_c;M,Q)\to-\infty$. Near extremality, \eqref{eq:cj-b} gives $b\simeq(\beta_L+\beta_R)(r_+-r_-)/(4R_0)$, where $R_0=\cQ$ is the extremal radius, so the wormhole becomes very thin, and its energy approaches $E_{\rm ext}(Q)=\cQ/G_4=\Phi_{\rm ext}|Q|$.

The integral \eqref{eq:cj-ZZ} has a similar form to the integral over the neck area $\cA$ for the constrained wormholes of Section~\ref{sec:2bdry-saddles}, but the exponents differ by the contribution $-\lambda\cA$ of the constraint brane to the action. Still, the cutoff argument of Section~\ref{sec:moments} depends only on the circumference of the time circle at the neck and on the curvature there, of order $1/R_0^2$. Therefore, it applies without change, and we impose the same cutoff on the circumference of the throat, $b\geq b_c(Q)=K_cG_4^{1/4}R_0^{1/2}$. The integral \eqref{eq:cj-ZZ} is then dominated by its lower limit $b_c(Q)$, and
\begin{equation}\label{Zconnected}
\overline{Z(\beta_L,Q)Z(\beta_R,Q)}_{\rm connected}\simeq P(Q)\,e^{-(\beta_L+\beta_R)E(b_c(Q))},
\end{equation}
where $P(Q)$ grows slower than exponentially with $|Q|$. For $\beta_L=\beta_R=\beta$, after including the factors $e^{\beta\Phi Q}$ from the two boundaries and summing over $Q$, this has the form \eqref{eq:kappa2-cutoff}, with $I_c(Q)$ replaced by $2\beta E(b_c(Q))$. Moreover, by \eqref{eq:cj-b} and \eqref{smoothness2}, $b/\beta$ depends only on $r_+/R_0$ and, logarithmically, on $r_c/R_0$. Hence, at fixed $r_+/R_0$, the cutoff gives $R_0\propto\beta^2/\lpl$, and the exponent, which is proportional to $\beta R_0/G_4$, grows as $(\beta/\lpl)^3$. Therefore, $\log\kappa_2$ grows as $(\beta/\lpl)^3$ as $\lpl/\beta\to0$, as in \eqref{eq:logkappa2} for $D=4$.

\section{Explicit construction of a genus-zero 3-boundary wormhole core}
\label{app:three-boundary-core}

In this appendix we provide an explicit construction of a genus-zero three-boundary wormhole core $\Sigma_{0,3}$. As explained in Section~\ref{sec:multi-bdry-saddles}, the metric on $\Sigma_{0,3}$ has constant Gaussian curvature $-a^2$, and our goal is to find three boundary circles $C_i$, $i=1,2,3$, with prescribed lengths $b_i$ and constant geodesic curvatures $k_i\equiv\hat k_i$, given by \eqref{eq:bis} and \eqref{eq:ncorek_Ddim}. The normal vectors point outward from the core, so matching with the RN--AdS exteriors requires $k_i<0$.

Curves of constant geodesic curvature $k$ in the hyperbolic plane $\mathbb H^2$ are orbits of one-parameter groups of isometries. They come in three types: equidistant curves, at fixed distance from a geodesic, for $|k|<a$; horocycles for $|k|=a$; and geodesic circles for $|k|>a$. Correspondingly, near $C_i$ the core is a piece of Euclidean AdS$_2$ in one of its three standard slicings,
\begin{equation}\label{eq:core-three-slicings}
\dd s^2=\dd\rho^2+\cosh^2(a\rho)\,\dd u^2,\qquad
\dd s^2=\dd\rho^2+e^{2a\rho}\,\dd u^2,\qquad
\dd s^2=\dd\rho^2+\frac{\sinh^2(a\rho)}{a^2}\,\dd\phi^2,
\end{equation}
namely the double trumpet, the cusp, and the disk. In each case $C_i$ sits at constant $\rho$, with outward normal $-\partial_\rho$. All three cases use the same cut-and-paste construction. We take a hyperbolic polygon, glue two copies of it along three geodesic sides (the seams), and cut out a neighborhood of each of the three remaining ``vertices'' (the other three sides of a right-angled hexagon, or the vertices of a triangle) along a curve of constant curvature. Since the cuts meet the seams at right angles, each cut joins its mirror image into a smooth closed curve. The construction is well defined as long as the cuts fit inside the polygon and do not overlap.

\paragraph{Equidistant boundaries: \texorpdfstring{$|k_i|<a$}{}.}
Let $\mathsf H$ be a right-angled geodesic hexagon whose alternate sides have lengths $L_1/2$, $L_2/2$ and $L_3/2$. Such a hexagon exists for any $L_i>0$ and is unique up to isometry \cite[Ch.~2]{Buser:1992}. Gluing two copies of $\mathsf H$ along the other three sides (the seams) gives the hyperbolic pair of pants $P$ with geodesic boundaries $\gamma_i$ of lengths $L_i$. Near $\gamma_i$ we use Gaussian normal coordinates $(\rho_i, u_i)$, where $\rho_i$ is the proper distance from $\gamma_i$ along the geodesics orthogonal to it and $u_i\sim u_i+L_i$ is the proper length along $\gamma_i$; the hyperbolic metric takes the form
\begin{equation}\label{eq:core-geodesic-metric}
    \dd s^2=\dd\rho_i^2+\cosh^2(a\rho_i)\,\dd u_i^2.
\end{equation}
From each boundary we remove the strip $0\leq\rho_i<s_i$. The new boundary $C_i=\{\rho_i=s_i\}$ is an equidistant curve with outward normal $n_i=-\partial_{\rho_i}$. Its length and geodesic curvature are
\begin{equation}\label{eq:core-geodesic-data}
    b_i=L_i\cosh(as_i),\qquad k_i=-a\tanh(as_i).
\end{equation}
Conversely, prescribed $b_i>0$ and $-a<k_i\leq0$ determine the geodesic lengths and the widths of the removed strips, $L_i=b_i\sqrt{1-k_i^2/a^2}$ and $s_i=\frac1a\operatorname{arctanh}\left(-k_i/a\right)$.

The construction is well defined provided the strips are embedded and mutually disjoint, so it can fail in one of two ways. The strip around $\gamma_i$ can reach its own mirror image across the seam opposite to $\gamma_i$, or it can reach the strip around $\gamma_j$. In $\mathsf H$, the first happens when $s_i$ equals the distance $h_i$ between the side of length $L_i/2$ and the opposite seam. The second happens when $s_i+s_j$ equals the length $d_{ij}$ of the seam joining the sides of lengths $L_i/2$ and $L_j/2$, which is also the distance between $\gamma_i$ and $\gamma_j$ in $P$. Hence the construction is well defined if and only if
\begin{equation}\label{eq:core-geodesic-disjoint}
    s_i<h_i,\qquad s_i+s_j<d_{ij},\qquad \{i,j,k\}=\{1,2,3\}.
\end{equation}
The seam length $d_{ij}$ can be calculated from the cosine rule for right-angled hexagons \cite[Thm.~2.4.1]{Buser:1992}, and $h_i$ from cutting $\mathsf H$ along the common perpendicular of the side $L_i/2$ and the opposite seam into two right-angled pentagons, in which $\cosh(ah_i)=\sinh(ad_{ij})\sinh\frac{aL_j}{2}$ \cite[Ch.~2]{Buser:1992}:
\begin{equation}\label{eq:core-hexagon}
\begin{aligned}
\cosh(a d_{ij})&=\frac{\cosh\frac{aL_k}{2}+\cosh\frac{aL_i}{2}\cosh\frac{aL_j}{2}}{\sinh\frac{aL_i}{2}\,\sinh\frac{aL_j}{2}},\\
\sinh^2(ah_i)\,\sinh^2\frac{aL_i}{2}&=\cosh^2\frac{aL_j}{2}+\cosh^2\frac{aL_k}{2}+2\cosh\frac{aL_i}{2}\cosh\frac{aL_j}{2}\cosh\frac{aL_k}{2}.
\end{aligned}
\end{equation}
Since $\cosh x>1$ for $x>0$, these imply $h_i>w(L_i)$ and $d_{ij}>w(L_i)+w(L_j)$, where
\begin{equation}\label{eq:core-collar-bound}
w(L)=\frac1a\operatorname{arcsinh}\left(\frac1{\sinh(aL/2)}\right).
\end{equation}
The second inequality uses $\cosh\big(a\,w(L_i)+a\,w(L_j)\big)=\big(1+\cosh\frac{aL_i}{2}\cosh\frac{aL_j}{2}\big)/\big(\sinh\frac{aL_i}{2}\sinh\frac{aL_j}{2}\big)$. Thus \eqref{eq:core-geodesic-disjoint} holds whenever $s_i<w(L_i)$ for all $i$. This is the collar lemma \cite{Keen:1974,Buser:1978} for a pair of pants \cite[Prop.~3.1.8]{Buser:1992}.

\paragraph{Horocyclic boundaries: $|k_i|=a$.}
In the upper half-plane $z=x+\i y$, $y>0$, with metric $\dd s^2=\frac{1}{a^2}\frac{\dd z \dd \bar{z}}{|{\rm Im}(z)|^2},$ let $\Delta$ be the ideal triangle with vertices $z=0,1,\infty$, bounded by the geodesics $x=0$, $x=1$ and $|z-\frac12|=\frac12$. The segment of the line $y=1/\nu$ between $x=0$ and $x=1$ meets both sides orthogonally, has length $\nu/a$, and has geodesic curvature $\nabla_\mu n^\mu=-a$ with respect to the upward unit normal $n=ay\,\partial_y$. We cut off the vertex at infinity along this segment with $\nu=\nu_3$, and the vertices $0$ and $1$ along its images with $\nu=\nu_1$ and $\nu=\nu_2$ under the isometries $z\mapsto1/(1-z)$ and $z\mapsto1-1/z$, which map $\Delta$ to itself and $\infty$ to $0$ and $1$, respectively. These images are arcs of the circles $|z-\i\nu_1/2|=\nu_1/2$ and $|z-1-\i\nu_2/2|=\nu_2/2$, so the excised regions are disjoint and do not reach the opposite sides if
\begin{equation}\label{eq:core-horocycle-bound}
    \nu_i<2,\qquad \nu_i\nu_j<1\quad(i\ne j).
\end{equation}
Gluing two copies of the truncated triangle along the remaining parts of its sides gives $\Sigma_{0,3}$ with
\begin{equation}\label{eq:core-horocycle-data}
    b_i=\frac{2\nu_i}{a},\qquad k_i=-a.
\end{equation}
In terms of the boundary lengths, \eqref{eq:core-horocycle-bound} reads $ab_i<4$ and $a^2b_ib_j<4$; in particular, $ab_i<2$ is sufficient.

\paragraph{Circular boundaries: $|k_i|>a$.}
Consider a hyperbolic triangle $\Delta$ with geodesic edges and
interior angles $A_i$ satisfying
\begin{equation}\label{eq:core-triangle-angles}
    A_i>0,\qquad A_1+A_2+A_3<\pi.
\end{equation}
Gluing two identical copies along the corresponding edges gives a
topological sphere $\hat\Sigma_{0,0}$ with three cone points $p_i$.
This is the sphere with three conical defects familiar from JT gravity \cite{Mertens:2019tcm,Maxfield:2020ale,Witten:2020wvy}, with cone angle $2A_i$ at $p_i$.
Locally around $p_i$, its metric is
\begin{equation}\label{eq:core-cone-metric}
    \dd s^2=\dd r_i^2+
    \frac{\sinh^2(ar_i)}{a^2}\,\dd\phi_i^2,
    \qquad \phi_i\sim\phi_i+2A_i.
\end{equation}
Remove the circular neighborhoods $D_i(s_i)=\{0\leq r_i<s_i\}$.
The remaining surface
\begin{equation}
    \Sigma_{0,3}=\hat\Sigma_{0,0}\backslash\bigcup_{i=1}^3D_i(s_i)
\end{equation}
has outward normals $n_i=-\partial_{r_i}$ and boundary data
\begin{equation}\label{eq:core-circular-data}
    b_i=\frac{2A_i}{a}\sinh(as_i),\qquad
    k_i=-a\coth(as_i)<-a.
\end{equation}
The prescribed lengths and curvatures therefore determine the lengths $s_i$ and angles $A_i$.

Let $c_i$ be the side length opposite vertex $i$. The hyperbolic
cosine law gives
\begin{equation}\label{eq:core-cosine-law}
    \cosh(ac_i)=
    \frac{\cos A_i+\cos A_j\cos A_k}{\sin A_j\sin A_k},
    \qquad \{i,j,k\}=\{1,2,3\}.
\end{equation}
Sufficient conditions for the excised neighborhoods to be embedded
and mutually disjoint are
\begin{equation}\label{eq:core-circular-disjoint}
    s_i<h_i,\qquad s_i+s_j<c_k,
    \qquad \{i,j,k\}=\{1,2,3\},
\end{equation}
where $h_i$ is the distance from vertex $i$ to the complete geodesic
containing the opposite side. Explicitly,
\begin{equation}\label{eq:core-altitudes}
    \sinh(ah_i)=\sinh(ac_j)\sin A_k
              =\sinh(ac_k)\sin A_j.
\end{equation}
The first inequality in \eqref{eq:core-circular-disjoint} keeps each cut within its triangle sector, while the second keeps different excisions disjoint.

\paragraph{Existence conditions and relation to the RN--AdS exteriors.}
For all three constructions, the Gauss--Bonnet theorem for $\Sigma_{0,3}$, whose Euler characteristic is $-1$, gives
\begin{equation}\label{eq:core-GB}
-a^2 A+\sum_{i=1}^3 k_i b_i=-2\pi,
\qquad A=\int_{\Sigma_{0,3}}\sqrt g\,\dd^2x.
\end{equation}
For $k_i=0$, this is the familiar area $2\pi/a^2$ of the pair of pants in JT gravity. Positive area therefore requires
\begin{equation}\label{eq:core-positive-area}
2\pi+\sum_i k_i b_i>0.
\end{equation}
This condition is necessary in every regime, but it is not sufficient. The embedding conditions \eqref{eq:core-geodesic-disjoint}, \eqref{eq:core-horocycle-bound} and \eqref{eq:core-circular-disjoint} are sufficient for the respective constructions; in the equidistant regime, \eqref{eq:core-geodesic-disjoint} is also necessary.

The gluing conditions fix the boundary data in terms of the RN--AdS
exteriors,
\begin{equation}\label{eq:glued_surface_params}
    b_i=\beta_i\sqrt{f_i(R_0)},\qquad
    k_i=-\frac{D-1}{2R_0}\sqrt{f_i(R_0)},\qquad
    a^2=\frac{(D-3)^2}{R_0^2}
          +\frac{(D-1)(D-2)}{\ell^2}.
\end{equation}
It follows that the equidistant, horocyclic, and circular constructions apply, respectively, when all three $f_i(R_0)$ lie below, at, or above $\frac{4a^2 R_0^2}{(D-1)^2}$, subject to their corresponding existence conditions.\footnote{Mixed cases, with the $f_i(R_0)$ on different sides of $\frac{4a^2 R_0^2}{(D-1)^2}$, can be built in the same way from a polygon with the corresponding mixture of vertex types; however, for brevity, we do not address these cases.} In every case,
\eqref{eq:core-positive-area} reads
\begin{equation}\label{positive_area_constraint_v2}
    \sum_{i=1}^3\beta_i f_i(R_0)<\frac{4\pi R_0}{D-1}.
\end{equation}
In particular, the construction with equidistant boundaries covers the near-extremal regime. Writing $\epsilon_i=f_i(R_0)$, the boundary data $b_i=\beta_i\sqrt{\epsilon_i}$ and $|k_i|=\frac{D-1}{2R_0}\sqrt{\epsilon_i}$ both vanish as $\epsilon_i\to0^+$. Hence, for fixed $\beta_i$, $|k_i| < a$ and $s_i \to 0$ while $w(L)\to \infty$ (see Eq.~\eqref{eq:core-collar-bound}), so $s_i < w(L_i)$ and the existence conditions \eqref{eq:core-geodesic-disjoint} are satisfied. Therefore, three-boundary wormhole cores exist in the near-extremal limit.

\section{Axio-dilaton wormholes}\label{app:timelike-reduction}
In this appendix, we relate the constrained two-boundary wormholes of Section~\ref{sec:2bdry-saddles} to the axio-dilaton wormholes of \cite{Maldacena:2026jqd}. We reduce the Einstein--Maxwell theory on the Euclidean time circle and show that, near extremality, the neck of the constrained wormhole approaches the axio-dilaton wormhole, whose uplift is an $\mathrm{AdS}_2\times S^{D-2}$ throat. At fixed boundary temperatures this limit is degenerate; a nondegenerate throat, with vanishing constraint-brane tension, also requires $\beta_L,\beta_R\to\infty$. We focus on the asymptotically flat case, $\ell\to\infty$, and use subscripts $D$ and $D-1$ for the parent and the reduced metrics.

\paragraph{Dimensional reduction of Einstein--Maxwell theory.}
We reduce the Einstein--Maxwell theory \eqref{eq:os-EM-action} on the circle with the ansatz
\begin{equation}\label{eq:timelike-ansatz}
    \dd s_D^2=e^{-2\nu/(D-3)}\dd s_{D-1}^2+e^{2\nu}\dd\tau^2,
    \qquad
    A=\frac{\hat\kappa\tilde\alpha}{4\sqrt{\pi G_D}}\dd\tau,
    \qquad
    \hat\kappa^2=\frac{2(D-2)}{D-3},
\end{equation}
with $\tau \sim \tau + 1.$ The bulk action of the corresponding $(D-1)$-dimensional theory is\footnote{With nonzero cosmological constant $\Lambda$, the reduced bulk action also contains the term $2\Lambda e^{-2\nu/(D-3)}$.}
\begin{equation}\label{eq:timelike-bulk}
    I_{D-1,\mathrm{bulk}}
    =\frac{1}{16\pi G_D}\int\dd^{D-1}x\sqrt{g_{D-1}}
    \left[-R_{D-1}+\frac{\hat\kappa^2}{2}
    \left((\nabla\nu)^2+e^{-2\nu}(\nabla\tilde\alpha)^2\right)\right].
\end{equation}
We call $\nu$ the dilaton, since it measures the size of the thermal circle, and $\tilde\alpha$ the axion. The two scalars are coordinates on a moduli space with metric $\dd s_{\mathcal M}^2=\hat\kappa^2(\dd\nu^2+e^{-2\nu}\dd\tilde\alpha^2)$. On our saddles the Euclidean electric field is imaginary, so $\i\tilde\alpha$ is real, and along such configurations the moduli-space metric becomes $\hat\kappa^2\bigl(\dd\nu^2-e^{-2\nu}\dd(\i\tilde\alpha)^2\bigr)$, i.e.\ two-dimensional Lorentzian AdS.

Upon dimensional reduction, Dirichlet boundary conditions for the metric become Dirichlet boundary conditions for the dilaton and the $(D-1)$-dimensional metric. Fixed electric flux corresponds to Neumann boundary conditions for the axion. The boundary terms in \eqref{eq:os-EM-action} reduce accordingly. The GHY term becomes the $(D-1)$-dimensional GHY term plus a term proportional to $n\cdot\nabla\nu$. That extra term cancels the boundary contribution of the total derivative $\propto\nabla^2\nu$ generated by the reduction of $R_D$. The Maxwell boundary term \eqref{eq:INeumann} becomes
\begin{equation}
I_N=-\frac{\hat\kappa^2}{16\pi G_D}\int_{\partial M}\dd^{D-2}x\sqrt{h_{D-1}}\,e^{-2\nu}\,\tilde\alpha\,n\cdot\nabla\tilde\alpha,
\end{equation}
which is precisely the boundary term that turns the Dirichlet problem for $\tilde\alpha$ into the Neumann problem at fixed axion charge.

\paragraph{Constraint brane and junction conditions.}
Let us assume that the constraint brane wraps the Kaluza--Klein (KK) circle, so that the constrained surface in the lower-dimensional theory is $\cS/S^1$. If $\sigma_{D-1}$ is the metric induced on $\cS/S^1$, the brane action becomes
\begin{equation}\label{eq:timelike-constraint}
    \lambda\bigl(\operatorname{vol}(\cS)-\zeta\bigr)
    =\lambda\left[
    \int_{\cS/S^1}\dd^{D-2}x\sqrt{\sigma_{D-1}}\,
             e^{-\nu/(D-3)}-\zeta\right].
\end{equation}
We see that the reduced brane sources both the metric and the dilaton, but carries no axion charge. Let $n_L,n_R$ be the unit normals in the KK-reduced theory. Then the Israel junction conditions of the $D$-dimensional theory become
\begin{align}
    \sum_{i=L,R}\left(K^{(D-1)}_{i,ab}
                   -K^{(D-1)}_i(\sigma_{D-1})_{ab}\right)
    &=8\pi G_D\lambda e^{-\nu/(D-3)}(\sigma_{D-1})_{ab},
    \label{eq:timelike-metric-junction}\\
    \sum_{i=L,R}n_i\cdot\nabla\nu_i
    &=-\frac{8\pi G_D\lambda}{D-2}e^{-\nu/(D-3)}.
    \label{eq:timelike-scalar-junction}
\end{align}
The electric flux conservation condition at the constraint surface reduces to
\begin{equation}
    \sum_{i=L,R}e^{-2\nu} n_i\cdot\nabla\tilde\alpha_i=0.\label{eq:axion-flux-continuity}
\end{equation}
All fields on the right-hand sides of equations \eqref{eq:timelike-metric-junction}, \eqref{eq:timelike-scalar-junction} and \eqref{eq:axion-flux-continuity} are evaluated at the junction.

\paragraph{The reduced constrained saddles.}
The two RN exteriors reduce to
\begin{equation}\label{eq:timelike-metric}
    \dd s_{D-1,(i)}^2
    =(\beta_i^2f_i(r))^{1/(D-3)}
      \left(\frac{\dd r^2}{f_i(r)}+r^2\dd\Omega_{D-2}^2\right),
    \qquad e^{\nu_i}=\beta_i\sqrt{f_i(r)}.
\end{equation}
With \eqref{eq:timelike-metric}, the unit normals are $n_i=e^{-\nu/(D-3)}\sqrt{f_i(r)}\,\partial_r$, and the condition \eqref{eq:timelike-scalar-junction} reduces to the parent junction condition \eqref{eq:2bdry_wormhole_masses_eq} upon using the value of $\lambda$ found in Section~\ref{sec:2bdry-saddles}.
Upon fixing the common axion value at $r=r_0$ to be zero, dimensional reduction of the ${\rm U}(1)$ gauge field gives
\begin{equation}\label{eq:timelike-axions}
    \i\tilde\alpha_L=\beta_L\mathcal Q
       \left(\frac{1}{r^{D-3}}-\frac{1}{r_0^{D-3}}\right),
    \qquad
    \i\tilde\alpha_R=-\beta_R\mathcal Q
       \left(\frac{1}{r^{D-3}}-\frac{1}{r_0^{D-3}}\right).
\end{equation}
The opposite signs reflect the orientation of the flux through the wormhole.

On each exterior, the scalars move along a geodesic of the moduli space. Indeed, on the background \eqref{eq:timelike-metric}, the radial scalar action is proportional to
\begin{equation}\label{eq:timelike-geodesic-action}
    \int\dd r\,r^{D-2}f_i(r)
       \left[(\partial_r\nu_i)^2
                  +e^{-2\nu_i}(\partial_r\tilde\alpha_i)^2\right].
\end{equation}
In terms of an affine parameter $s$ with $\dd s\propto\dd r/[r^{D-2}f_i(r)]$, and with dots denoting $\dd/\dd s$, it becomes proportional to $\int\dd s\,(\dot\nu_i^2+e^{-2\nu_i}\dot{\tilde\alpha}_i^2)$, the action of a free particle on AdS$_{2}$. Its equations of motion are the geodesic equations
\begin{equation}\label{eq:timelike-geodesic-equations}
    \ddot\nu_i+e^{-2\nu_i}\dot{\tilde\alpha}_i^{\,2}=0,
    \qquad
    \ddot{\tilde\alpha}_i-2\dot\nu_i\dot{\tilde\alpha}_i=0.
\end{equation}
The full constrained solution is therefore piecewise geodesic: at the brane, the dilaton momentum changes according to \eqref{eq:timelike-scalar-junction}, while the axion momentum remains continuous.

\paragraph{Extremal limit and the \texorpdfstring{AdS$_2\times S^{D-2}$}{AdS2 x S(D-2)} throat.} With $\ell\to\infty$, the gluing conditions of Section~\ref{sec:2bdry-saddles} give
\begin{equation}\label{eq:timelike-neck}
    \begin{gathered}
    e^{2\nu(r_0)}=\beta_L\beta_R\frac{D-3}{D-1}
       \left(1-\frac{\mathcal Q^2}{r_0^{2(D-3)}}\right),\\
    \lambda=-\frac{D-2}{8\pi G_D\,r_0}e^{\nu(r_0)}
       \left(\frac1{\beta_L}+\frac1{\beta_R}\right),
    \qquad
    \zeta=\Omega_{D-2}r_0^{D-2}e^{\nu(r_0)}.
    \end{gathered}
\end{equation}
At fixed positive $\beta_L,\beta_R$, taking $r_0\to R_0$, where $R_0^{D-3}=\mathcal Q$, sends $\lambda,\zeta,e^{\nu(r_0)}$ to zero. Thus, the junction surface shrinks to zero, and the scalar fields $\tilde\alpha,\nu$ obey $\i\tilde\alpha_L=-e^{\nu_L}$ and $\i\tilde\alpha_R=e^{\nu_R}$, so their trajectories become null.

To obtain a regular wormhole throat in the limit $r_0 \to R_0$, we have to simultaneously take $\beta_L,\beta_R\to \infty$. To see this, let us introduce coordinate $\rho$ via
\begin{equation}\label{eq:timelike-neck-coordinate}
    r^{D-3}=
    \begin{cases}
    r_0^{D-3}\left(1-\dfrac{e^{\nu(r_0)}}{\beta_L}\sinh\rho\right),
    &\rho\leq0,\\[2mm]
    r_0^{D-3}\left(1+\dfrac{e^{\nu(r_0)}}{\beta_R}\sinh\rho\right),    &\rho\geq0.
    \end{cases}
\end{equation}
Substituting in the metric and scalars, and taking $r_0\to R_0$ at fixed $\rho$, gives
\begin{align}
    \frac{\dd s_{D-1}^2}{r_0^2e^{2\nu(r_0)/(D-3)}}
    &\longrightarrow
    \cosh^{2/(D-3)}\rho
       \left(\frac{\dd\rho^2}{(D-3)^2}+\dd\Omega_{D-2}^2\right),
    \label{eq:timelike-throat-metric}\\
    \frac{e^\nu}{e^{\nu(r_0)}}&\longrightarrow\cosh\rho,
    \qquad
    \frac{\i\tilde\alpha}{e^{\nu(r_0)}}\longrightarrow\sinh\rho.
    \label{eq:timelike-throat-scalars}
\end{align}
The scalar rescaling in these formulas is the moduli-space isometry $\nu\mapsto\nu-\nu(r_0)$, $\tilde\alpha\mapsto e^{-\nu(r_0)}\tilde\alpha$. These are the axio-dilaton wormhole profiles of \cite{Maldacena:2026jqd}. In terms of $\tilde s$ defined by $\sinh\rho=\tan\tilde s$, they read $e^\nu=e^{\nu(r_0)}/\cos\tilde s$ and $\i\tilde\alpha=e^{\nu(r_0)}\tan\tilde s$. This is a timelike geodesic of the moduli metric with proper time $s=\hat\kappa\tilde s$, along which the dilaton diverges at both ends, $\tilde s\to\pm\pi/2$. Their uplift is
\begin{equation}\label{eq:timelike-throat-uplift}
    \dd s_D^2\simeq
    \frac{R_0^2}{(D-3)^2}\dd\rho^2
      +e^{2\nu(r_0)}\cosh^2\rho\,\dd\tau^2
      +R_0^2\dd\Omega_{D-2}^2.
\end{equation}
After rescaling the time coordinate by $\hat\tau=(D-3)e^{\nu(r_0)}\tau/R_0$, this is locally $\mathrm{AdS}_2\times S^{D-2}$, with $\mathrm{AdS}_2$ radius $R_0/(D-3)$.

At fixed $\beta_{L,R}$, however, the period $(D-3)e^{\nu(r_0)}/R_0$ of $\hat\tau$ also shrinks to zero. A nondegenerate throat is obtained by simultaneously taking $\beta_L,\beta_R\to\infty$, with fixed ratio and fixed $e^{\nu(r_0)}$, so that $r_0\to R_0$ by \eqref{eq:timelike-neck}. Then $\lambda\to0$, while the size of the throat and $\zeta$ stay finite, and the boundary values of the dilaton, $e^{\nu}=\beta_{L,R}$, diverge, as in \cite{Maldacena:2026jqd}. The unconstrained $\mathrm{AdS}_2\times S^{D-2}$ wormhole arises as a near-horizon scaling limit of our constrained wormholes.

\section{Minisuperspace quantization of two-boundary wormholes}\label{app:minisuperspace}
In this appendix, we perform a minisuperspace approximation of the Einstein--Maxwell theory by restricting it to static, spherically symmetric configurations. The theory then reduces to a worldline theory with a three-dimensional target space, whose coordinates are determined by the sizes of the $(D-2)$-sphere and of the thermal circle, and by the holonomy of the gauge field around the thermal circle, that is, by the dilaton and the axion of Appendix~\ref{app:timelike-reduction}. We define the path integral of this worldline theory, fixing its integration contour with the help of the Lorentzian path integral of the $D$-dimensional theory, and discuss the area constraint that leads to the constrained two-boundary wormholes of Section~\ref{sec:2bdry-saddles}. As in Appendix~\ref{app:timelike-reduction}, we focus on the asymptotically flat case. 

\paragraph{The minisuperspace approximation.}
For now, we consider Dirichlet boundary conditions for the gauge field, and hence for $\tilde\alpha$; we turn to Neumann boundary conditions later in this appendix. Consider Einstein--Maxwell configurations of the form
\begin{equation}\label{eq:mini-ansatz}
    \dd s_D^2=N(x)^2\dd x^2+e^{2\nu(x)}\dd\tau^2+r(x)^2\dd\Omega_{D-2}^2,\qquad A=\sqrt{\frac{D-2}{8\pi(D-3)G_D}}\,\tilde\alpha(x)\,\dd\tau,
\end{equation}
with $\tau\sim\tau+1$. In terms of the variables
\begin{equation}\label{eq:mini-variables}
    R=r^{D-3}e^{\nu},\qquad \CN=N e^{\nu}r^{D-4},
\end{equation}
the bulk part of the Einstein--Maxwell action, together with the GHY term, is
\begin{equation}\label{eq:mini-action}
    I=\frac{D-2}{D-3}\frac{\Omega_{D-2}}{16\pi G_D}\int\dd x\left[\frac{1}{\CN}\left(-\dot R^2+R^2\left(\dot\nu^2+e^{-2\nu}\dot{\tilde\alpha}^2\right)\right)-(D-3)^2\CN\right],
\end{equation}
where $\dot f=\dd f/\dd x$. This action describes a relativistic particle moving in a three-dimensional space with coordinates $X^A=(R,\nu,\tilde\alpha)$ and metric
\begin{equation}\label{eq:mini-metric}
    G_{AB}\dd X^A\dd X^B=-\dd R^2+R^2\left(\dd\nu^2+e^{-2\nu}\dd\tilde\alpha^2\right),
\end{equation}
where $\CN\dd x$ is an einbein on the worldline. The metric \eqref{eq:mini-metric} is flat: in the coordinates
\begin{equation}\label{eq:mini-cartesian}
    X^0=\frac{R}{2}e^{-\nu}\left(1+e^{2\nu}+\tilde\alpha^2\right),\qquad X^1=Re^{-\nu}\tilde\alpha,\qquad X^2=\frac{R}{2}e^{-\nu}\left(1-e^{2\nu}-\tilde\alpha^2\right),
\end{equation}
it becomes $G_{AB}=\eta_{AB}$, where $\eta_{AB}=\mathrm{diag}(-1,1,1)$. The equations of motion for $X^A$ and $\CN$ are
\begin{equation}\label{eq:mini-eom}
    \frac{\dd \pi_A}{\dd x}=0,\qquad \eta^{AB}\pi_A\pi_B=-\left(\frac{(D-2)\,\Omega_{D-2}}{8\pi G_D}\right)^2,
\end{equation}
where $\pi_A=\frac{D-2}{D-3}\frac{\Omega_{D-2}}{8\pi G_D}\frac{1}{\CN}\,G_{AB}\frac{\dd X^B}{\dd x}$ are the momenta conjugate to $X^A$. 
Clearly, the on-shell trajectories are timelike straight lines. One can readily verify that these straight lines correspond precisely to the Reissner--Nordstr\"om exteriors. In particular, $\tilde\alpha$ is a cyclic coordinate, so $\pi_{\tilde\alpha}$ is conserved, which corresponds to electric flux conservation in the higher-dimensional theory.

The minisuperspace approximation naturally describes two-boundary geometries, with the two ends of the worldline corresponding to the two boundaries. With Dirichlet boundary conditions $X'$ and $X''$ fixing the radius of the sphere, the length of the thermal circle and the value of the gauge field at the two boundaries, the minisuperspace path integral is
\begin{equation}\label{eq:mini-PI}
    G(X'',X')=\int_{X'}^{X''}\mathcal DX\,\mathcal D\CN\,e^{-I(X,\CN)},
\end{equation}
with $I(X,\CN)$ given by \eqref{eq:mini-action}. To define $G$ properly, we need to specify the integration contour for $X^A$ and $\CN$. Note that the real contour for the fluctuations of $X^0$ does not give a well-defined path integral: the integral over $X^0$ diverges. From the perspective of the $D$-dimensional theory, this divergence stems from the conformal mode problem \cite{Gibbons:1978ac,Mottola:1995sj,Liu:2023jvm}.

To find an appropriate contour, we need to go back to the Lorentzian path integral of the underlying $D$-dimensional theory. The Wick rotation of the thermal circle relates the Euclidean (E) and Lorentzian (L) variables as
\begin{equation}
    \nu_{\rm E}=\nu_{\rm L}+\frac{\i\pi}{2},\qquad R_{\rm E}=\i R_{\rm L},\qquad \mathcal N_{\rm E}=\i\mathcal N_{\rm L},\qquad N_{\rm E}=N_{\rm L},\qquad \tilde\alpha_{\rm E}=\tilde\alpha_{\rm L}.
    \label{ga:thermal-rotation}
\end{equation}
In terms of the Lorentzian variables, the Euclidean action \eqref{eq:mini-action} takes the form $I=-\i S$, where $S$ is the Lorentzian bulk Einstein--Maxwell action of the $D$-dimensional theory, together with the GHY term. Omitting the subscript L, it reads
\begin{equation}\label{eq:mini-lorentzian-action}
    S=\frac{D-2}{D-3}\frac{\Omega_{D-2}}{16\pi G_D}\int\dd x\left[\frac{1}{\CN}\left(\dot R^2-R^2\dot\nu^2+R^2e^{-2\nu}\dot{\tilde\alpha}^2\right)+(D-3)^2\CN\right].
\end{equation}
In the first-order formalism,
\begin{equation}\label{eq:mini-first-order}
    S=\int\dd x\,\bigl(p_A\dot X^A-\CN H\bigr),
\end{equation}
where $p_A$ are the momenta conjugate to $X^A$, and
\begin{equation}\label{eq:mini-H}
    H=\frac{D-3}{D-2}\frac{4\pi G_D}{\Omega_{D-2}}\left[p_R^2-\frac{1}{R^2}\,p_\nu^2+\frac{1}{R^2}\,e^{2\nu}p_{\tilde\alpha}^2-\left(\frac{(D-2)\,\Omega_{D-2}}{8\pi G_D}\right)^2\right].
\end{equation}
We can gauge-fix the reparametrization symmetry by setting $x\in(0,1)$ and $\CN=\mathrm{const}$.

\paragraph{Group-averaged propagator.}
We define
\begin{align}
    G(X'',X')&=\int_{X'}^{X''}\frac{\mathcal DX\,\mathcal Dp\,\mathcal D\CN}{\mathrm{Diff}_0}\,\exp\left[\i\int_0^1\dd x\,\bigl(p_A\dot X^A-\CN H\bigr)\right]\nonumber\\
    &=\lim_{\epsilon\to0^+}\int_{-\infty}^{\infty}\dd T\,e^{-\epsilon|T|}\,K_T(X'',X').\label{eq:mini-group-averaged}
\end{align}
Here $\mathrm{Diff}_0$ denotes the group of worldline reparametrizations. The second line is obtained by gauge-fixing $x\in(0,1)$ and $\CN=\mathrm{const}$, with $T=\int_0^1 \dd x\,\CN$. Note that in the standard formulation of worldline theory, one integrates only over positive proper times $T$. For the contour of the gravitational path integral, however, $T$ can also be negative. This implements the so-called group-averaging quantization procedure of \cite{Chen:2026mar,Held:2025mai}.

The kernel $K_T(X'',X')$ is
\begin{align}
    K_T(X'',X')&=\int_{X(0)=X'}^{X(1)=X''}\mathcal DX\,\mathcal Dp\,\exp\left[\i\int_0^1\dd x\,\bigl(p_A\dot X^A-TH\bigr)\right]\nonumber\\
    &=\int_{X'}^{X''}\mathcal DX\,\exp\left\{\i\,\tfrac{D-2}{D-3}\tfrac{\Omega_{D-2}}{16\pi G_D}\int_0^1\dd x\left[\frac{1}{T}\left(\dot R^2-R^2\dot\nu^2+R^2e^{-2\nu}\dot{\tilde\alpha}^2\right)+(D-3)^2T\right]\right\},\label{eq:mini-kernel}
\end{align}
where the configuration space path integral measure is\footnote{The minisuperspace metric in \eqref{eq:mini-kernel}, $G_{AB}\dd X^A\dd X^B=\dd R^2-R^2\dd\nu^2+R^2e^{-2\nu}\dd\tilde\alpha^2$, is flat: in the coordinates $Y^0=R\cosh\nu-\frac12\tilde\alpha^2Re^{-\nu}$, $Y^1=R\sinh\nu-\frac12\tilde\alpha^2Re^{-\nu}$ and $Y^2=\tilde\alpha Re^{-\nu}$, it becomes $(\dd Y^0)^2-(\dd Y^1)^2+(\dd Y^2)^2$. Hence, \eqref{eq:mini-kernel} describes a free relativistic particle in three-dimensional Minkowski space, and \eqref{eq:mini-measure} is its flat measure $\prod_x\dd^3Y(x)$ written in the coordinates $(R,\nu,\tilde\alpha)$, since $\dd^3Y=\sqrt{-\det G_{AB}}\,\dd R\,\dd\nu\,\dd\tilde\alpha$. More generally, the measure induced by the configuration-space metric, $\sqrt{|\det G_{AB}|}\prod_A\dd X^A$, is the standard choice for path integrals on curved configuration spaces \cite{DeWitt:1957at} and in minisuperspace models \cite{Halliwell:1988wc}. The part of $G_{AB}$ corresponding to gravitational degrees of freedom is the DeWitt supermetric \cite{DeWitt:1967yk} restricted to the minisuperspace ansatz.}
\begin{equation}\label{eq:mini-measure}
    \mathcal DX=\prod_{x\in(0,1)}\dd\mu\bigl(X(x)\bigr),\qquad \dd\mu(X(x))=R(x)^2e^{-\nu(x)}\,\dd R(x)\,\dd\nu(x)\,\dd\tilde\alpha(x).
\end{equation}

Up to the standard ordering ambiguity, we can define the quantum Hamiltonian operator as
\begin{align}
    \widehat H={}&-\frac{4\pi G_D(D-3)}{(D-2)\,\Omega_{D-2}}\left[\partial_R^2+\frac{2}{R}\,\partial_R+\frac{1}{R^2}\left(-\partial_\nu^2+\partial_\nu+e^{2\nu}\partial_{\tilde\alpha}^2\right)\right]\nonumber\\
    &-\frac{(D-2)(D-3)\,\Omega_{D-2}}{16\pi G_D}.\label{eq:mini-Hhat}
\end{align}
Then
\begin{equation}\label{eq:mini-kernel-operator}
    K_T(X'',X')=\langle X''|e^{-\i T\widehat H}|X'\rangle,
\end{equation}
and, as in \cite{Chen:2026mar,Held:2025mai,Halliwell:1988wc}, the integral over $T$ implements the Hamiltonian constraint,
\begin{equation}\label{eq:mini-delta-H}
    G(X'',X')=2\pi\,\langle X''|\delta(\widehat H)|X'\rangle.
\end{equation}

\paragraph{Neumann boundary conditions for the axion.}
Evaluating the Maxwell boundary term $I_N$ on the Euclidean minisuperspace ansatz \eqref{eq:mini-ansatz} gives
\begin{equation}\label{eq:mini-IN}
    I_N=-\int_{\partial M}\dd^{D-1}x\sqrt h\,n_\mu A_\nu F^{\mu\nu}=-\bigl[\pi_{\tilde\alpha,\rm E}\,\tilde\alpha_{\rm E}\bigr]_{X'}^{X''}.
\end{equation}
Upon the Wick rotation \eqref{ga:thermal-rotation}, $\pi_{\tilde\alpha,\rm E}=-\i p_{\tilde\alpha,\rm L}$, so that $e^{-I_N}=e^{-\i\left(p''_{\tilde\alpha}\tilde\alpha''-p'_{\tilde\alpha}\tilde\alpha'\right)}.$ Its insertion into the path integral gives
\begin{align}
    K_T(R'',\nu'',p''_{\tilde\alpha};R',\nu',p'_{\tilde\alpha})&=\int\frac{\dd\tilde\alpha''\,\dd\tilde\alpha'}{2\pi}\,e^{-\i\left(p''_{\tilde\alpha}\tilde\alpha''-p'_{\tilde\alpha}\tilde\alpha'\right)}\,K_T(R'',\nu'',\tilde\alpha'';R',\nu',\tilde\alpha')\nonumber\\
    &=\delta(p''_{\tilde\alpha}-p'_{\tilde\alpha})\,K_{T,Q}(R'',\nu'';R',\nu'),\label{eq:mini-fixed-charge-kernel}
\end{align}
where $Q$ is the electric flux of the $D$-dimensional theory through the sphere,
\begin{equation}\label{eq:mini-p-alpha}
    Q=\int_{S^{D-2}}\star F=\frac{\Omega_{D-2}\,r^{D-2}}{N e^{\nu}}\,\dot A_\tau=\Omega_{D-2}\sqrt{\frac{D-2}{8\pi(D-3)G_D}}\,\frac{R^2e^{-2\nu}\dot{\tilde\alpha}}{\CN}=\sqrt{\frac{8\pi(D-3)G_D}{D-2}}\,p_{\tilde\alpha},
\end{equation}
which fixes the momentum to $p''_{\tilde\alpha}=p'_{\tilde\alpha}=Q\sqrt{\frac{D-2}{8\pi(D-3)G_D}}$, and
\begin{equation}\label{eq:mini-KQ}
    K_{T,Q}(R'',\nu'';R',\nu')=\langle R'',\nu''|e^{-\i T\widehat H_Q}|R',\nu'\rangle,
\end{equation}
with
\begin{align}
    \widehat H_Q={}&-\frac{4\pi G_D(D-3)}{(D-2)\,\Omega_{D-2}}\left[\partial_R^2+\frac{2}{R}\,\partial_R+\frac{1}{R^2}\left(-\partial_\nu^2+\partial_\nu-\frac{(D-2)\,Q^2}{8\pi(D-3)\,G_D}\,e^{2\nu}\right)\right]\nonumber\\
    &-\frac{(D-2)(D-3)\,\Omega_{D-2}}{16\pi G_D}.\label{eq:mini-HQ}
\end{align}

\paragraph{Constraint branes.}
Consider now inserting the constraint
\begin{equation}\label{eq:mini-constraint}
    \delta(\cA-\zeta)=\int_{\mathbb R}\frac{\dd\lambda}{2\pi}\,e^{-\i\lambda(\cA-\zeta)},\qquad \cA=\Omega_{D-2}\,r^{D-2}(x_c)\,e^{\nu(x_c)},
\end{equation}
into the Lorentzian path integral \eqref{eq:mini-group-averaged}. In the $D$-dimensional theory, this corresponds precisely to the insertion of a codimension-one constraint brane at $x=x_c\in(0,1)$. It modifies the Lorentzian action by
\begin{equation}\label{eq:mini-constraint-action}
    S_\cC=-\lambda(\cA-\zeta).
\end{equation}
The corresponding constrained kernel is
\begin{align}
    K_{\zeta,x_c,T}(X'',X')&=\int_{X'}^{X''}\mathcal DX\,\mathcal Dp\,e^{\i\int_0^1\dd x\,(p_A\dot X^A-TH)}\,\delta\bigl(\cA(X(x_c))-\zeta\bigr)\nonumber\\
    &=\int\dd\mu(X_c)\,K_{(1-x_c)T}(X'',X_c)\,\delta\bigl(\cA(X_c)-\zeta\bigr)\,K_{x_cT}(X_c,X'),\label{eq:mini-constrained-kernel}
\end{align}
with $\dd\mu(X_c)$ given by \eqref{eq:mini-measure}. We have
\begin{equation}\label{eq:mini-composition}
    \int_{\mathbb R}\dd\zeta\,K_{\zeta,x_c,T}(X'',X')=\int\dd\mu(X_c)\,K_{(1-x_c)T}(X'',X_c)\,K_{x_cT}(X_c,X')=K_T(X'',X').
\end{equation}

The constraint does not depend on $\tilde\alpha$, so we can similarly define the constrained kernel with fixed momentum $p_{\tilde\alpha}$,
\begin{align}
    K_{\zeta,x_c,T,Q}(R'',\nu'';R',\nu')=\int\dd\mu(R_c,\nu_c)\,&K_{(1-x_c)T,Q}(R'',\nu'';R_c,\nu_c)\nonumber\\
    &\times\delta\bigl(\cA(R_c,\nu_c)-\zeta\bigr)\,K_{x_cT,Q}(R_c,\nu_c;R',\nu').\label{eq:mini-constrained-kernel-Q}
\end{align}

\paragraph{Wick rotation of the constraint.}
As explained in Section~\ref{sec:constr-instantons}, the constrained-instanton method does not change the path integral. Inserting $1=\int\dd\zeta\,\delta(\cA-\zeta)$, with \eqref{eq:mini-constraint}, only postpones the integral over one of its modes, the area of the brane, to the end. Therefore, the integration contour of $\zeta$ can be rotated like that of any other mode. By \eqref{ga:thermal-rotation}, the Euclidean area of the brane is $\cA_{\rm E}=\i\cA$, which is real for the constrained wormholes of Section~\ref{sec:2bdry-saddles}. Thus, to go to Euclidean theory, we rotate the contour of $\zeta$ so that $\zeta_{\rm E}=\i\zeta$ is real, and, correspondingly, rotate the contour of $\lambda$ to the imaginary axis, so that
\begin{equation}\label{eq:mini-constraint-rotation}
    \int_{\mathbb R}\dd\zeta\int_{\mathbb R}\frac{\dd\lambda}{2\pi}\,e^{\i S-\i\lambda(\cA-\zeta)}
    =\int_{\mathbb R}\dd\zeta_{\rm E}\int_{c + \i\mathbb R}\frac{\dd\lambda_{\rm E}}{2\pi\i}\,e^{-I-\lambda_{\rm E}(\cA_{\rm E}-\zeta_{\rm E})}.
\end{equation}

\paragraph{Constrained saddles.}
We can now discuss the saddles of the Euclidean constrained path integral. We will work in the ensemble with fixed momentum $p_{\tilde\alpha}$, related to the $D$-dimensional electric flux by \eqref{eq:mini-p-alpha}.  Away from the constraint brane, the saddles satisfy \eqref{eq:mini-eom}, so they are geodesics in the target space. Varying the modes localized on the brane, $X(x_c)$, gives the junction conditions
\begin{equation}\label{eq:mini-junction}
    [\pi_A]=\lambda\,\partial_A\cA,\qquad \cA=\zeta,
\end{equation}
which determine the jump $[\pi_A]=\lim_{x\to x_c^+}\pi_A-\lim_{x\to x_c^-}\pi_A$ of the momenta across the brane. Here, $\cA$, $\zeta$ and $\lambda$ stand for $\cA_{\rm E}$, $\zeta_{\rm E}$ and $\lambda_{\rm E}$. Since $\cA=\Omega_{D-2}R^{\frac{D-2}{D-3}}e^{-\frac{\nu}{D-3}}$, the junction conditions give
\begin{equation}\label{eq:mini-momentum-jump}
    [\pi_R]=\frac{D-2}{D-3}\,\frac{\lambda\zeta}{R},\qquad [\pi_\nu] =-\frac{\lambda\zeta}{D-3},\qquad [\pi_{\tilde\alpha}]=0.
\end{equation}
In particular, the electric flux is continuous across the brane. These are the junction conditions \eqref{eq:timelike-metric-junction}--\eqref{eq:axion-flux-continuity} restricted to the minisuperspace, and the corresponding saddles are the constrained two-boundary wormholes of Section~\ref{sec:2bdry-saddles}.

To describe them explicitly, we will now parametrize the worldline with the radial coordinate $r$. The asymptotic end of the worldline is at the cutoff surface $r=r_c$, and the constraint brane is located at $r = r_0$. Adding the boundary term \eqref{eq:mini-IN} to the action \eqref{eq:mini-action} and integrating out $\tilde\alpha$, we find that the action of half of the worldline is
\begin{equation}\label{eq:mini-IQ}
    I_Q=-\frac{(D-2)\,\Omega_{D-2}}{16\pi G_D}\int_{r_0}^{r_c}\dd r\left[\frac{r^{D-4}\,w'}{\CN}+(D-3)\left(1-\frac{\mathcal Q^2}{r^{2(D-3)}}\right)\,\CN\right],
\end{equation}
where $w=r^{D-3}e^{2\nu}=Re^\nu$, primes denote $\dd/\dd r$, and $\cQ^2=\frac{8\pi G_DQ^2}{(D-2)(D-3)\,\Omega_{D-2}^2}$, where $Q$ is given by \eqref{eq:mini-p-alpha}. Its saddles are the asymptotically flat Reissner--Nordstr\"om exteriors or equivalently the AdS-RN saddles in \eqref{eq:RN-metric} with AdS length $\ell\to\infty$. The two halves of the constrained wormhole, $i=L,R$, with periods $\beta_L$ and $\beta_R$, are
\begin{equation}\label{eq:mini-RN}
    \CN=\beta_i\,r^{D-4},\qquad w=\beta_i^2r^{D-3}f_i(r),\qquad f_i(r)=1-\frac{2M_i}{r^{D-3}}+\frac{\cQ^2}{r^{2(D-3)}},
\end{equation}
where the masses $M_i$, the radius $r_0$ and the brane tension are fixed by
\begin{align}
    &\beta_i^2f_i(r_0)=\frac{\zeta^2}{\Omega_{D-2}^2r_0^{2(D-2)}},\label{eq:mini-neck}\\
    &\lambda=-\frac{(D-2)(\beta_L+\beta_R)\,\zeta}{8\pi G_D\Omega_{D-2}\beta_L\beta_R\,r_0^{D-1}},\qquad \frac{\zeta^2}{\Omega_{D-2}^2\beta_L\beta_R}=\frac{D-3}{D-1}\,r_0^2\left(r_0^{2(D-3)}-\cQ^2\right)\label{eq:mini-tension},
\end{align}
where the first equation follows from continuity and is valid off-shell and the second line is valid at the saddle point from junction conditions on momenta \eqref{eq:mini-momentum-jump}. These equations agree with \eqref{eq:2bdry_tautau_continuity}, \eqref{eq:2bdry_wormhole_masses_eq} and \eqref{eq:2bdry_tension}.

\paragraph{Stability of the constrained wormholes.} We now show that fluctuations around the constrained two-boundary wormholes have a positive quadratic action on the integration contour inherited from the Lorentzian path integral. We follow the approach of~\cite{Marolf2022thermal}, in which $\operatorname{Tr}e^{-\beta H}$ is computed by a real-time path integral whose saddles are Euclidean black holes, and its extension to Einstein--Maxwell theory in~\cite{Chen:2025leq}. That is, we start from the Lorentzian path integral, deform its contour to pick up the Euclidean constrained saddles, and then study fluctuations about these saddles along the original Lorentzian contour. To this end, we expand the variables around their saddle values in terms of Lorentzian fluctuations. By \eqref{ga:thermal-rotation},
\begin{equation}\label{eq:mini-lorentzian-fluctuations}
    w=w_\star-\delta w_{\rm L},\qquad \CN=\CN_\star+\i\,\delta\CN_{\rm L},\qquad r_0=r_{0\star}+\delta r_{0,\rm L},
\end{equation}
where the saddle-point values $w_\star$ and $\CN_\star$ are given by \eqref{eq:mini-RN}, $r_{0\star}$ by \eqref{eq:mini-neck}. We set the Lorentzian fluctuations $\delta w_{\rm L}$, $\delta\CN_{\rm L}$ and $\delta r_{0,\rm L}$ to be real. The value of $\zeta$ is fixed, and the lapse $\CN$ is integrated over on each half separately, including its proper time.

We can decompose the fluctuations into ones that vanish on the constraint surface and ones that do not. Let us denote the fluctuations of the fields collectively by $\phi=(\delta w_{\rm L},\delta\CN_{\rm L})$, and write their quadratic action as
\begin{equation}\label{eq:mini-quadratic-operator}
    I^{(2)}[\phi]=\frac12\int_{r_0}^{r_c}\dd r\,\phi\,\mathcal O\phi+\text{boundary terms},
\end{equation}
where $\mathcal O$ is the symmetric operator of the linearized equations of motion. We write $\phi=\phi_0+\tilde\phi$, where $\phi_0$ satisfies $\mathcal O\phi_0=0$ away from the brane and carries fluctuations on the constraint surface, while $\tilde\phi$ vanishes at $r_0$. The mixed terms in $I^{(2)}[\phi_0+\tilde\phi]$ are $\int\dd r\,\tilde\phi\,\mathcal O\phi_0=0$ and, by definition, the boundary terms depend only on $\phi_0$. Hence,
\begin{equation}\label{eq:mini-decomposition}
    I^{(2)}[\phi_0+\tilde\phi]=I^{(2)}_{\rm constr}[\phi_0]+I^{(2)}_{\rm bulk}[\tilde\phi],
\end{equation}
so we can separately check stability against $\phi_0$ and $\tilde \phi$.

Expanding \eqref{eq:mini-IQ} around the saddle \eqref{eq:mini-RN} to second order in $\tilde\phi$, we find
\begin{align}
    I^{(2)}[\tilde\phi]=\frac{(D-2)\,\Omega_{D-2}}{16\pi G_D}\int_{r_0}^{r_c}\dd r\,r^{D-4}\left[\frac{w_\star'}{\CN_\star^3}\left(\delta\CN_{\rm L}-\frac{\i\,\CN_\star}{2w_\star'}\,\delta w_{\rm L}'\right)^2+\frac{(\delta w_{\rm L}')^2}{4\CN_\star w_\star'}\right],\label{eq:mini-IQ2}
\end{align}
where $\CN_\star=\beta_ir^{D-4}$ and
\begin{equation}\label{eq:mini-wprime}
    w_\star'=(D-3)\,\beta_i^2r^{D-4}\left(1-\frac{\cQ^2}{r^{2(D-3)}}\right)
\end{equation}
are positive, since $r\ge r_0$ and, by \eqref{eq:mini-tension}, $r_0^{D-3}>|\cQ|$. Hence, the integral over the real $\delta\CN_{\rm L}$ is a convergent Gaussian integral. Integrating out $\delta \CN_L$ gives us
\begin{equation}\label{eq:mini-bulk}
    I^{(2)}_{\rm bulk}=\sum_{i=L,R}\frac{(D-2)\,\Omega_{D-2}}{64\pi(D-3)\,G_D\beta_i^3}\int_{r_0}^{r_c}\dd r\,\frac{(\delta w_{{\rm L},i}')^2}{r^{D-4}\left(1-\cQ^2/r^{2(D-3)}\right)},
\end{equation}
where $\delta w_{{\rm L},i}$ is the fluctuation on the $i$-th half. The action \eqref{eq:mini-bulk} is positive-definite, so the bulk modes $\tilde{\phi}$ are stable.

We now turn to $\phi_0$. Since $\phi_0$ satisfies the linearized equations of motion, $I^{(2)}_{\rm constr}[\phi_0]$ is the second-order variation of the on-shell action of the two wormhole exteriors, as a function of their field data at the constraint brane. We therefore need to check stability against varying these data. At fixed $\mathcal Q$, they are the radius $r_0$ and the area $\cA=\Omega_{D-2}r_0^{D-2}e^{\nu(r_0)}$ of the brane. For given $(\cA,r_0)$, the classical solutions on the two halves are \eqref{eq:mini-RN}, with the mass parameters fixed by the continuity condition \eqref{eq:mini-neck}; we do not assume the conditions \eqref{eq:mini-tension}, as they are valid for on-shell boundary modes. Evaluating \eqref{eq:mini-IQ} on \eqref{eq:mini-RN}, we find
\begin{equation}\label{eq:mini-Iext}
    I_{\rm ext}(\cA,r_0)=\frac{(D-2)\,\Omega_{D-2}(\beta_L+\beta_R)}{16\pi G_D}\left[r_0^{D-3}+\frac{\cQ^2}{r_0^{D-3}}+\frac{\cA^2}{\Omega_{D-2}^2\beta_L\beta_R\,r_0^{D-1}}\right]+I_{\rm cutoff},
\end{equation}
where $I_{\rm cutoff}$ is a contribution of the cutoff surface, which diverges as $r_c\to\infty$ but is independent of $\cA$ and $r_0$, and can be canceled by a counterterm. The area is fixed by the constraint, $\cA=\zeta$, so we only vary $r_0$. At the saddle point, where the second relation in \eqref{eq:mini-tension} holds, we find
\begin{equation}\label{eq:mini-Iext2}
    \partial_{r_0}^2I_{\rm ext}(\zeta,r_0)=\frac{(D-2)(D-3)\,\Omega_{D-2}(\beta_L+\beta_R)}{8\pi G_D}\,r_0^{D-5}\left(D-2-\frac{\cQ^2}{r_0^{2(D-3)}}\right)>0.
\end{equation}
The inequality holds since $r_0^{2(D-3)}>\cQ^2$, so the term in parentheses is larger than $D-3$. Hence, $I^{(2)}_{\rm constr}[\phi_0]=\frac12\,\partial_{r_0}^2I_{\rm ext}\,(\delta r_{0,\rm L})^2$ is positive, and the integral over the real $\delta r_{0,\rm L}$ is convergent. Together with \eqref{eq:mini-bulk}, this shows that the constrained two-boundary wormholes are stable saddles of the constrained minisuperspace path integral, i.e., against static, spherically symmetric fluctuations.

\section{Picard--Lefschetz analysis for single-boundary saddles}\label{app:picard-lefschetz}

In this appendix, we determine which saddles contribute to the single-boundary integral \eqref{eq:1bdry-flat-integral} both in the canonical ensemble (with fixed electric charge) and in the grand canonical ensemble (with fixed electric potential). We also calculate $\kappa_1$ using saddle-point approximation of the charge sum \eqref{eq:1bdry-charge-sum}.

\subsection{Saddles in the canonical ensemble}

We want to analyze the saddle-points of the integral
\begin{align}
 Z_{\rm 1\text{-}bdry}(\beta,Q)&\simeq\int_\Gamma\dd r_+\,e^{-I(r_+)},
 \label{ta:canonical-integral}
\end{align}
where
\begin{equation}
    I(r_+)=\beta E(Q,r_+)-\frac{\Omega_{D-2}r_+^{D-2}}{4G_D},\qquad E(Q,r_+)=\frac{(D-2)\Omega_{D-2}}{16\pi G_D}\left(r_+^{D-3}+\frac{R_0^{2(D-3)}}{r_+^{D-3}}\right).
\label{ta:action}
\end{equation}
Here, $R_0$ is the horizon radius of the extremal black hole with charge $Q$, $R_0^{D-3}=|\mathcal Q|$. The contour $\Gamma$ starts at $r_+=0$ and approaches $\infty\cdot e^{-\i\pi/(D-2)}$. We will assume that $\Gamma$ lies inside the wedge
\begin{equation}
 W=\left\{r_+\;\middle|\; -\frac{\pi}{D-2}<\arg(r_+)<0\right\}.
 \label{ta:wedge}
\end{equation}

The saddle points are the points where the $r_+$-derivative of $I$ vanishes,
\begin{equation}
 4\pi r_+^{2D-5}-\beta(D-3)r_+^{2D-6}
             +\beta(D-3)R_0^{2(D-3)}=0,
 \label{ta:saddle-polynomial}
\end{equation}
which is equivalent to \eqref{eq:os-1bdry-beta}. Equation \eqref{ta:saddle-polynomial} has $2D-5$ solutions. For $\beta>\beta_{\min}(Q)$, see \eqref{eq:os-beta-min}, there are three real solutions, one negative, $\hat r_{<}$, and two positive,
\begin{equation}
 \hat r_{<}<0<R_0<r_s<r_\star<r_\ell,\qquad
 r_\star=R_0\,(2D-5)^{1/(2D-6)},
 \label{ta:real-roots}
\end{equation}
and $D-4$ pairs of complex solutions. $r_s,r_\ell$ are the flat-space limit of the small and intermediate RN--AdS BHs. When $\beta$ approaches $\beta_{\min}$ from above, $r_s$ and $r_\ell$ both approach $r_\star$. For $\beta<\beta_{\min}$, $r_s$ and $r_\ell$ become complex saddles conjugate to one another. In particular, at small $\beta$ with fixed $R_0$, the saddles satisfy
\begin{equation}
 4\pi r_+^{2D-5}\simeq-\beta(D-3)R_0^{2(D-3)},
 \label{ta:large-R0}
\end{equation}
so that
\begin{equation}
 r_+=\hat r_j\simeq
 \left(\frac{\beta(D-3)R_0^{2(D-3)}}{4\pi}\right)^{\!1/(2D-5)}
 e^{-\frac{\i\pi}{2D-5}(2j+1)},\qquad j=0,1,\ldots,2D-6 .
 \label{ta:root-asymptotics}
\end{equation}
Saddle $j=D-3$ is real and negative, and the other saddles form complex-conjugate pairs, $\hat r_j^*=\hat r_{2D-6-j}$. We can continue these saddles to all positive values of $\beta$. If two saddle points merge, not only the polynomial in \eqref{ta:saddle-polynomial}, but also its 1st derivative vanishes,
\begin{equation}
 r_+^{2D-7}\left[4\pi(2D-5)r_+-2(D-3)^2\beta\right]=0 .
 \label{ta:collision}
\end{equation}
For $\beta>0$, we see that the only collision occurs at $r_+=r_\star$, $\beta=\beta_{\min}$. Thus, only the saddles closest to the real axis, $\hat r_0$ and $\hat r_{2D-6}=\hat r_0^*$, can merge, and they merge at $\beta=\beta_{\min}$, $r=r_\star$. For $\beta>\beta_{\min}$, $\hat r_0$ and $(\hat r_0)^*$ become real, equal to $r_s$ and $r_\ell$.

Denote the steepest-descent contour corresponding to saddle $\hat r_j$ as
$\mathcal J_j$, and its dual as $\mathcal K_j$. We want to show that
\begin{equation}
 \Gamma=\mathcal J_0.
 \label{ta:decomposition}
\end{equation}

\paragraph{High temperatures.} For generic $R_0$ and $\beta < \beta_{\rm min}$, the actions of different saddles have different imaginary parts,
\begin{equation}
 \operatorname{Im}I(\hat r_j)\ne\operatorname{Im}I(\hat r_k)
 \qquad\text{for }j\ne k.
 \label{ta:distinct-phases}
\end{equation}
Since $\operatorname{Im}I$ is constant along $\mathcal J_j$ and $\mathcal K_j$, the dual thimbles $\mathcal K_{j\ne0}$ do not intersect $\mathcal J_0$. Hence, if \eqref{ta:decomposition} holds, the saddles with $j\ne0$ do not contribute.

Substituting $r_+=\rho e^{-\i\varphi}$ into \eqref{ta:saddle-polynomial} and separating the real and imaginary parts, we find
\begin{equation}
 \left(\frac{\rho}{R_0}\right)^{2(D-3)}
   =\frac{\sin[(2D-5)\varphi]}{\sin\varphi},
 \qquad
 \beta=\frac{4\pi\rho}{D-3}
       \frac{\sin[(2D-5)\varphi]}{\sin[2(D-3)\varphi]}.
 \label{ta:saddle-angle}
\end{equation}
In $W$, the first relation requires $\sin[(2D-5)\varphi]>0$. Hence, all saddles in $W$ lie in the narrower wedge $0<\varphi<\pi/(2D-5)$, where both right-hand sides of \eqref{ta:saddle-angle} are positive. As $\varphi$ increases from $0$ to $\pi/(2D-5)$, $\rho$ decreases monotonically from $r_\star$ to $0$, and $\beta$ decreases from $\beta_{\min}$ to $0$. The latter decrease is also monotonic, since along the solutions of \eqref{ta:saddle-polynomial}, $\dd\beta/\dd r_+$ vanishes only at
\begin{equation}
 r_+^{2(D-3)}=(2D-5)R_0^{2(D-3)},
 \label{ta:turning-point-condition}
\end{equation}
which has no solutions in the narrower wedge. Therefore, for $0<\beta<\beta_{\min}$, the only saddle in $W$ is $\hat r_0$. The action of this saddle has a positive imaginary part. Indeed, since $\partial I/\partial r_+$ vanishes at $\hat r_0$,
\begin{equation}
 \frac{\dd}{\dd\beta}\operatorname{Im}I(\hat r_0)
 =\operatorname{Im}E(Q,\hat r_0)
 =\frac{(D-2)\Omega_{D-2}}{16\pi G_D}\sin[(D-3)\varphi]
   \left(\frac{R_0^{2(D-3)}}{\rho^{D-3}}-\rho^{D-3}\right).
 \label{ta:imag-derivative}
\end{equation}
As $\beta$ increases from $0$ to $\beta_{\min}$, $\rho$ increases monotonically from $0$ to $r_\star>R_0$. Hence, $\operatorname{Im}I(\hat r_0)$ increases until $\rho=R_0$ and decreases afterwards. It vanishes at both ends: at $\beta=\beta_{\min}$, where $\hat r_0=r_\star$ is real, and at $\beta\to0$, where $\hat r_0\to0$ and $I(\hat r_0)\to0$. Therefore, $\operatorname{Im}I(\hat r_0)>0$ for $0<\beta<\beta_{\min}$.

Along $\mathcal J_0$, only the real part of $I$ changes, so that
\begin{equation}
 I(r_+)=I(\hat r_0)+t,\qquad t\geq0,
 \label{ta:inverse-ray}
\end{equation}
or, equivalently,
\begin{equation}
 4\pi r_+^{2D-5}-(D-2)\beta r_+^{2D-6}
 +\frac{16\pi G_D}{\Omega_{D-2}}\bigl(I(\hat r_0)+t\bigr)r_+^{D-3}
 -(D-2)\beta R_0^{2(D-3)}=0 .
 \label{ta:inverse-polynomial}
\end{equation}
As $t$ changes, the roots of \eqref{ta:inverse-polynomial} cannot cross the upper boundary of $W$, since $I$ is real on the positive real axis, while $\operatorname{Im}I(\hat r_0)>0$. On the lower boundary of $W$,
\begin{equation}
 \operatorname{Im}I\bigl(\rho e^{-\i\pi/(D-2)}\bigr)
 =\frac{(D-2)\Omega_{D-2}\beta}{16\pi G_D}\sin\!\left(\frac{\pi}{D-2}\right)
  \left(\frac{R_0^{2(D-3)}}{\rho^{D-3}}-\rho^{D-3}\right),
 \label{ta:lower-edge-phase}
\end{equation}
which decreases monotonically with $\rho$. Since $\dd r_+/\dd t=1/I'(r_+)$, a root on this boundary moves as
\begin{equation}
 \frac{\dd}{\dd t}\arg(r_+)=\operatorname{Im}\frac{1}{r_+ I'(r_+)}
 =-\frac{\partial_\rho\operatorname{Im}I\bigl(\rho e^{-\i\pi/(D-2)}\bigr)}
         {\rho\,|I'(r_+)|^2}>0,
 \label{ta:inward-flow}
\end{equation}
that is, into $W$. At $t=0$, the two branches of $\mathcal J_0$ start from the double root $r_+=\hat r_0$, which lies strictly inside $W$. Hence, they stay inside $W$ for all $t>0$. They cannot meet again or end on another saddle, since $\hat r_0$ is the only saddle in $W$. As $t\to+\infty$, exactly two roots of \eqref{ta:inverse-polynomial} lie in $W$,
\begin{align}
 r_+&=
 \left[\frac{(D-2)\Omega_{D-2}\beta R_0^{2(D-3)}}
  {16\pi G_D\bigl(I(\hat r_0)+t\bigr)}\right]^{\!1/(D-3)}\bigl[1+O(t^{-2})\bigr],\label{ta:small-inverse-root}\\
 r_+&=\left[\frac{4G_D\operatorname{Re}\bigl(I(\hat r_0)+t\bigr)}{\Omega_{D-2}}\right]^{\!1/(D-2)}
 e^{-\i\pi/(D-2)}+\frac{\beta}{4\pi}+O\bigl(t^{-1/(D-2)}\bigr).\label{ta:large-inverse-root}
\end{align}
The small root \eqref{ta:small-inverse-root} approaches the real axis from below, since $\operatorname{Im}I(\hat r_0)>0$, and the large root \eqref{ta:large-inverse-root} approaches the ray $\arg r_+=-\pi/(D-2)$ from above. Hence, one branch of $\mathcal J_0$ ends at $r_+=0$, and the other one goes to $\infty\cdot e^{-\i\pi/(D-2)}$. Thus, $\mathcal J_0$ has the same endpoints as $\Gamma$, and equation~\eqref{ta:decomposition} holds.

\paragraph{Low temperatures.}

For $\beta>\beta_{\min}$, there are two positive real saddles, $r_s$ and $r_\ell$, corresponding to the small and large flat space black holes. The steepest-descent contour for the $r_+ = r_s$ saddle, $\mathcal{J}_s$, runs along the real axis from $r_+ = 0$ through the small black hole at $r_s$ to the large black hole at $r_\ell$; the two saddles are connected by a Stokes line. Continuing the $\mathcal{J}_s$ path from $r_\ell$ along the steepest-descent path $\mathcal{J}_\ell$ to infinity at $\arg r_+ = -\pi/(D-2)$ gives a contour equivalent to $\Gamma$. Along this contour, the action $I(r_+)$ is minimized at $r_+ = r_s$, so the small black hole dominates and we can approximate $Z_{\rm 1\text{-}bdry}(\beta,Q)\simeq e^{-I(r_s)}$. 

For $\beta+\i0^+$, the saddles $r_s$ and $r_\ell$ acquire small negative and positive imaginary parts, respectively, and the argument for $\beta<\beta_{\min}$ applies. Therefore, the small black hole thimble $\mathcal{J}_s$ itself can be deformed to $\Gamma$. For $\beta-\i0^+$, $r_s$ moves into the upper half-plane and $r_\ell$ into the lower half-plane. Then $\mathcal{J}_s$ ends at $\infty\cdot e^{\i\pi/(D-2)}$, and the decomposition of $\Gamma$ contains both the small and the large black hole thimbles. 

\begin{figure}
    \centering
    \includegraphics[width=0.8\linewidth]{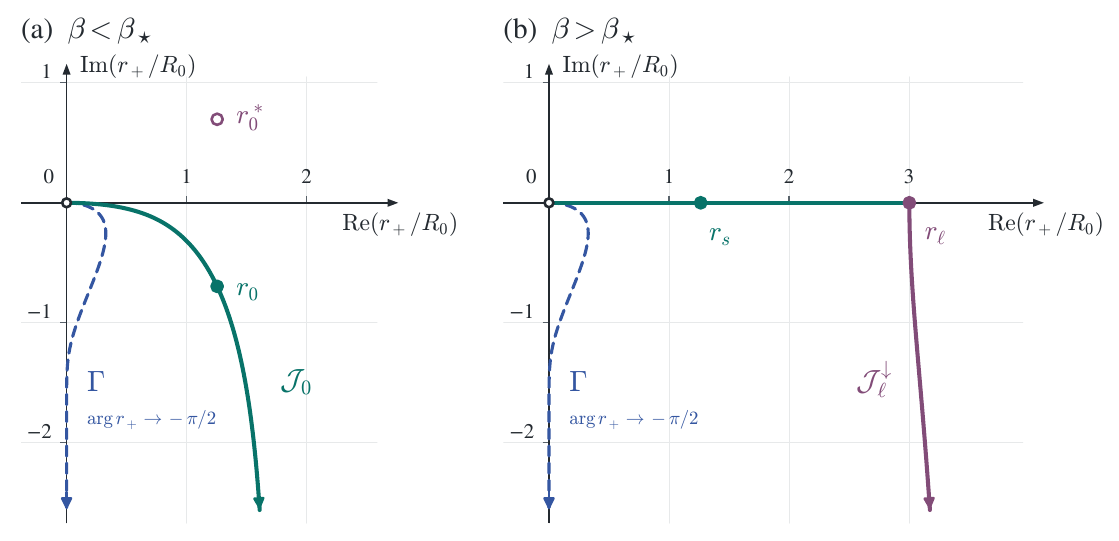}
    \caption{Lefschetz thimbles for the saddle points $\hat r_0$, $r_s$ and $r_\ell$ of the fixed-charge integral \eqref{ta:canonical-integral} in $D=4$. $\beta_\star\equiv\beta_{\min}(Q)$ is given by \eqref{eq:os-beta-min}. The dashed blue contour $\Gamma$ starts at $r_+=0$ along the positive real axis and approaches $\arg r_+=-\pi/2$ at infinity. (a) For $\beta=0.65\,\beta_\star$, $\Gamma$ can be deformed to the thimble $\mathcal J_0$ (green) of the saddle $\hat r_0$. The open purple circle marks the conjugate saddle $\hat r_0^*$, which does not contribute. (b) For $\beta=1.30\,\beta_\star$, the real saddles $r_s$ and $r_\ell$ are connected by a Stokes line. The thimble $\mathcal J_s$ (green) runs along the real axis from $r_+=0$ through $r_s$ to $r_\ell$. Its continuation along the lower branch $\mathcal J_\ell^\downarrow$ of the thimble of $r_\ell$ (purple) gives a contour equivalent to $\Gamma$.}
    \label{fig:pl-fixed-charge-contours}
\end{figure}

\subsection{Saddles in the grand canonical ensemble}
In this subsection we first evaluate the grand canonical partition function by summing over charges the small black hole saddles. Next, using Poisson summation we rewrite the sum over charges as a sum over imaginary shifts of the electric potential. We determine which saddles contribute to that sum using Picard--Lefschetz analysis and verify that it agrees with the saddle-point approximation for the sum over charges.

\paragraph{Sum over charges.}
The grand canonical partition function, or equivalently the first cumulant, is given by \eqref{eq:1bdry-charge-sum}. Approximating $Z_{\rm 1\text{-}bdry}(\beta,Q)$ by the small black hole saddles, we obtain, up to loop corrections,
\begin{equation}
\kappa_1\simeq1+\sum_{\substack{Q\in e\mathbb Z\\ Q\neq0}}e^{\beta\Phi Q-I_{\rm 1\text{-}bdry}(\beta,Q)}.
\label{ta:grand-charge-sum}
\end{equation}
Here, $1$ schematically denotes the hot flat space contribution, which arises as the zero-charge limit of the small black hole saddles, and $I_{\rm 1\text{-}bdry}(\beta,Q)$ is the action of the small black hole $r_s$ for $\beta>\beta_{\min}(Q)$, and of the complex saddle $\hat r_0$ for $\beta<\beta_{\min}(Q)$ (see Eqs.~\eqref{ta:real-roots} and \eqref{ta:decomposition}). The real part of $I_{\text{1-bdry}}(\beta,Q)$ grows faster than linearly with $|Q|$, and thus, the sum \eqref{eq:1bdry-charge-sum} converges for every $\Phi$. By charge conjugation, we can take $\Phi>0$ without loss of generality.

We now evaluate the sum \eqref{ta:grand-charge-sum} in the saddle-point approximation. Writing $Q=(D-3)\Omega_{D-2}\Phi_{\rm ext}R_0^{D-3}$ for $Q>0$, we use \eqref{ta:saddle-polynomial} to express the charge in terms of the radius of the saddle,
\begin{equation}
Q=(D-3)\Omega_{D-2}\Phi_{\rm ext}\,r_+^{D-3}\sqrt{1-\frac{4\pi r_+}{(D-3)\beta}},
\label{ta:charge-from-radius}
\end{equation}
where the square root equals $(R_0/r_+)^{D-3}$ and is continued from the small black hole. Since $\partial_{r_+}I(r_+)$ vanishes at the saddle, $\partial_QI_{\rm 1\text{-}bdry}(\beta,Q)=\bigl(\partial_Q+\frac{\partial r_+}{\partial Q}\,\partial_{r_+}\bigr)I(r_+)=\beta Q/[(D-3)\Omega_{D-2}r_+^{D-3}]$, where $I(r_+)$ is given by \eqref{ta:action}. Extremizing the absolute value of the summand in \eqref{ta:grand-charge-sum} with respect to $Q$, we obtain $\operatorname{Re}\partial_QI_{\rm 1\text{-}bdry}(\beta,Q)=\beta\Phi$. With $r_+=\rho e^{-\i\varphi}$, \eqref{ta:charge-from-radius} and \eqref{ta:saddle-angle} give
\begin{align}
\frac{\Phi}{\Phi_{\rm ext}}&=\operatorname{Re}\sqrt{1-\frac{4\pi r_+}{(D-3)\beta}}=\left(\frac{R_0}{\rho}\right)^{D-3}\cos[(D-3)\varphi]=\cos[(D-3)\varphi]\sqrt{\frac{\sin\varphi}{\sin[(2D-5)\varphi]}},
\label{ta:charge-peak-radius}
\end{align}
with $0<\varphi<\frac{\pi}{2D-5}$. Using \eqref{ta:saddle-polynomial} and \eqref{ta:charge-peak-radius}, we find that the absolute value of the summand at this stationary point
\begin{equation}
\left|e^{\beta\Phi Q_*-I_{\rm 1\text{-}bdry}(\beta,Q_*)}\right|=\exp\left[-\frac{\Omega_{D-2}}{4(D-3)G_D}\operatorname{Re}\bigl(r_+^{D-2}\bigr)\right].
\label{ta:summed-rate}
\end{equation}

The charged saddles compete with the hot flat space contribution when the exponent in \eqref{ta:summed-rate} becomes positive. For real $r_+$, the exponent is negative, and the hot flat space dominates the path integral. For $r_+=\rho e^{-\i\varphi}$, the sign of the exponent is that of $-\cos[(D-2)\varphi]$, so it vanishes at $\varphi=\pi/[2(D-2)]$, which lies in the range $0<\varphi<\pi/(2D-5)$, and it is positive for larger $\varphi$. At $\varphi=\pi/[2(D-2)]$, equation \eqref{ta:charge-peak-radius} gives us
\begin{equation}
\Phi|_{\varphi = \pi/[2(D-2)]}=\Phi_{\rm ext}\sin(\frac{\pi}{2(D-2)}).
\label{ta:charge-threshold}
\end{equation}
Hence, below the threshold \eqref{ta:charge-threshold}, $\kappa_1$ is of order one and dominated by the flat space contribution. Thus, above the threshold, $\kappa_1$ is dominated by the charges near the stationary point. From \eqref{ta:saddle-angle} and \eqref{ta:charge-peak-radius}, we see that $r_+=\rho e^{-\i\varphi}$ is proportional to $\beta$ and $\varphi$ is determined by $\Phi/\Phi_{\rm ext}$. Therefore, by \eqref{ta:summed-rate}, $\log\kappa_1$ grows as $\beta^{D-2}/G_D=(\beta/\lpl)^{D-2}$.

\paragraph{Picard--Lefschetz in the grand canonical ensemble.}
Using the Poisson resummation, we can rewrite the sum over charges \eqref{eq:1bdry-charge-sum} as
\begin{align}
\kappa_1&\simeq \sum_{n\in\mathbb Z}\int_\Gamma\dd r_+\,e^{-I_{\Phi_n}(r_+)},\label{ta:image-sum}\\
I_{\Phi_n}(r_+)&=\frac{\Omega_{D-2}}{4G_D}\left[\,\frac{(D-2)\beta}{4\pi}\left(1-\frac{\Phi_n^2}{\Phi_{\rm ext}^2}\right)\,r_+^{D-3}-r_+^{D-2}\right],\label{ta:dirichlet-polynomial}
\end{align}
where $\Phi_n = \Phi + \frac{2\pi \i n}{\beta e}$. In \eqref{ta:image-sum} we dropped the measure factors of the $r_+$ integral.

For fixed $n$, the $r_+$ integral has only one nonzero saddle point,
\begin{equation}
    r_+ = \tilde r_n = \frac{(D-3)\beta}{4\pi}\left(1-\frac{\Phi_n^2}{\Phi_{\rm ext}^2}\right).\label{eq:saddle-pt-grand-can-appF}
\end{equation}
This saddle-point corresponds to a Reissner--Nordstr\"om black hole with electric potential $\Phi_n$. In the gauge in which $A_\tau\to\i\beta\Phi$ at the asymptotic boundary, $e A_\tau=2\pi n$ at the horizon, so the holonomy angle around the thermal circle at the horizon is $2\pi n$. The electric charge of these black holes is
\begin{equation}
    Q_n = (D-3)\,\Omega_{D-2}\,\Phi_n\,\bigl(\tilde r_n\bigr)^{D-3},
    \label{ta:image-charge}
\end{equation}
and their on-shell action is
\begin{equation}
    I_n\equiv I_{\Phi_n}(\tilde r_n) = \frac{\Omega_{D-2}}{4(D-3)G_D}\bigl(\tilde r_n\bigr)^{D-2}.
    \label{ta:image-action}
\end{equation}

Below, we use the Picard--Lefschetz analysis to determine which of these shifted saddles contribute. The analogous analysis in AdS, where the contour runs along the real $r_+$ axis, was carried out by Kolanowski and Marolf \cite{KM2026}.

We can deform $\Gamma$ to the ray $\gamma = \{r_+=\rho\,e^{-\i\pi/(D-2)} \;|\;0\leq\rho<\infty\}$. Consequently, saddle $\tilde r_n$ will be picked up by the $r_+$-integral only if the intersection number of $\gamma$ with the dual Lefschetz thimble $\mathcal{K}_n$ is nonzero, $\langle\Gamma,\mathcal{K}_n\rangle\neq 0$. Along $\mathcal K_n$, $\operatorname{Im}I_{\Phi_n}$ is constant and $\operatorname{Re}I_{\Phi_n}$ decreases. Thus, we can parametrize $\mathcal{K}_n$ with $t < 0$ such that 
\begin{equation}
    I_{\Phi_n}(r_+)=I_n+t.
\end{equation}
In terms of $u=r_+/\tilde r_n$, we can recast this equation as
\begin{equation}
u^{D-3}\bigl[(D-2)-(D-3)u\bigr]=1+\frac{t}{I_n}.
\label{ta:dirichlet-saddle-flow}
\end{equation}
The two branches of $\mathcal K_n$ leave the double root $u=1$ as
\begin{equation}
r_+(t)=\tilde r_n\pm\sqrt{\frac{8G_D|t|}{(D-2)\Omega_{D-2}\,\bigl(\tilde r_n\bigr)^{D-4}}}+\mathcal{O}(|t|),\qquad t\to0^-,
\label{ta:dirichlet-flow-roots}
\end{equation}
and at large negative $t$,
\begin{equation}
    r_+^{D-2}\overset{t\to -\infty}{\sim}4G_D|t|/\Omega_{D-2}\label{ta:dual-thimble-roots-grand-can}
\end{equation}
so their ends approach rays with angles $2\pi j/(D-2)$, $j\in\mathbb Z$. 

To determine which two solutions of \eqref{ta:dual-thimble-roots-grand-can} are continuously connected to the two roots \eqref{ta:dirichlet-flow-roots}, let $k$ be the integer such that $2\pi(k-1)/(D-2)<\arg \tilde r_n<2\pi k/(D-2)$. This covers all $\tilde r_n\neq0$ except those on the rays $\arg \tilde r_n=2\pi k/(D-2)$, which we discuss below. As $t$ decreases, the right-hand side of \eqref{ta:dirichlet-saddle-flow} moves along a straight line from $1$ in the direction $-1/I_n$. Since $I_n$ is proportional to $\bigl(\tilde r_n\bigr)^{D-2}$ with a positive coefficient, its argument starts at zero and tends to $(2k-1)\pi-(D-2)\arg \tilde r_n\in(-\pi,\pi)$. For non-real $I_n$, this line stays off the real axis at $t<0$, so the two roots that leave $u=1$ stay in opposite half-planes of $u$. The same holds for real $I_n<0$: the two roots are then complex conjugate and could meet only at a multiple root, but the derivative of the left-hand side of \eqref{ta:dirichlet-saddle-flow}, $(D-2)(D-3)u^{D-4}(1-u)$, excludes such a collision at $t<0$. On the branch of $\mathcal{K}_n$ with $\operatorname{Im}u>0$, the factor $(D-2)-(D-3)u$, whose imaginary part is $-(D-3)\operatorname{Im}u$, starts at $1$ and stays in the lower half-plane, so its argument approaches $\arg u-\pi$ as $|u|\to\infty$. On the branch with $\operatorname{Im}u < 0$, it approaches $\arg u+\pi$. Equating the arguments of both sides of \eqref{ta:dirichlet-saddle-flow} as $t\to-\infty$ gives $\arg u\to2\pi k/(D-2)-\arg \tilde r_n$ on the branch with $\operatorname{Im}u>0$ and $\arg u\to2\pi(k-1)/(D-2)-\arg \tilde r_n$ on the branch with $\operatorname{Im}u<0$. Therefore,
\begin{equation}
\lim_{t\to-\infty}\arg r_+(t)=
\begin{cases}
2\pi k/(D-2), & \operatorname{Im}u>0,\\
2\pi(k-1)/(D-2), & \operatorname{Im}u<0.
\end{cases}
\label{ta:dual-end-angles}
\end{equation}
We orient $\mathcal K_n$ from the $\operatorname{Im}u<0$ branch through $\tilde r_n$ to the $\operatorname{Im}u > 0$ branch, and count a counterclockwise crossing of the outward ray $\gamma$ as $+1$ and a clockwise crossing as $-1$. The net count is one if the direction $-\pi/(D-2)$ of $\gamma$ lies between the two end angles in \eqref{ta:dual-end-angles}, modulo $2\pi$, and zero otherwise. Hence, in the open sectors,
\begin{equation}
\langle\Gamma,\mathcal K_n\rangle=
\begin{cases}
1, & -2\pi/(D-2)<\arg \tilde r_n<0,\\
0, & \text{otherwise},
\end{cases}
\qquad -\pi<\arg \tilde r_n\leq\pi.
\label{ta:dirichlet-index}
\end{equation}
For $\Phi>0$, \eqref{ta:dirichlet-index} selects all $n>0$ in $D=4$, and, in $D>4$, those $n>0$ for which $\Phi_{\rm ext}^2-\Phi^2+(\frac{2\pi n}{\beta e})^2>\frac{4\pi n}{\beta e}\,\Phi\cot(\frac{2\pi}{D-2})$.

For $\arg \tilde r_n=2\pi k/(D-2)$, where $I_n>0$, the dual thimble reaches the endpoint $r_+=0$: as $t$ decreases from $0$ to $-I_n$, the real root with $0<u<1$ moves from the saddle to the endpoint. The other branch, with $u>1$, runs to infinity as $t\to-\infty$, so $\mathcal K_n$ is the ray $\arg r_+=2\pi k/(D-2)$. Since this ray meets $\gamma$ only at the endpoint $r_+=0$, the saddle $\tilde r_n$ does not contribute.

One can verify that the sum over $n\in \mathbb{Z}$ of the contributing saddles at fixed potential does match the sum over charges $Q\in e\mathbb{Z}$ of the small black hole saddles at fixed charge, as expected from Poisson summation.

\begin{figure}[H]
    \centering
    \includegraphics[width=0.8\linewidth]{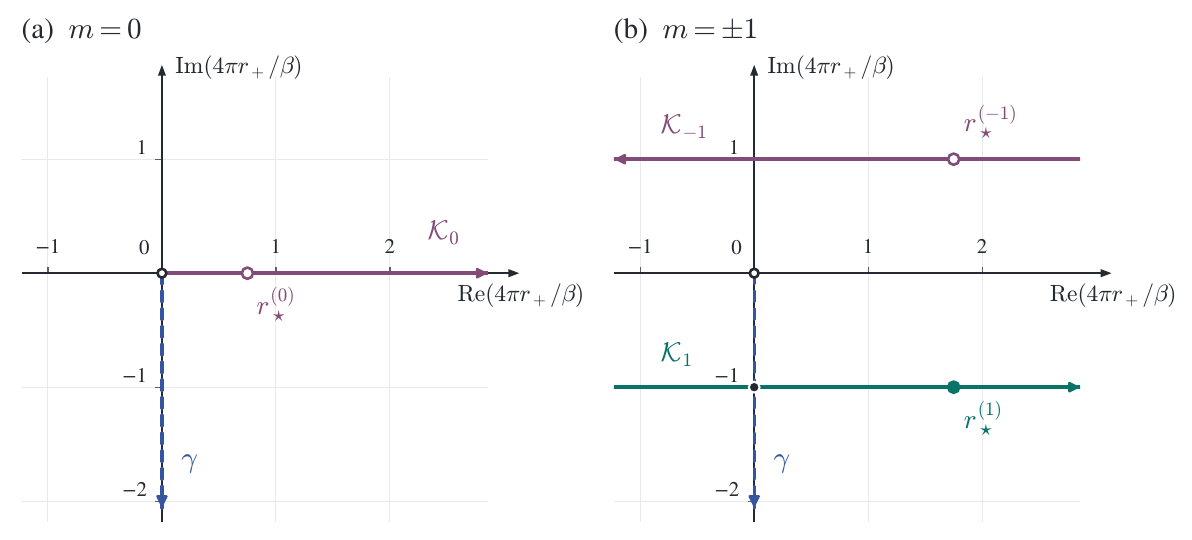}
    \caption{Dual thimbles of the saddle-point \eqref{eq:saddle-pt-grand-can-appF} in \(D=4\). The integer \(n\) labels the shifts \(\Phi_n=\Phi+2\pi i n/(\beta e)\). Here \(\Phi=\Phi_{\rm ext}/2\) and \(2\pi/(\beta e)=\Phi_{\rm ext}\). The dashed blue contour \(\gamma\) runs from \(r_+=0\) down the negative imaginary axis. In (a), the purple dual thimble \(\mathcal K_0\) through the unshifted saddle \(r_\star^{(0)}\) meets \(\gamma\) only at the endpoint \(r_+=0\). In (b), the black dot at \(4\pi r_+/\beta=-i\) marks the crossing of the green dual thimble \(\mathcal K_1\) with \(\gamma\). Its intersection number is \(+1\), so \(r_\star^{(1)}\) contributes. The purple dual thimble \(\mathcal K_{-1}\) lies above the real axis so that $\langle\gamma,\mathcal{K}_{-1}\rangle = 0$.}
    \label{fig:pl-grand-canonical-duals}
\end{figure}

\bibliographystyle{utphys2}
{\small \bibliography{Biblio}{}}

\end{document}